\documentclass{aa}
\usepackage{rotating}
\usepackage{graphicx}
\usepackage{txfonts}
\usepackage{natbib}

\begin{document}

\title{Intriguing X-ray and optical variations of the $\gamma$~Cas analog HD~45314\thanks{Based on observations collected with {\it XMM-Newton}, an ESA Science Mission with instruments and contributions directly funded by ESA Member States and the USA (NASA), and with the TIGRE telescope (La Luz, Mexico).}}
\author{G.\ Rauw\inst{1} \and Y.\ Naz\'e\inst{1}\fnmsep\thanks{Research Associate FRS-FNRS (Belgium).} \and M.A.\ Smith\inst{2} \and A.S.\ Miroshnichenko\inst{3} \and J.\ Guarro Fl\'o\inst{4} \and F.\ Campos\inst{5} \and P.\ Prendergast\inst{6} \and S.\ Danford\inst{3} \and J.N. Gonz\'alez-P\'erez\inst{7} \and A.\ Hempelmann\inst{7} \and M.\ Mittag\inst{7} \and J.H.M.M.\ Schmitt\inst{7} \and K.-P.\ Schr\"oder\inst{8} \and S.V.\ Zharikov\inst{9}}

\offprints{G.\ Rauw}
\mail{rauw@astro.ulg.ac.be}
\institute{Space sciences, Technologies and Astrophysics Research (STAR) Institute, Universit\'e de Li\`ege, All\'ee du 6 Ao\^ut, 19c, B\^at B5c, 4000 Li\`ege, Belgium \and National Optical Astronomical Observatory, 950 N.\ Cherry Ave., Tucson, AZ, USA
\and Department of Physics and Astronomy, University of North Carolina at Greensboro, Greensboro, NC 27412, USA
\and Balmes, 2, 08784 Piera, Barcelona, Spain
\and Observatori Puig d'Agulles, Passatge Bosc 1,E-08759 Vallirana, Barcelona, Spain
\and Kernersville Observatory, Kernersville, NC, USA
\and Hamburger Sternwarte, Universit\"at Hamburg, Germany
\and Departamento de Astronom\'{\i}a, Universidad de Guanajuato, Guanajuato, Mexico
\and Instituto de Astronom\'{\i}a, Universidad Nacional Aut\'onoma de M\'exico, Ensenada, 22800, Baja California, Mexico}
\date{Received date / Accepted date}
\abstract{A growing number of Be and Oe stars, named the $\gamma$~Cas stars, are known for their unusually hard and intense X-ray emission. This emission could either trace accretion by a compact companion or magnetic interaction between the star and its decretion disk.}{To test these scenarios, we carried out a detailed optical monitoring of HD~45314, the hottest member of the class of $\gamma$~Cas stars, along with dedicated X-ray observations on specific dates.}{High-resolution optical spectra were taken to monitor the emission lines formed in the disk, while X-ray spectroscopy was obtained at epochs when the optical spectrum of the Oe star was displaying peculiar properties.}{Over the last four years, HD~45314 has entered a phase of spectacular variations. The optical emission lines have undergone important morphology and intensity changes including transitions between single- and multiple-peaked emission lines as well as shell events, and phases of (partial) disk dissipation. Photometric variations are found to be anti-correlated with the equivalent width of the H$\alpha$ emission. Whilst the star preserved its hard and bright X-ray emission during the shell phase, the X-ray spectrum during the phase of (partial) disk dissipation was significantly softer and weaker.}{The observed behaviour of HD~45314 suggests a direct association between the level of X-ray emission and the amount of material simultaneously present in the Oe disk as expected in the magnetic star-disk interaction scenario.}
\authorrunning{G. Rauw et al.}
\titlerunning{HD~45314}
\maketitle
\section{Introduction}
Classical Be stars are non-supergiant B stars that have at least once displayed Balmer emission lines in their spectrum. Be stars are very rapidly rotating and often non-radially pulsating B stars surrounded by a Keplerian decretion disk \citep{Rivinius,Silaj14a}. Besides their fast, near critical rotation, most of their fundamental properties (effective temperature, gravity, chemical composition) are rather normal for their spectral types. Overall, about 20\% of the B stars in our Galaxy belong to this category. The incidence of the Be phenomenon is strongly peaked around spectral type B1-2 \cite[e.g.][]{Rivinius} and decreases sharply for spectral types earlier than B0. Long-term changes (e.g.\ transitions from a normal B-star phase to a Be phase) have been observed and are interpreted in terms of formation or dissipation of the disk. Changes in the emission line morphology are likewise attributed to changes in the structure of the disk \citep[e.g.][]{Silaj14a,Okazaki16,Rivinius}. 

Objects classified as Oe stars are considered to form an extension of the Be phenomenon into the temperature range of O stars \citep{CL,NSB,Sota}. These stars display emissions of H\,{\sc i}, He\,{\sc i,} and Fe\,{\sc ii}, but lack the emissions of He\,{\sc ii} $\lambda$\,4686 and N\,{\sc iii} $\lambda\lambda$\,4634 -- 4640 that are seen in earlier and more luminous Of stars. The fraction of O stars that belong to the Oe category is quite low in the Milky Way ($0.03 \pm 0.01$), but much higher in the Small Magellanic Cloud \citep[$0.26 \pm 0.04$,][]{GoldenMarx}.\\ 

In our Galaxy, about a dozen Be or Oe stars were found to belong to the so-called category of $\gamma$~Cas stars \citep[for a recent review see][]{gamCasrev}. The currently known $\gamma$~Cas stars have spectral types O9.7-B1.5e\,III-V. 
Their distinguishing characteristics concern their X-ray emission: they display an X-ray luminosity of $10^{32}$ -- $10^{33}$\,erg\,s$^{-1}$, about a factor of ten higher than those of normal stars of same spectral type, but significantly smaller than those of Be/X-ray binaries. The bulk of their X-ray emission \citep[more than 80\%,][]{gamCasrev} arises from a hot (kT $\geq 10$\,keV) thermal, optically thin plasma. The X-ray spectrum displays Fe\,{\sc xxv} and Fe\,{\sc xxvi} emission lines produced in this hot plasma, as well as fluorescent Fe K emission lines arising in a cooler, less ionized, medium \citep{gamCasrev}. X-ray variability occurs on various timescales: bursts (so-called shots) with durations of seconds to about a minute, variations of the basal flux that last a few hours, and long-term variations occurring on timescales of months to years \citep[e.g.][]{Robinson,Motch}. 

The origin of the $\gamma$~Cas phenomenon remains unclear. The most likely scenarios either imply accretion by a compact neutron star \citep[e.g.][]{White,Postnov1}, or white dwarf companion \citep[e.g.][]{Murakami,Kenji}, or a magnetic star-disk interaction \citep[e.g.][]{Smith98,Robinson,Motch}. Coordinated optical and X-ray observations of $\gamma$~Cas stars could help us gain further insight into the origin of the phenomenon. 

While simultaneous optical and X-ray variations have been reported for a few $\gamma$~Cas stars, observations have up to now sampled a relatively narrow range of disk strength. In particular, no campaign was ever obtained as the disk was dissipating, though this would provide a most critical test for any scenario aiming at explaining the $\gamma$~Cas phenomenon. In this context, we discuss here a new set of observations of the Oe star HD~45314 (= PZ~Gem), which was previously found to be a member of the class of $\gamma$~Cas stars \citep{Oeletter}. With an O9e -- O9.7e spectral type, HD~45314 has the earliest spectral type among the currently-known sample of $\gamma$~Cas stars.\footnote{Another $\gamma$~Cas star, HD~119682 has been classified as O9.7e or B0e \citep[see the discussion in][]{Rakowski}.} In previous optical studies of this star \citep{ibvs,Oepaper}, we found that the Oe emission lines undergo strong variations including several episodes of disk outbursts when the equivalent width (EW) of the H$\alpha$ line increased by a factor of two. In the present paper, we discuss the spectral changes undergone by HD~45314 over the past four years (2014 -- 2017), during which (partial) disk dissipation occurred. This paper is organized as follows. Section\,\ref{obs} presents our optical and X-ray observations. The X-ray spectra are analysed in Sect.\,\ref{Xspec}, whereas the radial velocities and line profile variations, measured on the optical data, are discussed in Sects.\,\ref{deltaRV} and \ref{varopt}, respectively. We discuss our findings in Sect.\,\ref{discuss} and present our conclusions in Sect.\,\ref{conclusion}. 

\section{Observations \label{obs}}
\subsection{X-ray observations}
HD~45314 was observed twice with {\it XMM-Newton} \citep{Jansen}. The first observation took place in April 2012 (ObsID 0670080301), when the star was in an intermediate emission state (EW(H$\alpha$) = $-22.3$\,\AA). This observation led to the discovery of the $\gamma$~Cas nature of HD~45314 \citep{Oeletter}. In March 2016, when the level of the Be-like emission lines reached a historical minimum, we triggered a target of opportunity (ToO) observation (ObsID 0760220601, see Table\,\ref{journal}).
For both observations, the European Photon Imaging Camera \citep[EPIC,][]{MOS,pn} instruments were operated in full frame mode and used with the thick filter to reject optical and UV photons. The data were processed with the Science Analysis System (SAS) software version 15.0. Both observations were affected by short background flares at the beginning of the exposures. These intervals were discarded from our subsequent analysis.

HD~45314 is clearly detected in both observations, but with very different count rates (see Table\,\ref{journal}). We used the {\it especget} routine to extract the X-ray spectra of HD~45314 over a circular region of radius 30\arcsec\ centred on the Simbad coordinates of the star. The background was extracted over a nearby source-free region of the same radius. Dedicated Auxiliary Response File (ARF) and Redistribution Matrix File (RMF) response files were also generated. The EPIC spectra from both epochs were grouped with the SAS command {\it specgroup} to obtain an oversampling factor of five and to ensure that at least a signal-to-noise ratio of three was reached in each spectral bin of the background-corrected spectra.
Given the hardness of the source and its moderate count rate, the Reflection Grating Spectrometer \citep[RGS,][]{RGS} did not provide useful data for HD~45314 in either of the two observations. 

\begin{figure*}[htb]
\begin{minipage}{8.5cm}
\begin{center}
\resizebox{8.5cm}{!}{\includegraphics{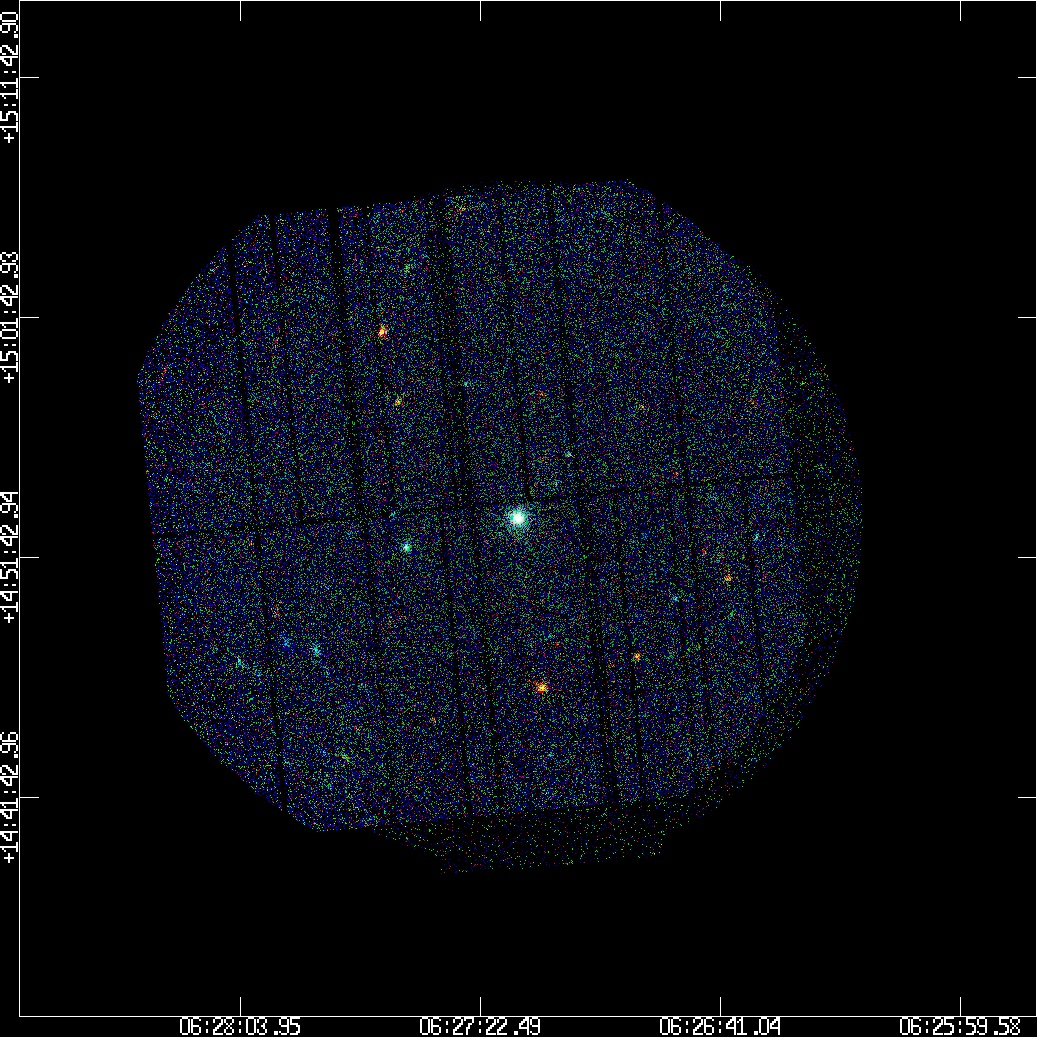}}
\end{center}
\end{minipage}
\hfill
\begin{minipage}{8.5cm}
\begin{center}
\resizebox{8.5cm}{!}{\includegraphics{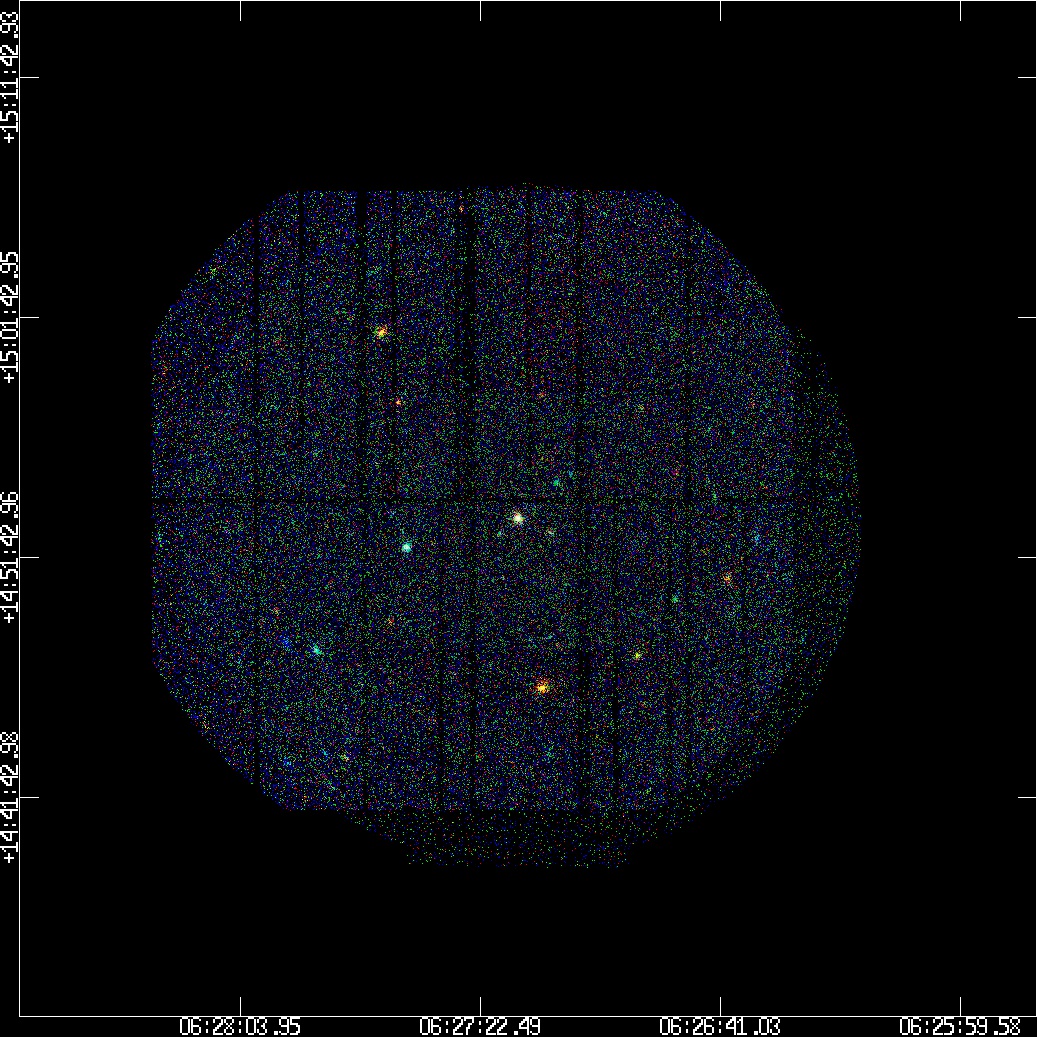}}
\end{center}
\end{minipage}
\caption{Energy-coded three-colour image of the {\it XMM-Newton} field of view around HD~45314. The left and right panels correspond to the observations of April 2012 and March 2016, respectively. HD~45314, the object in the centre of each panel, appears bluer and brighter in the April 2012 observation. For each epoch, the data from the three EPIC instruments were combined and exposure corrected. Red, green, and blue colours correspond to photon energies in the ranges $[0.5,1.0]$, $[1.0,2.0]$, and $[2.0,8.0]$\,keV, respectively. \label{3colour}}
\end{figure*}

\begin{table*}
\caption{Journal of the X-ray observations discussed in this paper.\label{journal}}
\begin{center}
\begin{tabular}{c c c c c c c }
\hline
\multicolumn{7}{c}{{\it XMM-Newton}}\\
\hline
Epoch & Date       & EW(H$\alpha$) & Eff.\ exposure & MOS1 & MOS2 & pn\\
      & HJD-2\,450\,000  &    (\AA)      & (ks) & (cts\,s$^{-1}$) & (cts\,s$^{-1}$) & (cts\,s$^{-1}$)\\
\hline 
April 2012 & 6031.941 & $-22.6$   & 21.1 & $0.075 \pm 0.002$ & $0.077 \pm 0.002$ & $0.225 \pm 0.004$ \\
March 2016 & 7455.896 &  $-8.5$   & 23.1 & $0.011 \pm 0.001$ & $0.010 \pm 0.001$ & $0.037 \pm 0.002$ \\
\hline
\end{tabular}
\begin{tabular}{c c c c c c c}
\multicolumn{7}{c}{{\it Suzaku}}\\
\hline
Epoch & Date       & EW(H$\alpha$) & Exposure & XIS0 & XIS1 & XIS3 \\
      & HJD-2\,450\,000  &    (\AA)& (ks)     & (cts\,s$^{-1}$) & (cts\,s$^{-1}$) & (cts\,s$^{-1}$)\\
\hline 
October 2014 & 6937.217 & $-7.9$   & 82.1 & $0.024 \pm 0.001$ & $0.021 \pm 0.002$ & $0.029 \pm 0.001$ \\
\hline
\end{tabular}
\tablefoot{The heliocentric Julian days are given at mid-exposure. The EW of the nearest-in-time H$\alpha$ measurement is given for each observation. In the top part of the table, Column 4 provides the effective exposure time of the EPIC-pn camera after discarding high background episodes. The net count rates for the EPIC instruments are given over the 0.4--10\,keV energy band. These count rates were obtained with the {\it edetectchain} command and correspond to equivalent on-axis count rates over the full PSF. In the lower part of the table, the net XIS count rates are given inside the source extraction region and over the full energy band of the XIS detectors.}
\end{center}
\end{table*}
Figure\,\ref{3colour} illustrates an energy-coded three-colour image of the field of view around HD~45314 as seen with {\it XMM-Newton} at the two epochs. The hardness of HD~45314's X-ray emission is clearly apparent from its blue colour. One also sees that the source was much fainter and softer in March 2016 than in April 2012. 

We further extracted and analysed an archival observation of HD~45314 obtained with the {\it Suzaku} X-ray satellite \citep{Suzaku}. This observation was performed in October 2014, when HD~45314 was in a shell phase (see below). The data taken with the X-ray Imaging Spectrometer \citep[XIS, see][]{XIS} detectors XIS0, XIS1, and XIS3 were processed using the { xselect} v2.4c interface of the { heasoft-6.18} software package. The spectra of the source were extracted over a circular area with a radius of 208\,arcsec. For the front-illuminated charge-coupled devices (CCDs) XIS0 and XIS3, the background spectrum was extracted over a nearby source-free box-shaped region. For the back-illuminated CCD (XIS1), the background was significantly higher and displayed more structures. For this instrument, we thus used instead an annular region (inner and outer radii of 260 and 312 arcsec). The XIS spectra were grouped in such a way as to have at least 25 counts per spectral bin.
To the best of our knowledge, HD~45314 is the only $\gamma$~Cas star that has been observed in X-rays while it was undergoing substantial changes of its optical disk emission.
 
\begin{figure*}[thb]
\begin{minipage}{8.5cm}
\begin{center}
\resizebox{8.5cm}{!}{\includegraphics{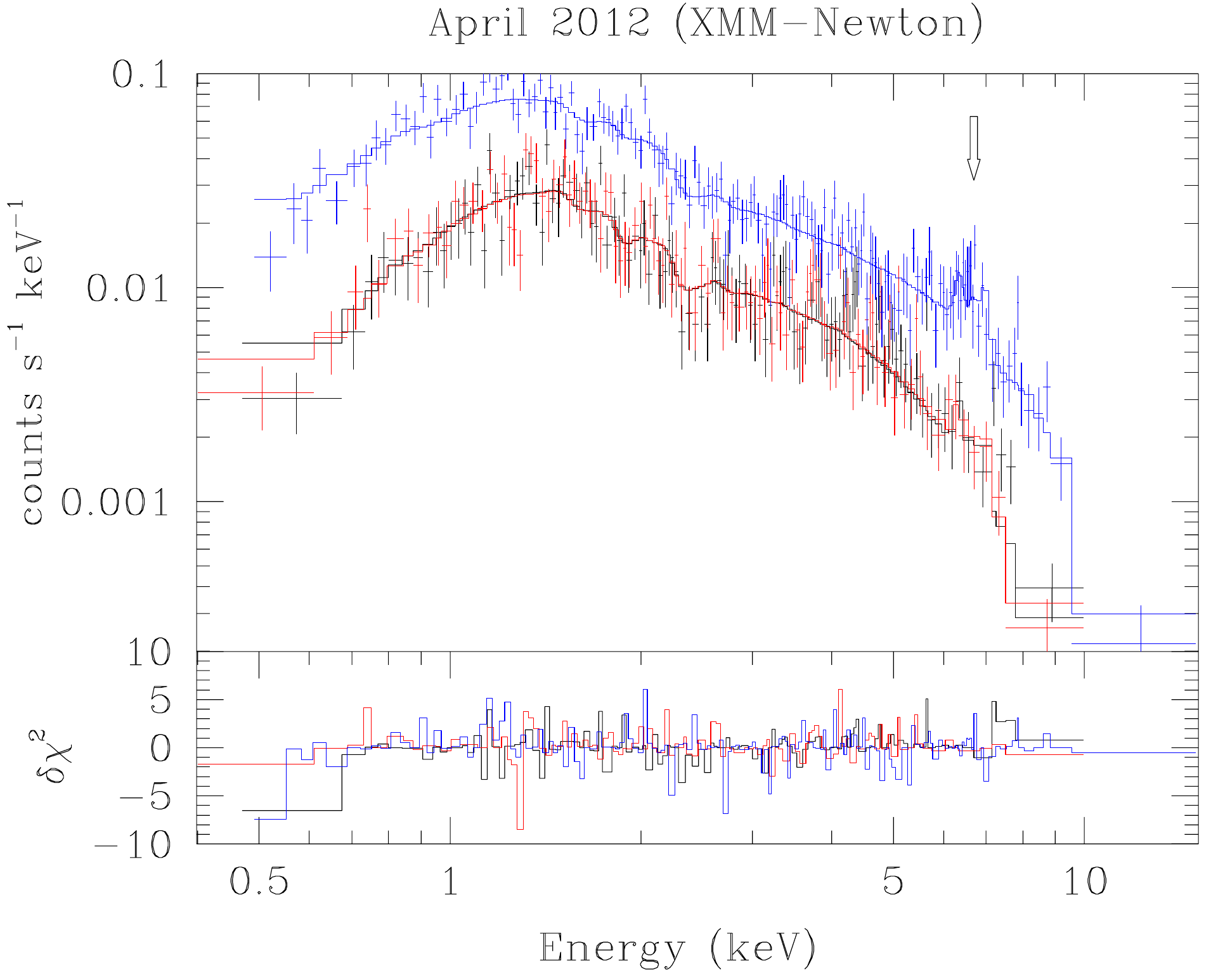}}
\end{center}
\end{minipage}
\hfill
\begin{minipage}{8.5cm}
\begin{center}
\resizebox{8.5cm}{!}{\includegraphics{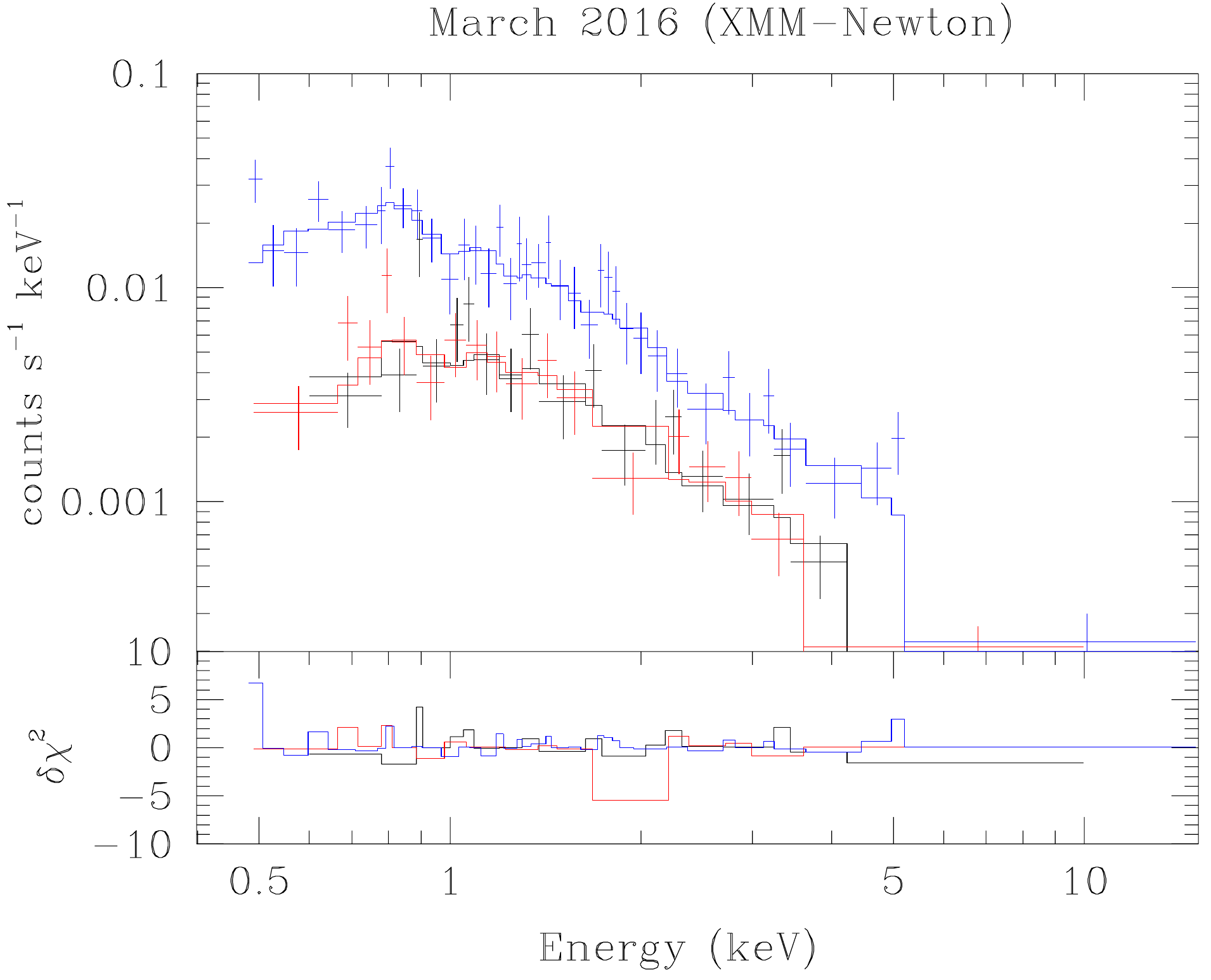}}
\end{center}
\end{minipage}
\begin{minipage}{8.5cm}
\begin{center}
\resizebox{8.5cm}{!}{\includegraphics{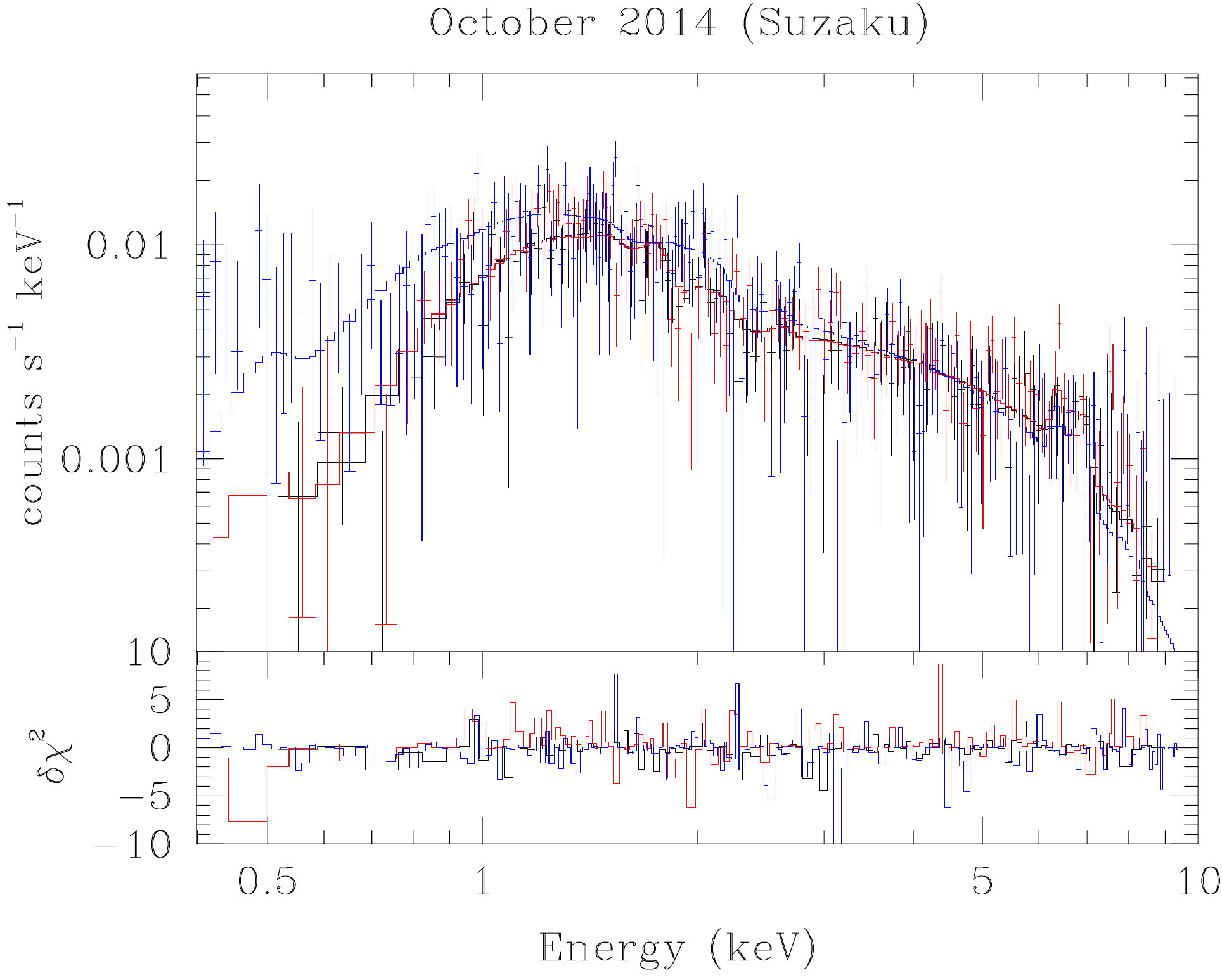}}
\end{center}
\end{minipage}
\hfill
\begin{minipage}{8.5cm}
\begin{center}
\resizebox{8.5cm}{!}{\includegraphics{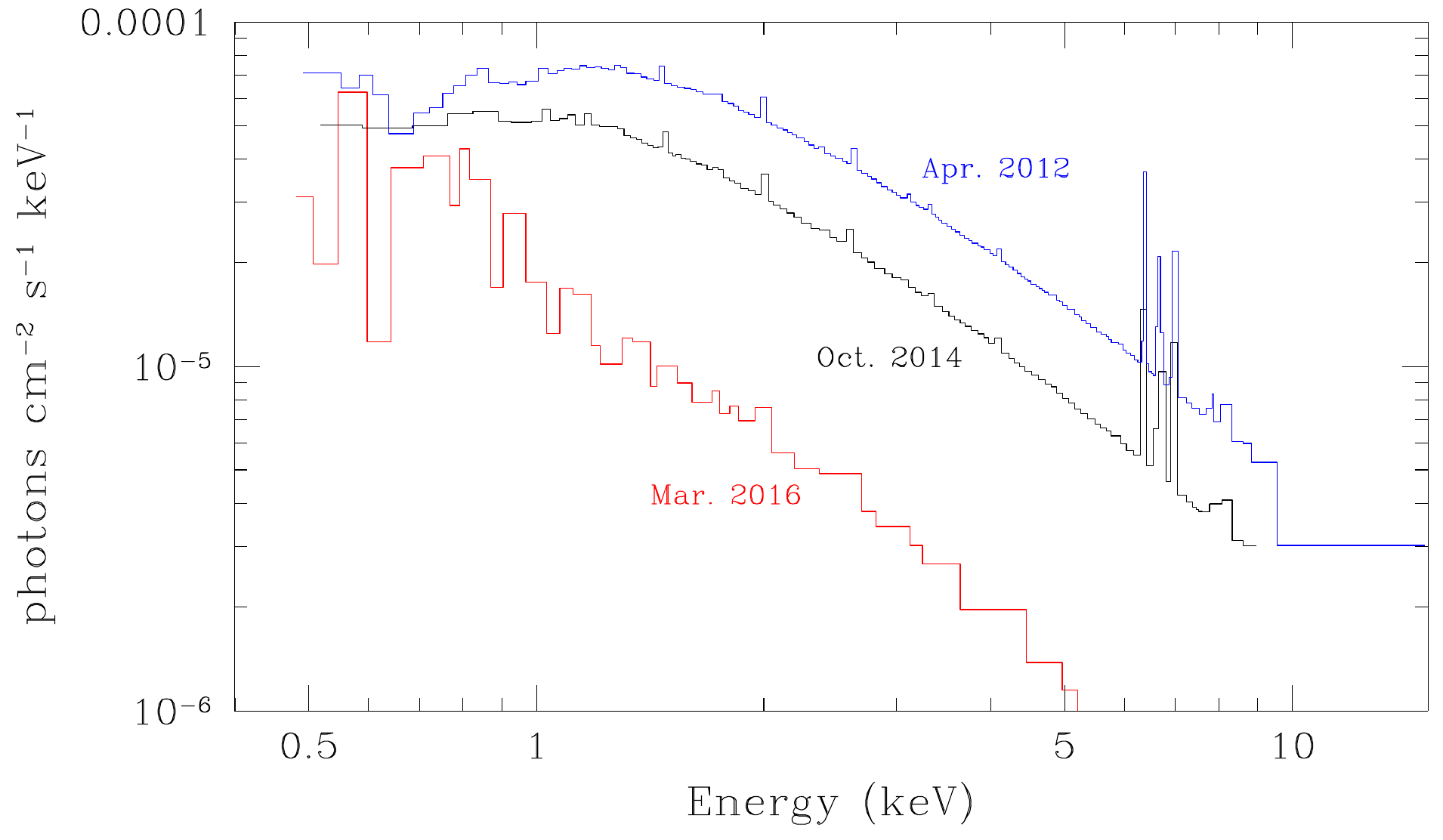}}
\end{center}
\end{minipage}
\caption{Top row: {\it XMM-Newton} EPIC spectra of HD~45314 in the high (left) and low (right) optical emission states. The black, red, and blue data points refer respectively to the MOS1, MOS2, and pn data. The continuous lines correspond to the best-fit 1-T thermal model including an extra Gaussian to fit the fluorescent Fe K$\alpha$ line for the high state and the best-fit 2-T thermal model for the low state (see Table\,\ref{fits}). The contributions of the various energy bins to the $\chi^2$ are given in the lower panels. The sign of $\delta \chi^2$ is set to the sign of the data minus the model. To ease comparison, both panels have the same horizontal and vertical axes. The arrow in the left panel points at the Fe K$\alpha$ + Fe\,{\sc xxv} and Fe\,{\sc xxvi} lines (see Fig.\,\ref{FeK}). Bottom row: The left panel shows the {\it Suzaku} XIS spectra of HD~45314 taken in October 2014. The black, red, and blue data points refer respectively to the XIS0, XIS3, and XIS1 data. The continuous lines correspond to the best-fit 1-T thermal model including an extra Gaussian to fit the fluorescent Fe K$\alpha$ line (see Table\,\ref{fits}). The right panel displays the unfolded best-fit models for the three epochs.\label{EPICspectra}}\end{figure*}

\subsection{Optical spectroscopy and photometry}
Our monitoring campaign of HD~45314 involves both professional and amateur observatories. In addition to the data already presented in \citet{ibvs,Oepaper}, we collected new data between autumn 2014 and spring 2017.
Twenty-three spectra were obtained with the 1.2~m Telescopio Internacional de Guanjuato Rob\'otico Espectrosc\'opico \citep[TIGRE, formerly known as the Hamburg Robotic Telescope,][]{Hempelmann,Schmitt} installed at La Luz Observatory near Guanajuato (Mexico). The telescope is operated in a fully robotic way and spectra are taken with the refurbished Heidelberg Extended Range Optical echelle Spectrograph \citep[HEROS,][]{Kaufer2}, which offers a spectral resolving power of 20\,000 over the full optical range, though with a small gap near 5800\,\AA. The HEROS data were reduced with the corresponding reduction pipeline \citep{Mittag,Schmitt}. 

Twenty-one spectra were obtained between November 2014 and March 2017 at the 0.81~m telescope of the Three College Observatory (TCO) near Greensboro (North Carolina, USA). The spectrograph is a commercial `Eshel' fiber-fed echelle spectrograph manufactured by Shelyak Instruments. It provides a spectral resolving power of $\sim 12\,000$ over the spectral range from about 4200 to 7800\AA. The detector is an Atik 460EX CCD camera with $2749 \times 2199$ pixels of $4.54 \times 4.54$\,$\mu$m$^2$. The spectra were taken with a $2 \times 2$ pixel binning. The data were reduced using the echelle package of the Image Reduction and Analysis Facility (IRAF) software.\footnote{IRAF is distributed by the National Optical Astronomy Observatories, which are operated by the Association of Universities for Research in Astronomy, Inc., under contract with the National Science Foundation.}

Two spectra were obtained with the REOSC (Recherche et Etude en Optique et Sciences Connexes) Cassegrain echelle spectrograph at the 2.12~m telescope at the Observatorio Astron\'omico Nacional San Pedro M\`artir (SPM, Baja California, Mexico). This instrument provides a resolving power of 18\,000. The data were also reduced with the IRAF echelle package.

The above data were complemented by amateur spectra taken by co-authors JGF, FC, and PP at their private observatories. Twenty-four spectra were obtained by JGF between December 2016 and April 2017 with a Meade 16-inch telescope equipped with a self-made echelle spectrograph that achieves a resolving power of 9000. The detector was an Atik 460EX CCD camera and the data were taken with a $2 \times 2$  binning. The data reduction was done using the ISIS software designed by C.\ Buil.\footnote{www.astrosurf.com/buil/isis/isis\_en.htm.} 

Another six spectra were obtained by FC using either a 0.20\,m f/4.7 Newtonian (focal length doubled to f/9.5 with a Barlow lens) or a 0.20\,m f/8 Ritchey-Chr\'etien telescope. The spectrograph was a Baader DADOS long-slit spectrograph, working with a 1200\,l\,mm$^{-1}$ grating and a 25\,$\mu$m slit. The CCD camera was an Atik 314L+ (Sony ICX-285AL). The raw data were reduced using MaxIm DL v5 and ISIS v5.5.1a. Six spectra were taken by PP with a 50~cm telescope located near Pilot Mountain (North Carolina, USA). The spectrograph and the data reduction were exactly the same as for the TCO data. 

Since the optical spectra were taken with a variety of instruments, one could wonder about the accuracy of the wavelength cross-calibration of the various spectra. To assess the typical calibration errors, we have measured the radial velocities of the interstellar Na\,{\sc i} D$_1$ and D$_2$ lines on all data where these lines fall within the spectral domain that was covered. The results of this test are presented in Table\,\ref{ISM}. The mean values of all subsets of data agree with the mean of the overall sample within 1.3\,km\,s$^{-1}$, which we thus take as our estimate of the typical calibration error. This indicates that there should be no significant bias between the velocity measurements performed with the different instruments.

\begin{table}
\caption{Radial velocities of the interstellar Na\,{\sc i} D$_1$ and D$_2$ lines\label{ISM}} 
\begin{center}
\begin{tabular}{c r c c}
\hline
Spectrograph & N & RV$_{D_1}$ &  RV$_{D_2}$ \\
             &   & (km\,s$^{-1}$) & (km\,s$^{-1}$) \\
\hline
FEROS        &  23 & $15.2 \pm 0.4$ & $14.7 \pm 0.4$ \\
FIES         &   5 & $15.3 \pm 0.3$ & $14.6 \pm 0.2$ \\
HEROS        &  32 & $14.4 \pm 0.8$ & $13.8 \pm 1.0$ \\
TCO          &  21 & $15.5 \pm 1.1$ & $15.0 \pm 1.1$ \\
JGF          &  24 & $13.5 \pm 0.4$ & $13.2 \pm 0.5$ \\
PP           &   6 & $15.5 \pm 2.8$ & $15.5 \pm 3.2$ \\
\hline
All data     & 121 & $14.8 \pm 1.2$ & $14.3 \pm 1.3$ \\
\hline
\end{tabular}
\end{center}
\tablefoot{The second column lists the number of spectra covering the Na\,{\sc i} lines obtained with a given instrument. All spectrographs for which at least five spectra were obtained are listed. The last row lists the results for the full sample of measurements.}
\end{table}

For all optical spectra discussed above, we used the {\it telluric} tool within IRAF along with the list of telluric lines of \citet{Hinkle} to remove the telluric lines in the He\,{\sc i} $\lambda$\,5876 and H$\alpha$ regions. The spectra were continuum normalized using MIDAS (Munich Image Data Analysis System) routines and adopting the same set of continuum windows for all spectra to achieve self-consistent results. 

\begin{table*}[bth]
\caption{Results of the fits of the X-ray spectra of HD~45314.\label{fits}}
\begin{center}
\begin{tabular}{c c c c c c c c c c c c c}
\hline
\multicolumn{13}{c}{\tt tbabs*wind*(apec(2T)+gauss)}\\
\hline
Epoch & $\log{N_{\rm wind}}$ &  kT$_h$ & norm$_1$ & kT$_s$ & norm$_2$ & E$_{\rm line}$ & $EW_{\rm line}$ & $\chi^2_{\nu}$ & d.o.f. & $f_X^{\rm soft}$ & $f_X^{\rm med}$ & $f_X^{\rm hard}$ \\
     & (cm$^{-2}$) & (keV) & (cm$^{-5}$) & (keV)  & (cm$^{-5}$) & (keV) & (keV) & & & \multicolumn{3}{c}{($10^{-14}$\,erg\,cm$^{-2}$\,s$^{-1}$)} \\  
\hline
\vspace*{-2mm}\\
Apr.\ 2012 & $21.38^{+.09}_{-.08}$ & $15.4^{+5.4}_{-4.5}$ & $(7.2^{+0.2}_{-0.2})\,10^{-4}$ & & &$6.35^{+.12}_{-.08}$ & $0.16 \pm 0.05$ & 1.06 & 387 & $3.9$ & $15.5$ & $95.3$ \\
\vspace*{-2mm}\\
Apr.\ 2012 & $21.65^{+.03}_{-.03}$ & $9.7^{+1.8}_{-1.3}$ & $(7.5^{+0.2}_{-0.2})\,10^{-4}$ & $0.27$ (fix.) & $8.9\,10^{-5}$ (fix.) & $6.33^{+.07}_{-.07}$ & $0.18 \pm 0.05$ & 1.16 & 387 & $4.0$ & $15.2$ & $93.6$ \\
\vspace*{-2mm}\\
Oct.\ 2014 & $21.01^{+.22}_{-.48}$ & $12.4^{+6.5}_{-3.0}$ & $(3.9^{+0.2}_{-0.2})\,10^{-4}$ & & &$6.36^{+.08}_{-.08}$ & $0.19 \pm 0.07$ & 0.99 & 456 & $3.1$ & $9.9$ & $52.4$ \\
\vspace*{-2mm}\\
Oct.\ 2014 & $21.53^{+.07}_{-.06}$ & $8.9^{+1.9}_{-1.5}$ & $(4.2^{+0.2}_{-0.2})\,10^{-4}$ & $0.27$ (fix.) & $8.9\,10^{-5}$ (fix.) & $6.35^{+.06}_{-.07}$ & $0.19 \pm 0.06$ & 1.05 & 456 & $3.6$ & $9.6$ & $52.1$ \\
\vspace*{-2mm}\\
Mar.\ 2016 & $21.58^{+.28}_{-.89}$ & $4.8^{+2.0}_{-1.3}$ & $(8.6^{+1.3}_{-1.2})\,10^{-5}$ & $0.27^{+.03}_{-.03}$ & $(8.9^{+15.0}_{-6.3})\,10^{-5}$ & & & 0.86 & 72 & $1.8$ & $2.4$ & $8.4$ \\
\vspace*{-2mm}\\
\hline
\end{tabular}
\begin{tabular}{c c c c c c c c c c c}
\hline
\multicolumn{11}{c}{\tt tbabs*wind*(power+gauss)}\\
\hline
Epoch & $\log{N_{\rm wind}}$ &  $\alpha$ & norm & E$_{\rm line}$ & $EW_{\rm line}$ & $\chi^2_{\nu}$ & d.o.f. & $f_X^{\rm soft}$ & $f_X^{\rm med}$ & $f_X^{\rm hard}$ \\
     & (cm$^{-2}$) & & (photons\,keV$^{-1}$\,cm$^{-2}$\,s$^{-1}$) & (keV) & (keV) & & & \multicolumn{3}{c}{($10^{-14}$\,erg\,cm$^{-2}$\,s$^{-1}$)} \\  
\hline
\vspace*{-2mm}\\
Apr.\ 2012 & $21.46^{+.08}_{-.09}$ & $1.46^{+.07}_{-.07}$ & $(1.69^{+.17}_{-.14})\,10^{-4}$ & $6.56^{+.14}_{-.27}$ & $0.15 \pm 0.06$  & 1.10 & 387 & $3.9$ & $15.3$ & $98.6$ \\
\vspace*{-2mm}\\
Oct.\ 2014 & $21.23^{+.21}_{-.34}$ & $1.53^{+.11}_{-.11}$ & $(1.02^{+.16}_{-.13})\,10^{-4}$ & $6.39^{+.47}_{-.10}$ & $0.16 \pm 0.06$ & 1.00 & 456 & $3.2$ & $9.9$ & $54.0$ \\
\vspace*{-2mm}\\
Mar.\ 2016 & & $2.03^{+.13}_{-.13}$ & $(2.85^{+.21}_{-.21})\,10^{-5}$ & & & 1.02 & 75 & $1.6$ & $2.7$ & $6.9$ \\
\vspace*{-2mm}\\
\hline
\end{tabular}
\end{center}
\tablefoot{All fits assume an interstellar neutral hydrogen column density of $1.9 \times 10^{21}$\,cm$^{-2}$. For the April 2012 (high state) and October 2014 (shell state) observations, two types of thermal models are considered: single temperature models as well as two temperature models where the properties of the cooler component were set to those derived in the March 2016 (low state) observation. The normalization of the {\tt apec} models corresponds to $\frac{10^{-14}\,\int\,n_e\,n_H\,dV}{d^2}$ , where $d$ is the distance of the source (in cm), and $n_e$ and $n_H$ are the electron and hydrogen number densities of the source (in cm$^{-3}$). The photon index of the power law is labelled $\alpha$, and the normalization of the power law is given for a photon energy of 1\,keV. The soft, medium, and hard energy domains correspond to the 0.5 -- 1.0, 1.0 -- 2.0, and 2.0 -- 10.0\,keV bands, respectively. The fluxes in the last three columns correspond to observed values, that is\ they are not corrected for interstellar absorption.}
\end{table*}

We have retrieved the visual magnitudes of the star as listed in the database of the American Association of Variable Star Observers \citep[AAVSO,][]{Kafka}. We also retrieved $V$-band photometry of HD~45314 taken from the All Sky Automated Survey \citep[ASAS-3,][]{ASAS}. This survey was carried out from the Las Campanas Observatory in Chile using two wide-field ($8.8^{\circ} \times 8.8^{\circ}$) telescopes, each equipped with a 200/2.8 Minolta telephoto lens and a $2048 \times 2048$ pixels AP-10 CCD camera. These instruments were complemented by a narrow-field ($2.2^{\circ} \times 2.2^{\circ}$) 25\,cm Cassegrain telescope equipped with the same type of CCD camera. The ASAS-3 photometric catalogue provides magnitude measurements performed with five different apertures varying in diameter from two to six pixels. Substantial differences between large- and small-aperture magnitudes indicate either contamination by close neighbours or saturation. The ASAS-3 data are not uniform in terms of exposure time and thus in terms of saturation limit, as the exposure time was changed in the course of the project from 180\,s (saturation near $V \sim 7.5$) to 60\,s (saturation around $V \sim 6$). Since HD~45314 has a magnitude close to the latter saturation limit, we have filtered the observations, keeping only data with a dispersion between the various aperture photometric measurements less than or equal to the mean error on the photometry.

\section{X-ray spectral analysis \label{Xspec}}
The EPIC and XIS spectra of HD~45314 were analysed with version 12.9.0i of the { xspec} software \citep{Arnaud}. Optically thin thermal plasma emission models usually provide a good description of CCD-type X-ray spectra of massive stars. For the hard emission components of $\gamma$~Cas stars, the choice of these models is at first sight less obvious, as their spectra exhibit very few lines. We have thus also tested various combinations of thermal and non-thermal models to fit our data. 

In our spectral fits, we considered only bins with an energy above 0.3\,keV as the lower energy channels are subject to larger calibration uncertainties. For the XIS spectra, we also explicitly restricted the fit to energies below 10.0\,keV.
Estimates of the observed X-ray fluxes and of their errors were obtained using the {\it flux err} command under{ xspec}. 

The emission from collisionally-ionized-equilibrium, optically-thin thermal plasma was modelled using {\tt apec} models \citep{apec} computed  with ATOMDB v2.0.2. The plasma abundances were taken to be solar \citep{Asplund}. The absorption by the interstellar medium (ISM) was modelled using the T\"ubingen-Boulder model \citep[{\tt tbabs},][]{Wilms}. The value of the total neutral hydrogen column density was set to $1.9 \times 10^{21}$\,cm$^{-2}$ according to the compilation of interstellar column densities of \citet{ISM}. On top of the absorption by the ISM, the X-ray spectra of massive stars can be absorbed by the ionized stellar wind or other intervening circumstellar gas. To model such an absorption, we imported the stellar wind absorption model of \citet{HD108} into{ xspec} as a multiplicative tabular model (hereafter labelled {\tt wind}). 

For the bright emission state, observed in April 2012 with {\it XMM-Newton}, the best results are obtained for a single temperature thermal plasma model with an additional Gaussian to account for the small photon excess redwards of the Fe\,{\sc xxv} triplet and the Fe\,{\sc xxvi} Ly$\alpha$ line\footnote{Since the intrinsic width of the line is most probably less than the instrumental resolution and can thus not be constrained by our data, we set it to 1\,eV in our fits.} (see Fig.\,\ref{EPICspectra}). The parameters are given in the upper part of Table\,\ref{fits}. The energy of the Gaussian line is consistent with fluorescent emission from iron in a rather low ionization stage. Our best-fit plasma temperature is lower than reported in our initial study whilst our best-fit wind column density is larger than derived previously \citep{Oeletter}. These differences reflect some differences in data processing and analysis, though both results overlap within their error bars. In \citet{Oeletter}, we considered energy bins down to 0.1\,keV, whereas the lowest energy bin considered here has an energy of 0.3\,keV. We also adopted a different way of binning the data and we chose a different model for the absorption by the interstellar medium. Whilst these changes concern mostly the low-energy part of the spectrum, they have an indirect impact also on the value of the best-fit temperature via the degeneracy between the total (interstellar + circumstellar) column density and the plasma temperature to produce the same spectral hardness. Furthermore, our present analysis uses emission and absorption models with more up-to-date atomic parameters than in \citet{Oeletter}. We finally note that the model parameters are better constrained in the present analysis than in our previous work. 

\begin{figure}[h!tb]
\begin{center}
\resizebox{8cm}{!}{\includegraphics{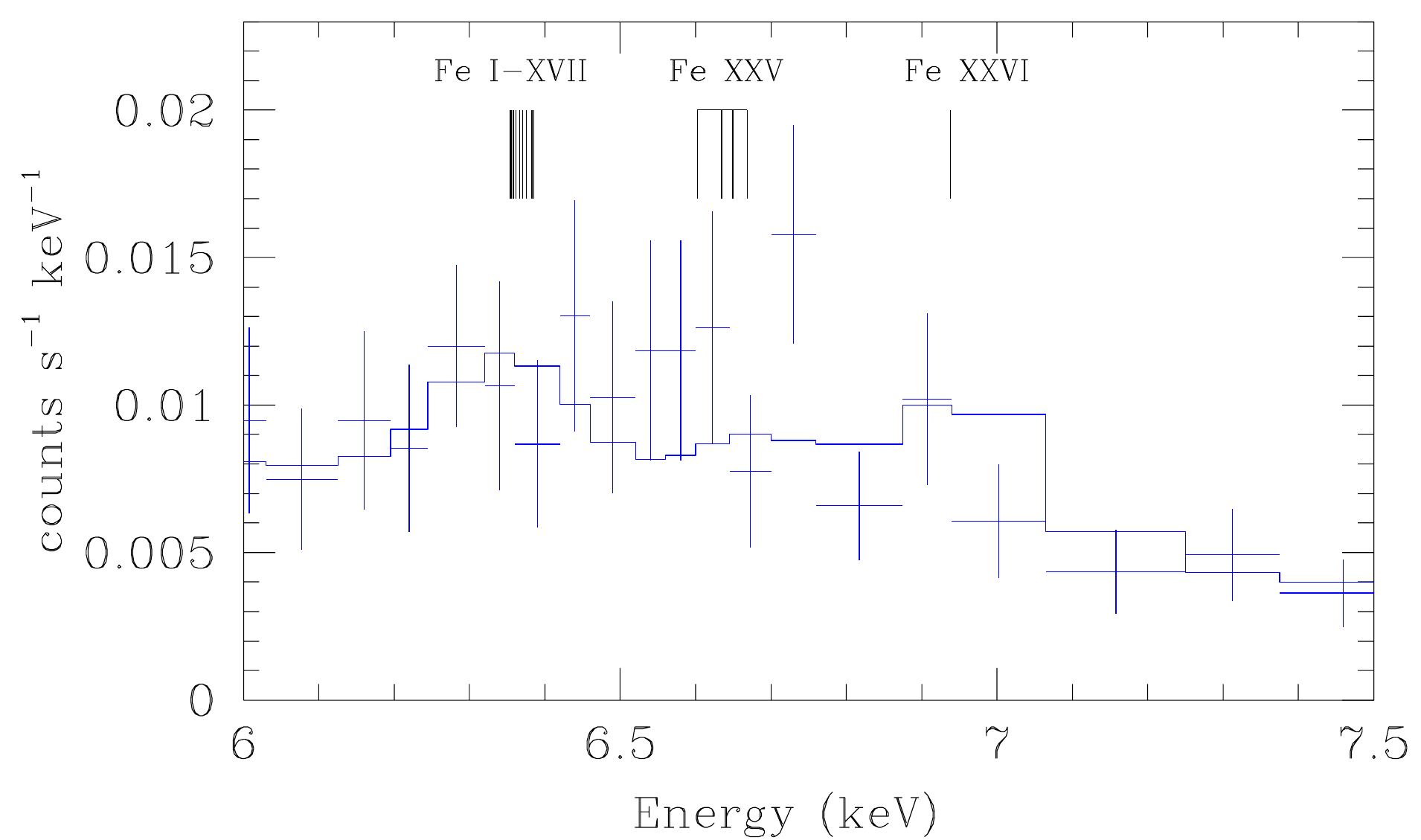}}
\end{center}
\caption{Zoom on the 6.0 - 7.5\,keV spectral band of the high-state EPIC-pn spectrum of HD~45314. The histogram yields the best-fit 1-T thermal model including an extra Gaussian to fit the fluorescent Fe K$\alpha$ line. The theoretical energies of the fluorescent lines of Fe\,{\sc i} - {\sc xvii} as well as the energies of the He-like lines of Fe\,{\sc xxv} and the H-like Fe\,{\sc xxvi} Ly$\alpha$ lines are indicated by the tick marks. The simultaneous presence of the latter features is the signature of a thermal plasma with $kT_h$ near 10\,keV.\label{FeK}}
\end{figure}
Figure\,\ref{FeK} illustrates the spectral region around the Fe\,{\sc xxv}, Fe\,{\sc xxvi,} and Fe K lines. The observed strength of the blend of the He-like lines is slightly ($1\,\sigma$) above our model suggesting that there could be another plasma component with a $kT$ near 5\,keV contributing to the emission. However, including such a component does not significantly improve the quality of the global fits. 
 
The Fe\,{\sc xxv} and Fe\,{\sc xxvi} lines provide the best evidence for the emission being thermal. Indeed, when we assume a non-thermal spectrum along with a fluorescent line, the fit quality somewhat degrades, though it remains formally acceptable. However, the Gaussian now has to account for both the fluorescent and the thermal line, which shifts its centroid to higher energies (see lower half of Table\,\ref{fits}).

Very similar conclusions are reached for the fit of the {\it Suzaku} spectrum taken during the shell phase in October 2014 (see Fig.\,\ref{EPICspectra} and Table\,\ref{fits}). The main difference compared to the first {\it XMM-Newton} observation concerns the fluxes, which are lower by a factor of 1.76 in October 2014 than in April 2012. We note that the parameters of the fluorescent line agree very well between the high-state {\it XMM-Newton} and the {\it Suzaku} spectra, though they were fitted independently.  

In the low emission state, observed with {\it XMM-Newton} in March 2016, a single temperature thermal plasma model is no longer sufficient to achieve a good fit. The best results for single temperature models have $\chi^2_{\nu} \geq 1.5$, with large residuals especially over the lower energy part of the spectrum. Adding a second thermal plasma component considerably improves the situation (see Table\,\ref{fits}). Although the uncertainty on the emission measure of this second plasma is quite large, we stress that its inclusion leads to a much lower $\chi^2_{\nu}$: comparing the $\chi^2$ values, the F statistics value is 34.2, indicating a significantly better fit. With $kT_s \simeq 0.27$\,keV, this second thermal component is much softer than the hard plasma component ($kT_h \sim 4.8$\,keV) and is quite typical of the emission of OB-type stars \citep[e.g.][]{Naze,CygOB2}. The ISM-corrected flux in the 0.5 -- 10\,keV band of this soft component amounts to $3.5 \times 10^{-14}$\,erg\,cm$^{-2}$\,s$^{-1}$. Comparison with the bolometric flux of HD~45314 \citep[see][]{Oeletter} yields $\log{\frac{L_{\rm X,s}}{L_{\rm bol}}} = -7.68$. The soft plasma component could thus (mostly) arise from the intrinsic wind emission of the Oe star. It is likely though that this soft component represents only part of the wind emission. Indeed, \citet{Naze} found that, for O-type stars, plasma emission with $kT$ near 0.3\,keV is frequently associated with another plasma component with $kT$ near 2\,keV. In our case, the latter component would be hidden by the stronger residual $\gamma$~Cas-like emission.  
One could then expect this wind emission to be present also in the high state, although it would be overwhelmed by the hard emission. We have tested this assumption by fitting the high state {\it XMM-Newton} and the {\it Suzaku} spectra with a model where we have fixed the parameters of the soft component to those obtained for the low emission state (see Table\,\ref{fits}). The result is a slightly poorer fit than for the single-temperature model, especially at low energies. Coming back to the low emission state spectrum, we note that the hard component has a temperature ($kT_h \sim 4.8$\,keV, see Table\,\ref{fits}) that, although significantly lower than in the high state, remains unusually high for a massive star. This suggests that even in the low state, the $\gamma$~Cas-like emission has not totally vanished. Still, considering the low emission state, we find that a single power-law model yields a slightly poorer fit than the two-temperature model, but remains acceptable. Finally, we note that the quality of the low-state spectrum is not sufficient to look for the presence of a fluorescent line.\\  

Our spectral fits indicate that the flux and the hardness of the X-ray emission of HD~45314 changed considerably between the high and low emission states. In the soft band (0.5 -- 1\,keV), the observed flux varied only by a factor of two, but much larger variations affected the medium (1 -- 2\,keV) and hard (2 -- 10\,keV) bands where the flux decreased by factors of six and eleven between the high state and the low state, respectively. 

The {\it Gaia} DR1 catalogue \citep{Gaia,GaiaDR1} quotes a parallax of $(1.17 \pm 0.47)$\,mas for HD~45314, which translates into a distance of $0.85$\,kpc. With this value, we obtain a 0.5 -- 10\,keV X-ray luminosity $L_{\rm X}$ of $1.05 \times 10^{32}$\,erg\,s$^{-1}$ in the high state. This places HD~45314 near the lower end of the X-ray luminosities of $\gamma$~Cas stars \citep{gamCasrev}, along with the B1.5\,Ve star HD~157832 \citep{LopesMotch}. The total X-ray luminosity of HD~45314 is reduced by a factor of eight in the low state. 

The relative uncertainty on the DR1 parallax is quite large, leading therefore to a substantial uncertainty on the distance and the absolute X-ray luminosity. We can nevertheless obtain a robust, distance-independent estimate of the $\frac{L_{\rm X}}{L_{\rm bol}}$ ratio. For this purpose, we use the bolometric fluxes evaluated in \citet{Oeletter} along with the above derived X-ray fluxes corrected for interstellar absorption. This leads to $\log{\frac{L_{\rm X}}{L_{\rm bol}}} = -6.13$ for the high state, $-6.37$ for the shell phase, and $-7.05$ for the low state. Whilst the former value is towards the lower end of $\gamma$~Cas stars, the latter value is fully in-line with the canonical relation for massive stars \cite[e.g.][and references therein]{Berghoefer,Naze}. However, the fact that the temperature of the hard component in the low state remained unusually high for an O-type star suggests that, even in this state, some fraction of the X-ray flux arises from the $\gamma$~Cas-like emission component.   

To this date, all existing X-ray data of HD~45314, except for the low-state {\it XMM-Newton} observation presented here, revealed the star in its $\gamma$~Cas-like state. Indeed, beside the {\it XMM-Newton} and {\it Suzaku} data discussed here, HD~45314 was also observed with the Medium Energy (ME) proportional counter instrument aboard {\it EXOSAT} on 19 October 1984 (i.e.\ on HJD~2\,445\,993.318\footnote{We could not find any report of optical spectroscopy for that epoch.}). As discussed in \citet{Oeletter}, this observation is consistent with the high-state April 2012 X-ray spectrum. 
 
We thus conclude that whilst the spectrum in the low state was still harder than what is expected for a normal early-type star, most of HD~45314's $\gamma$~Cas peculiarities had considerably faded. These are the first ever X-ray observations of a $\gamma$~Cas star that reveal a substantial reduction of the $\gamma$~Cas characteristics as the disk dissipates. On the other hand, during the shell episode, the X-ray flux and spectrum remained consistent with the $\gamma$~Cas classification of the star.     

\begin{figure*}[htb]
\begin{minipage}{8cm}
\begin{center}
\resizebox{8cm}{!}{\includegraphics{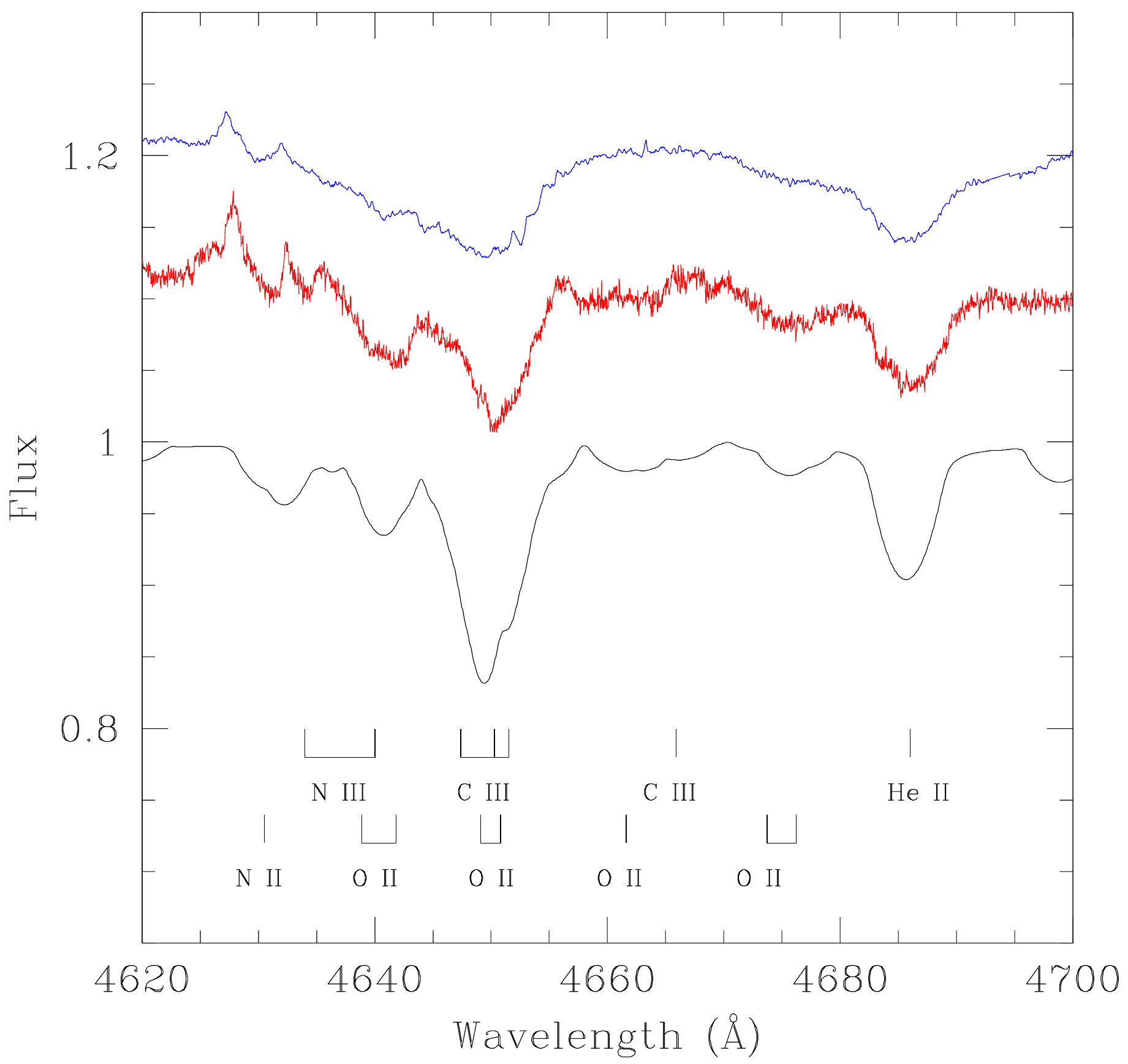}}
\end{center}
\end{minipage}
\hfill
\begin{minipage}{8cm}
\begin{center}
\resizebox{8cm}{!}{\includegraphics{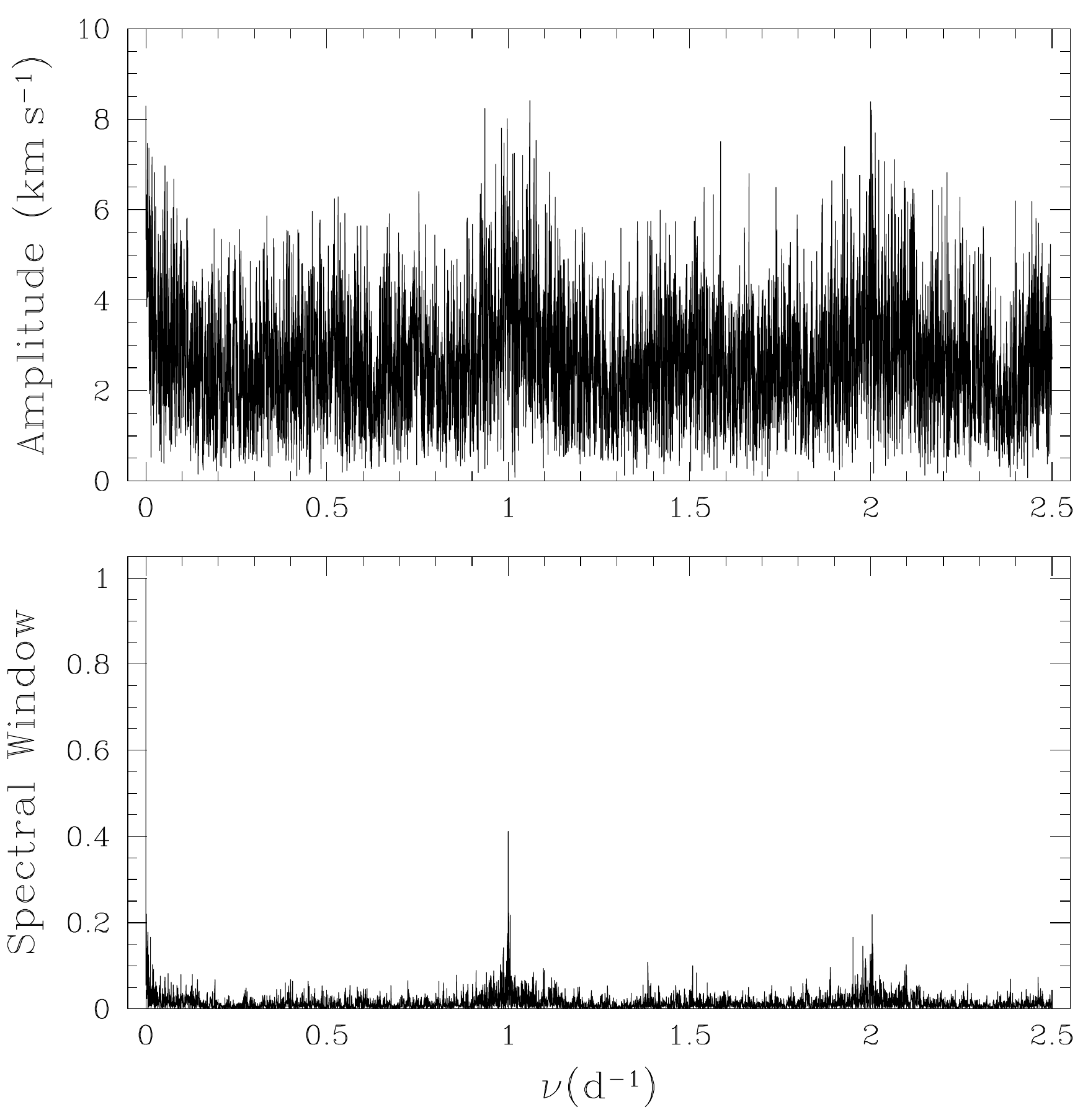}}
\end{center}
\end{minipage}
\caption{Left: Spectral region between 4620 and 4700\,\AA\ is shown in the synthetic spectrum (black line) as well as in two observations of HD~45314 taken with the FEROS instrument on HJD~2\,452\,336.596 (RV=$23.2$\,km\,s$^{-1}$, red line, shifted upward by 0.1 continuum unit) and 2\,453\,739.585 (RV=$-18.6$\,km\,s$^{-1}$, blue line, shifted upwards by 0.2 continuum unit). Right: Fourier analysis of the full sample of 87 RVs of HD~45314. The top panel illustrates the periodogram of the RVs, whilst the bottom panel provides the spectral window associated with our series of data.\label{RVs}}
\end{figure*}
\begin{figure*}
\begin{minipage}{8cm}
\begin{center}
\resizebox{8cm}{!}{\includegraphics{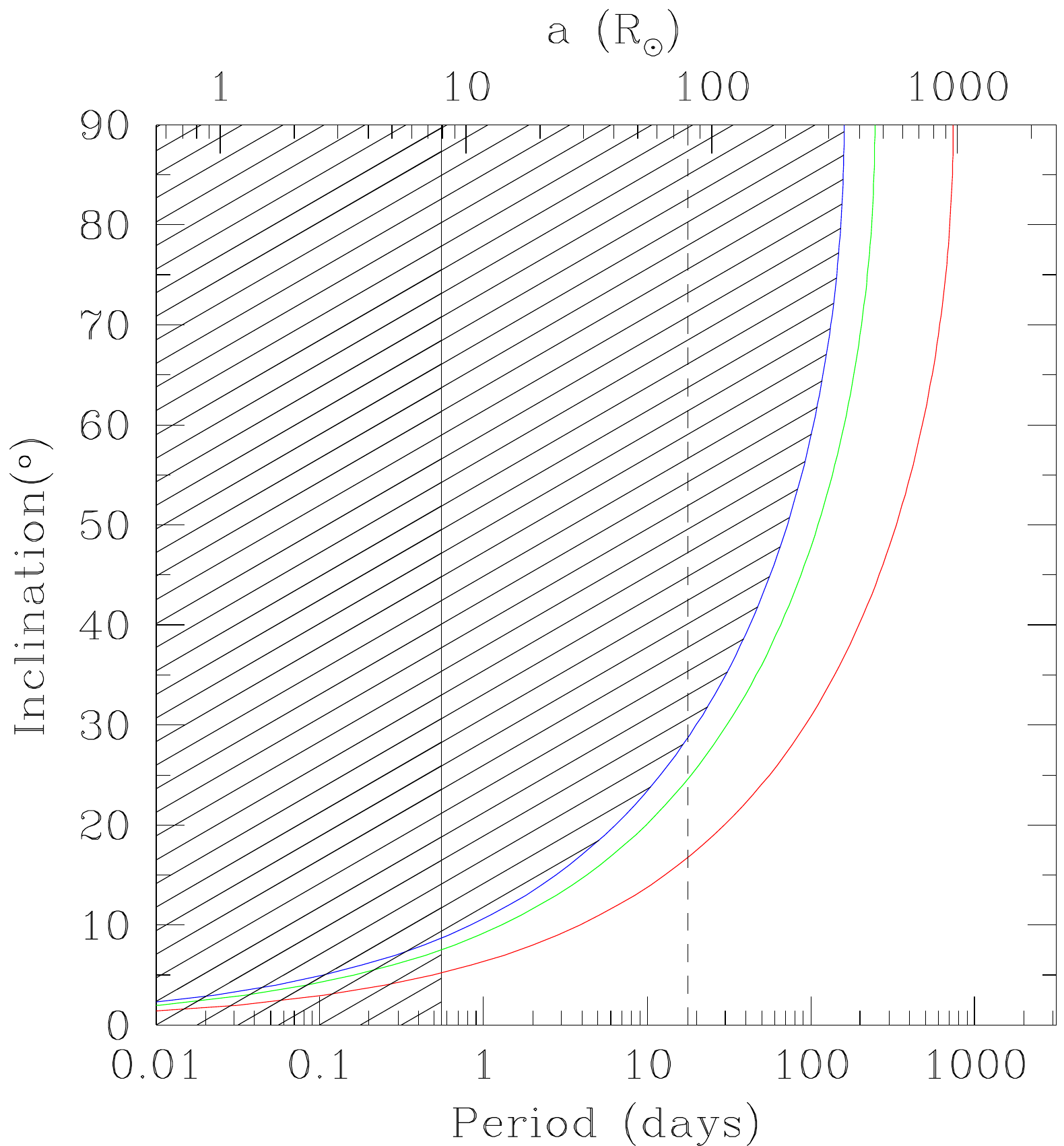}}
\end{center}
\end{minipage}
\hfill
\begin{minipage}{8cm}
\begin{center}
\resizebox{8cm}{!}{\includegraphics{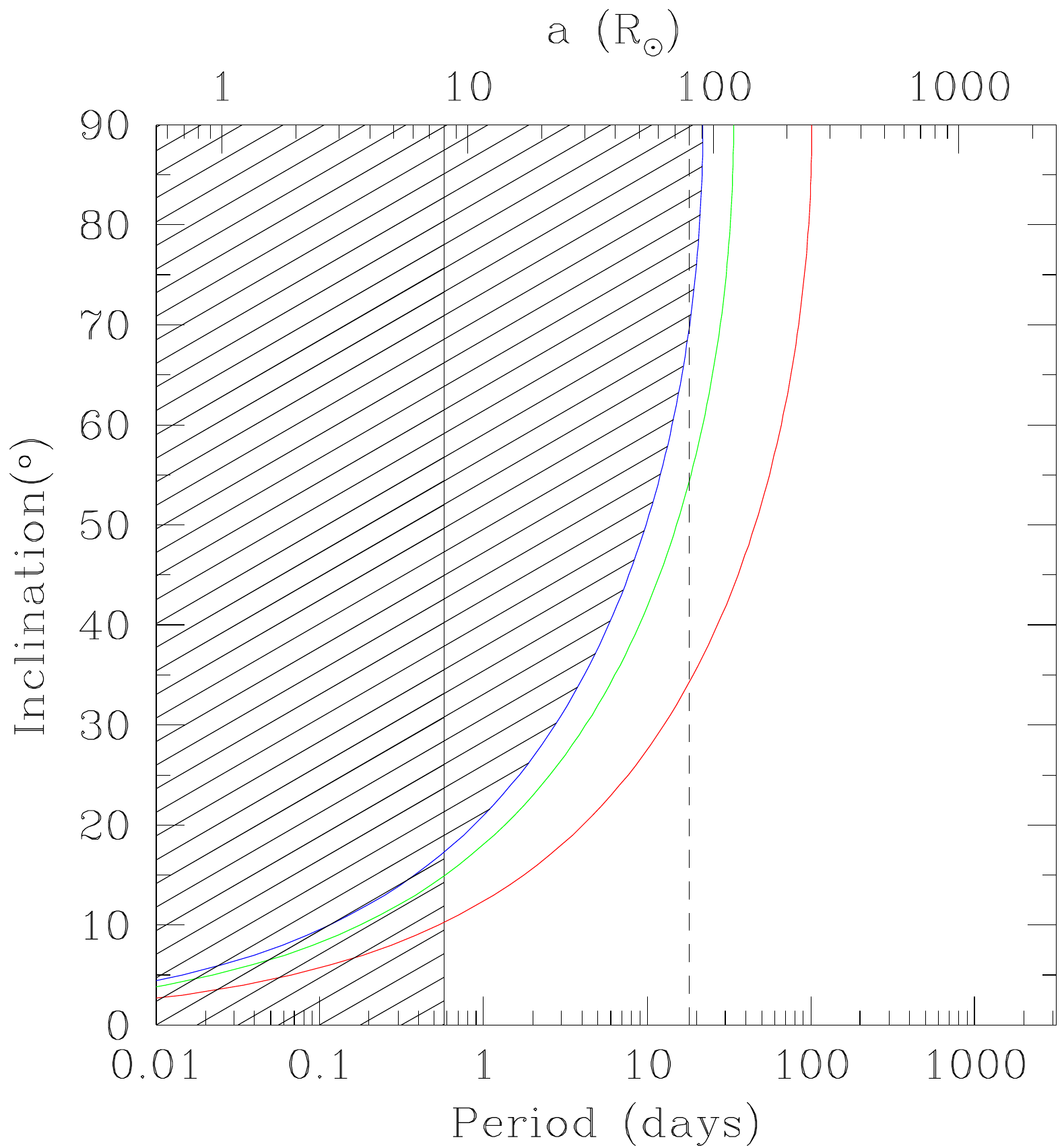}}
\end{center}
\end{minipage}
\caption{Constraints on the orbit for a putative compact companion of HD~45314 resulting from our upper limit of 10\,km\,s$^{-1}$ on $K$. We assume a mass of 20\,M$_{\odot}$ for the Oe star. In the left panel, we consider a 2.0\,M$_{\odot}$ companion, whereas the right panel corresponds to a 1\,M$_{\odot}$ companion. The blue, green, and red lines correspond to eccentricities of 0.0, 0.5, and 0.8. The vertical straight line corresponds to the adopted radius (8\,R$_{\odot}$) of the Oe star and the associated Keplerian rotation period. The dashed vertical line corresponds to the minimum outer radius of the Be disk that we have taken here as 10\,R$_*$ (see Sect.\,\ref{fitsHalpha}) and its associated Keplerian period. The hatched area indicates the part of the parameter space that is excluded from the upper limit on the RV amplitude. \label{fmass}}
\end{figure*}

From the various fits that we performed, we have estimated EWs of the fluorescent Fe K line of $0.16 \pm 0.05$\,keV and $0.18 \pm 0.06$\,keV for the high-state spectrum and the shell-episode spectrum of HD~45314, respectively. The fluorescent Fe K$\alpha$ line follows the removal of an electron from the K-shell. This can proceed either through photoionization by X-ray photons with energies above 7.11\,keV, that is,\ beyond the K-shell ionization edge of Fe, or by collisional excitation by electrons with energies in excess of the K-shell ionization energy.  Fluorescent Fe K$\alpha$ lines are commonly observed in the spectra of accreting binary systems and active galactic nuclei \citep{Sanford,Tanaka}. They have also been observed in the spectra of active low-mass stars and pre-main sequence stars \citep[e.g.][]{Hamaguchi,CS,Skinner}. For these objects, the fluorescent line is often, but not exclusively, observed in association with a flare in the coronae of the active cool stars \citep{CS}.

In the X-ray spectrum of $\gamma$~Cas, the strength of the Fe K fluorescent line appears to be correlated with the variable circumstellar column density \citep{gamCasrev}. From our data, it is unclear whether such a correlation also holds in the case of HD~45314. Our fits suggest a slightly (0.12 -- 0.37\,dex) smaller circumstellar column density for the shell-state spectrum compared to the high-state spectrum, whilst the Fe K line EW is unchanged within the uncertainties between the high state and the shell state. However, the uncertainties on the best-fit circumstellar column of the {\it Suzaku} spectrum are large. Moreover, these data were obtained with different instruments that have different sensitivities at low energies and this could lead to systematic errors on the circumstellar column density.    

Finally, as shown by \citet{Oeletter}, the X-ray flux of HD~45314 during the high-state observation displayed variations by about a factor of two (peak-to-peak) occurring on timescales of several hundred to a few thousand seconds. Because of the much lower mean flux in the low-state observation, the intra-pointing lightcurve cannot be analysed in detail. Therefore, the presence of such short-term variations cannot be assessed. It should be stressed though that the difference in the overall flux between the low and high states cannot be explained by short-term fluctuations as they would be washed out over the duration of the observation.

\section{Radial velocity variations in the optical spectrum\label{deltaRV}}
An important issue to address is whether or not HD~45314 could be a binary system. Indeed, several scenarios to explain the $\gamma$~Cas phenomenon involve the presence of a compact companion. Several claims of multiplicity for HD~45314 can be found in the literature. In 1994.87, \citet{Mason1} resolved HD~45314 in speckle interferometry, detecting a companion at a separation of $0\farcs05$ with a tentative orbital period of 30\,yrs (assuming a distance of 0.72\,kpc). Subsequent observations by \citet{Mason2} in 2005.86 and 2006.19 failed to detect the companion again and \citet{Mason2} accordingly suggested that it might have moved closer (to within less than $0\farcs03$) to the Oe star. No companion was detected with the Fine Guiding Sensor onboard the {\it Hubble Space Telescope} in 2008.77 for a range of separations from $0\farcs01$ to $1\farcs0$ and down to five magnitudes fainter than the Oe star \citep{FGS}. Whether or not the companion reported by \citet{Mason1} exists, it is very unlikely to be a compact object responsible for the $\gamma$~Cas-like behaviour, as such a compact companion would be too faint in the optical domain to be detected. 
 
\citet{Boyajian} reported a jump in the radial velocities of the H$\alpha$ line between two observing runs separated by two months, which could indicate binarity with a period of a few months. However, these authors caution that the H$\alpha$ line had changed its profile between the two epochs, casting doubt on the binary scenario. Based on 11 spectra taken between 2009 and 2012, \citet{Chini} classified HD~45314 as a single-lined spectroscopic binary, but did not provide an orbital solution, nor did they indicate how they reached this conclusion.\footnote{Line profile variability can sometimes be erroneously interpreted as genuine radial velocity changes due to binarity.}

Determining the radial velocities (RVs) of HD~45314 is a tedious task. The H$\alpha$ emission line bisector method, which is sometimes used to study RV variations of Be stars \citep[e.g.][]{Mirosh}, does not appear to be a good approach for HD~45314 as it could be biased by the strong profile variations displayed by this star. Furthermore, most absorption lines in the optical spectrum are polluted by the (variable) wings of nearby emission lines. In addition, the absorption lines are broad \citep[$v\,\sin{i} \simeq 210$\,km\,s$^{-1}$,][]{Oepaper} and diluted by continuum emission from the disk. This leaves very few lines suitable for RV measurements. The best candidates are found in the region between 4635 and 4695\,\AA. This spectral domain notably hosts several O\,{\sc ii} lines, the N\,{\sc iii} $\lambda\lambda$\,4634 -- 40, C\,{\sc iii} $\lambda\lambda$\,4647 -- 50, and He\,{\sc ii} $\lambda$\,4686 absorptions, and is mostly free of strong emissions (see Fig.\,\ref{RVs}). We have thus determined our estimates of the RVs of HD~45314 through a cross-correlation of the observed spectra and a synthetic TLUSTY spectrum \citep{LH} broadened to $v\,\sin{i} \simeq 210$\,km\,s$^{-1}$ over the above spectral region. The RVs were determined by fitting parabolas to the central part (between $-30$ and $+30$\,km\,s$^{-1}$ around the peak) of the correlation functions. Only the most reliable RV measurements (i.e.\ corresponding to spectra with good signal-to-noise ratios and showing no obvious signs of contamination of the absorption lines by neighbouring emissions) were retained. In this way, we ended up with 87 RVs spread over 6573\,days. We estimate an accuracy of about 10\,km\,s$^{-1}$ on individual measurements. Our RV measurements range from $-43$ to $+26$\,km\,s$^{-1}$. 

We searched for periodicities in the RV variations using the Fourier-method for uneven sampling of \citet{HMM}, modified by \citet{Gosset}. The resulting periodogram of the full time series is shown in the right panel of Fig.\,\ref{RVs}. We found no outstanding peaks in the Fourier analysis. The largest amplitudes did not exceed 8.0\,km\,s$^{-1}$ and the positions of the strongest peaks were found to be quite sensitive to the presence of one or the other data point in our time series. Moreover, the highest peak occurs for a period of about 2880\,days, whilst we regularly observe large RV changes (by up to 30\,km\,s$^{-1}$) in observations separated by a few days. In Sect.\,\ref{obs}, we showed that the combination of data collected with different instruments did not lead to significant biases in the wavelength calibration. To further check that there are no biases between the subsets of stellar RV data, we have also analysed those subsets of data associated with instruments for which more than 20 RV measurements are available. The resulting periodograms were in qualitative agreement with the results obtained for the full data set (absence of significant peaks), although with quantitative differences in the frequency and amplitude of the highest peaks. This behaviour shows that these periodograms actually reflect noise rather than a true periodicity. We also computed the average and the dispersions of the subsets and the full sample of RVs. These numbers are given in Table\,\ref{RVtable}. 

\begin{figure}
\begin{center}
\resizebox{8cm}{!}{\includegraphics{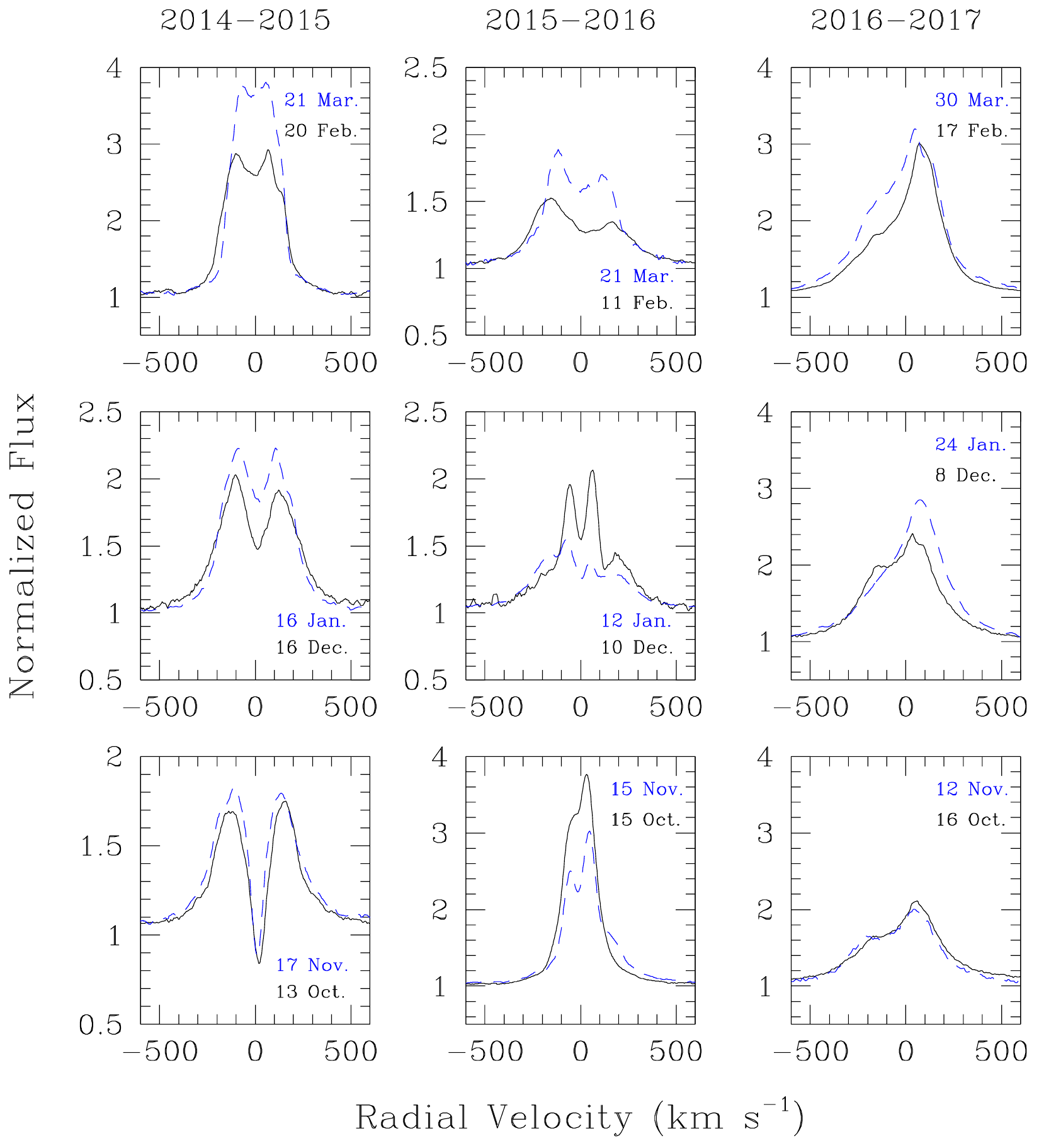}}
\end{center}
\caption{Variations of the H$\alpha$ line profile of HD~45314 over the past three observing seasons, with one season per column (the visibility window for HD~45314 ranges from early October to end of March). In each column, the older spectra (i.e.\ Oct-Nov) can be found at the bottom and the most recent ones (i.e.\ Feb-Mar) are shown at the top. For each epoch, the panels show representative pairs of observations separated by about one month (the dates are indicated by the labels) with the older spectrum shown in black and the more recent one in blue. To enhance the visibility, the vertical scale of some of the panels was changed, but the x-axis limits remained the same.\label{montageHaprofile}}
\end{figure}

From these results, we conclude that there is currently no evidence for periodic RV variations in HD~45314 with an amplitude above $\sim 10$\,km\,s$^{-1}$. Therefore, 10\,km\,s$^{-1}$ appears as a safe upper limit on the amplitude of a periodic RV modulation. The resulting constraints on the orbit for a putative compact companion of HD~45314 are shown in Fig.\,\ref{fmass}. The apparent RV variations that we have found might result from low-level line profile variations due for example\ to pulsations or contamination of the absorption lines by circumstellar emission.     
\begin{table}
\caption{Properties of the most important (sub)samples of RV measurements of HD5314.\label{RVtable}} 
\begin{center}
\begin{tabular}{c r c c}
\hline
Subsample & N & $<RV>$ &  $\sigma_{RV}$ \\
          &   & (km\,s$^{-1}$) & (km\,s$^{-1}$) \\
\hline
FEROS     &  22 & 1.8 & 11.3 \\
HEROS     &  22 & 0.2 & 14.9 \\
JGF       &  24 & 9.5 &  8.9 \\
\hline
All data  &  87 & 3.8 & 11.6 \\
\hline
\end{tabular}
\end{center}
\tablefoot{The second column lists the number of RV points. The third and fourth columns indicate the mean RV and its associated dispersion.}
\end{table}

\section{Optical variability \label{varopt}}
As shown in \citet{ibvs,Oepaper}, HD~45314 displays considerable variations of its emission lines. Previously, several episodes of rather abrupt increases of the strength of the Oe emission lines followed by more progressive decreases were recorded. Over the last four years though, the star displayed a behaviour that had not been observed before, with repeated short episodes of considerable fading of the disk emissions.

\begin{figure*}[h!tb]
\begin{center}
\resizebox{14cm}{!}{\includegraphics{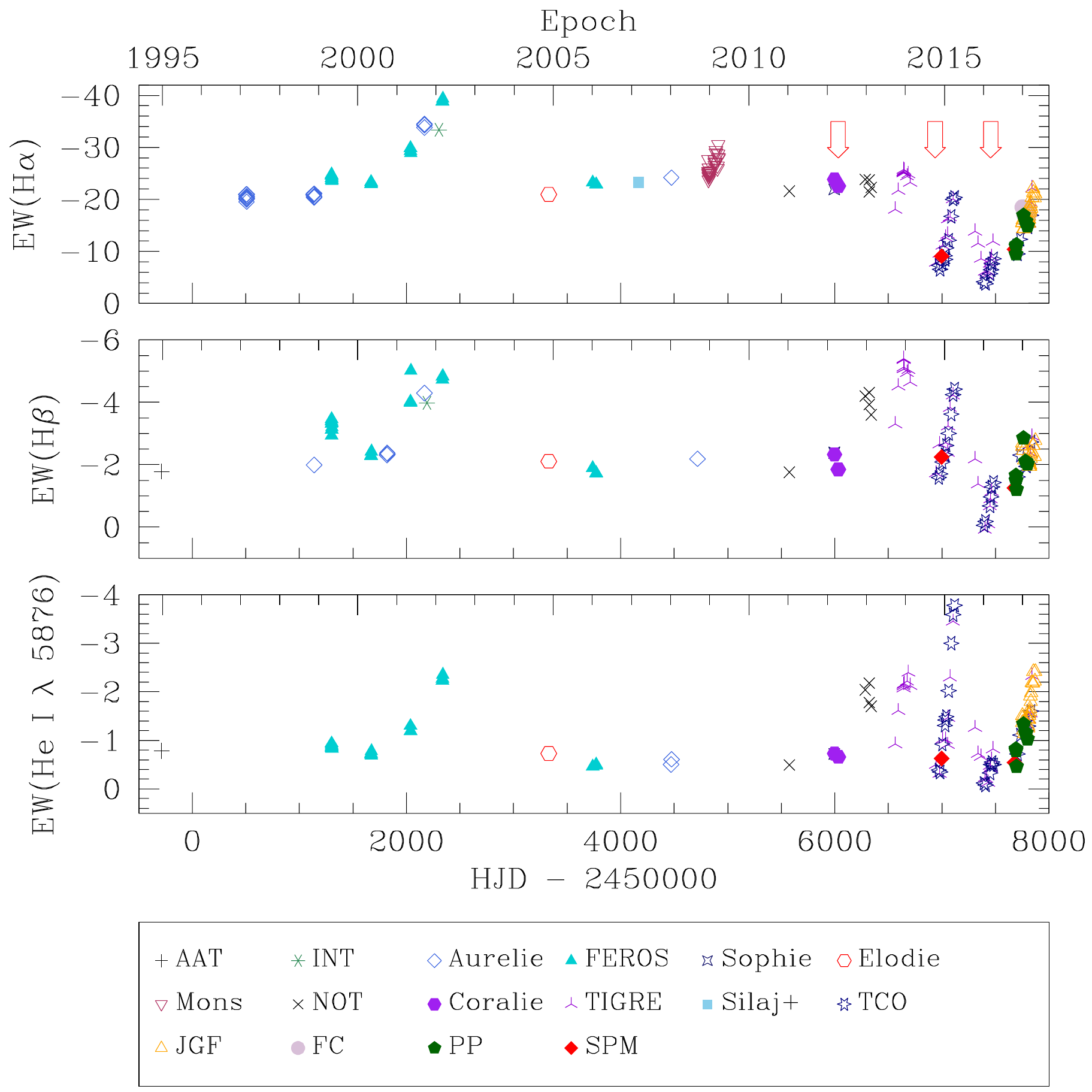}}
\end{center}
\caption{Variations of the EWs of the H$\alpha$, H$\beta,$ and He\,{\sc I} $\lambda$~5876 emission lines. The red arrows in the H$\alpha$ panel indicate the epochs of the X-ray observations. The various symbols stand for different instruments as identified in the box at the bottom.\label{EWall}}
\end{figure*}

\subsection{Emission line profile variability}
The strongest emission line in the optical spectrum of HD 45314 is H$\alpha$. This line has undergone very large variations over recent years (see Fig.\,\ref{montageHaprofile}). The plot of EW(H$\alpha$) as a function of time (Fig.\,\ref{EWall}) reveals three recent events between autumn 2014 and spring 2017 where the absolute value of the EW was well below the previously recorded lowest value of about 18\,\AA. On this figure, the three events look quite similar, but in reality they are very different as becomes clear when we consider the morphology of the line at these epochs (see Fig.\,\ref{montageHaprofile}). 

In fact, on the first observation of the 2014--2015 observing season (bottom left panel of Fig.\,\ref{montageHaprofile}), H$\alpha$ presents a shell-like profile with the central absorption reaching below the level of the continuum.\footnote{Be-shell stars are a subset of classical Be stars, which display a narrow absorption core on top of their highly rotationally-broadened emission lines. These features are usually understood as Be disks seen edge-on \citep[][see also Sect.\,\ref{shell}]{Silaj14b}.} In the following weeks, the level of the emission progressively increased and, from December 2014 onwards, the line was again in pure emission. At first, the line showed two well-distinguished emission peaks, which became more fuzzy as the strength of the line further increased.\\ 
At the start of the 2015--2016 observing season, the line displayed a completely different morphology: it appeared as a rather strong asymmetric emission, but much narrower than in previous observations. Over the following weeks, the core of the line weakened and up to three different peaks were seen simultaneously in the profile (middle panel of Fig.\,\ref{montageHaprofile}). The lowest emission level was reached in January 2016 (EW = $-4$\,\AA). From then on, the emission started to rise again with the line also becoming broader again. No shell-like event was observed during this observing season. \\
The 2016--2017 season started with a somewhat asymmetric (stronger red peak) weak H$\alpha$ emission. The strength of the emission line progressively increased, although not in a monotonic way, as did the asymmetry between the violet and red peak. At the end of the visibility period in the spring of 2017, EW(H$\alpha$) was back to $\sim -21$\,\AA,\ near the star's normal quiescent level observed at earlier epochs (Fig.\,\ref{EWall}).\\

\begin{figure*}[h!tb]
\begin{minipage}{8cm}
\resizebox{8cm}{!}{\includegraphics{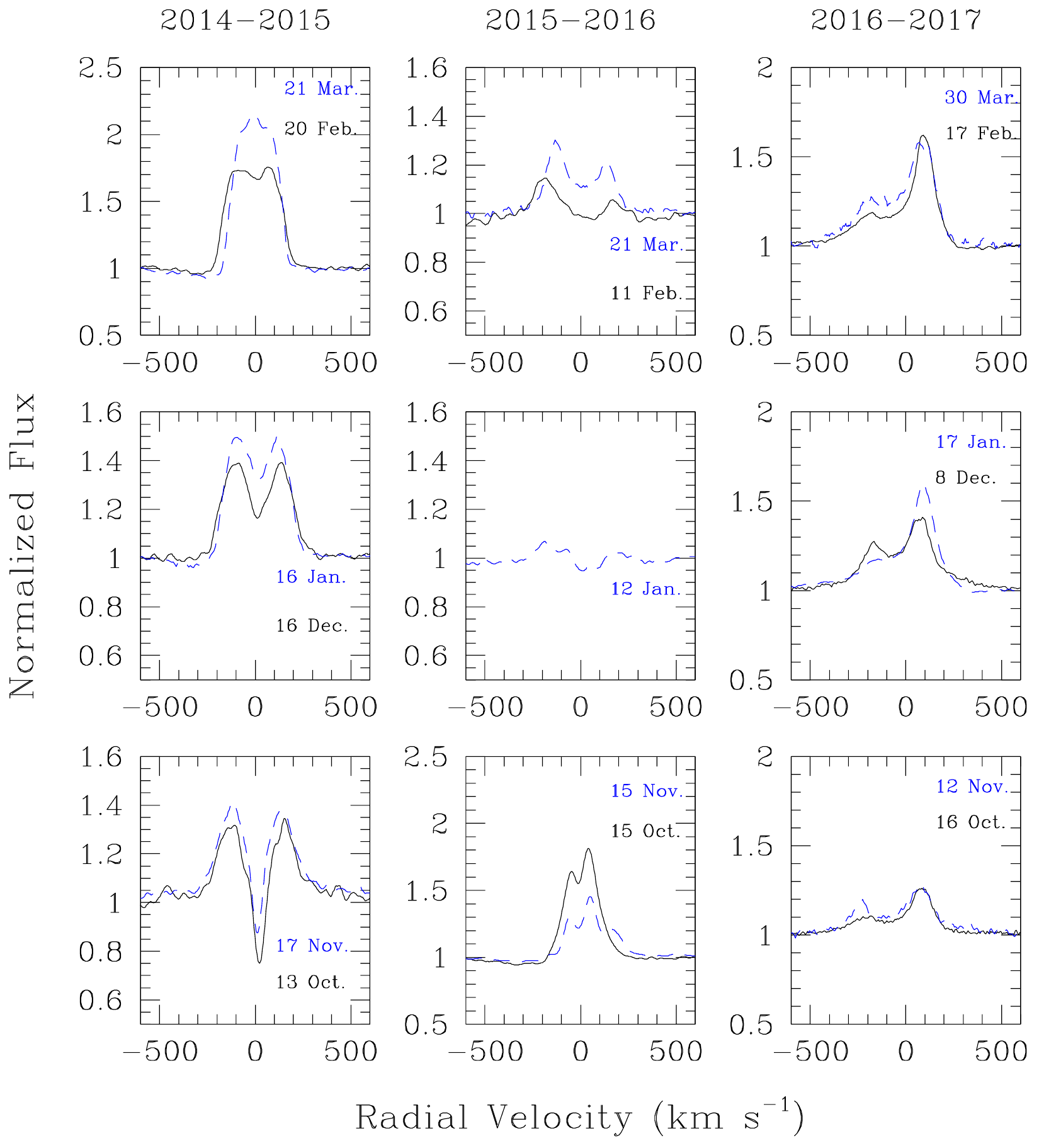}}
\end{minipage}
\hfill
\begin{minipage}{8cm}
\resizebox{8cm}{!}{\includegraphics{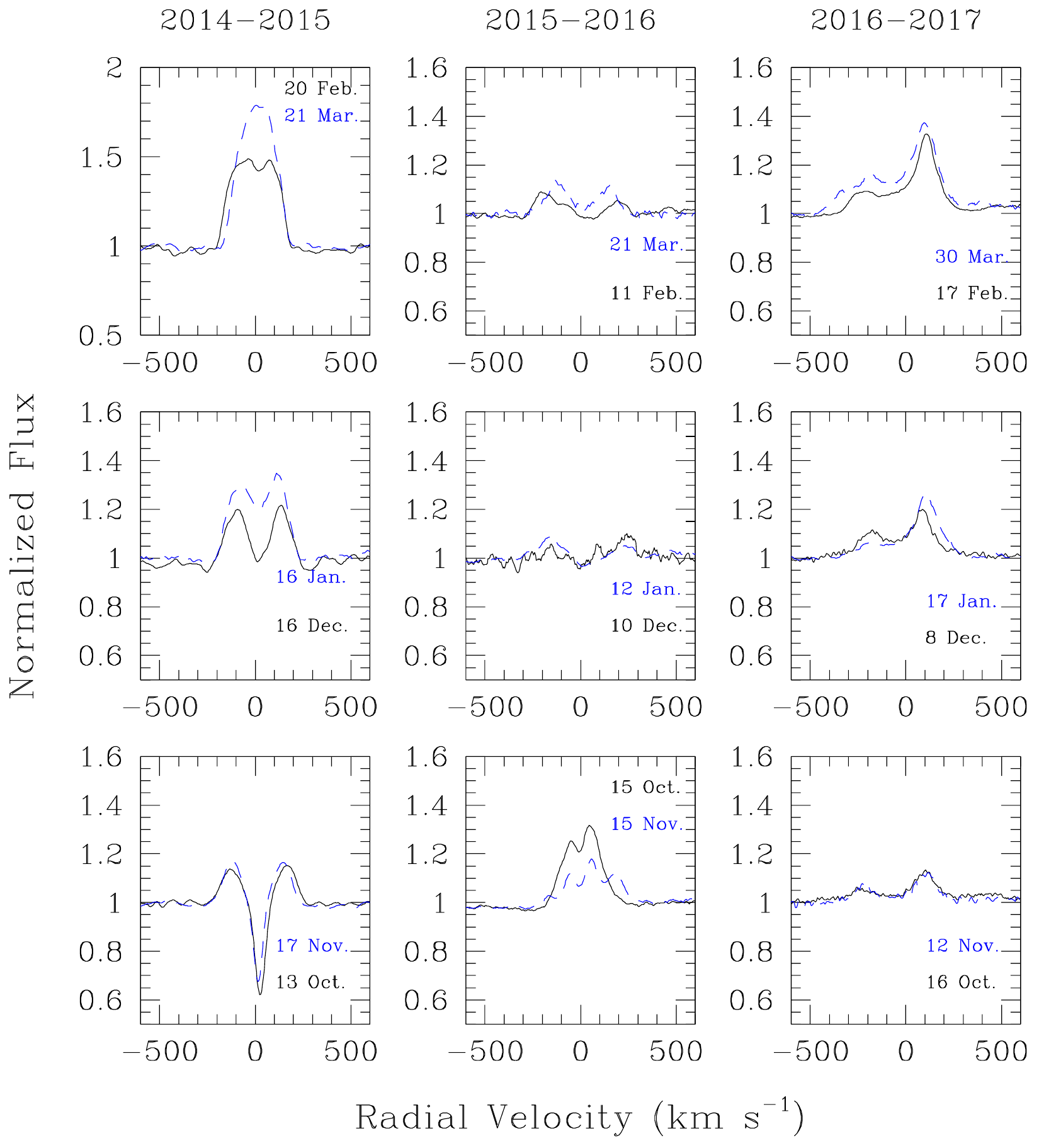}}
\end{minipage}
\caption{Same as Fig.\,\ref{montageHaprofile}, but for variations of the H$\beta$ (left) and He\,{\sc i} $\lambda$\,5876 (right) profiles in the spectrum of HD~45314 over the past three observing seasons. \label{montageHbHeIprofile}}
\end{figure*}

The H$\beta$ and He\,{\sc i} $\lambda$\,5876 lines display the same qualitative behaviour as the H$\alpha$ line as far as the line profile variations are concerned (see Fig.\,\ref{montageHbHeIprofile}). However, there appear substantial differences when the EWs are considered. Indeed, during the spring of 2015 and 2017, the strength of the He\,{\sc i} $\lambda$\,5876 line increased well beyond its nominal quiescence level. Both times, EW(He\,{\sc i} $\lambda$\,5876) reached values that had only been reached during the 2002 eruptive event or had even never been reached before. Indeed the most extreme strength of the He\,{\sc i} $\lambda$\,5876 line occurred in April 2015, after the recovery from the shell phase. The behaviour of EW(H$\beta$) is somewhat intermediate between that of EW(He\,{\sc i} $\lambda$\,5876) and EW(H$\alpha$), but actually more similar to the former. 
  
\subsection{Photometry}
Beside the spectroscopic variability, HD~45314 also displays photometric variations. \cite{SG} listed HD~45314 as a very variable star in multi-band photometry collected between 1980 and 1983. Based on {\it Hipparcos} photometry, \cite{Lefevre} classified HD~45314 as an irregular variable of early spectral type. 

\begin{figure*}[h!tb]
\begin{center}
\resizebox{14cm}{!}{\includegraphics{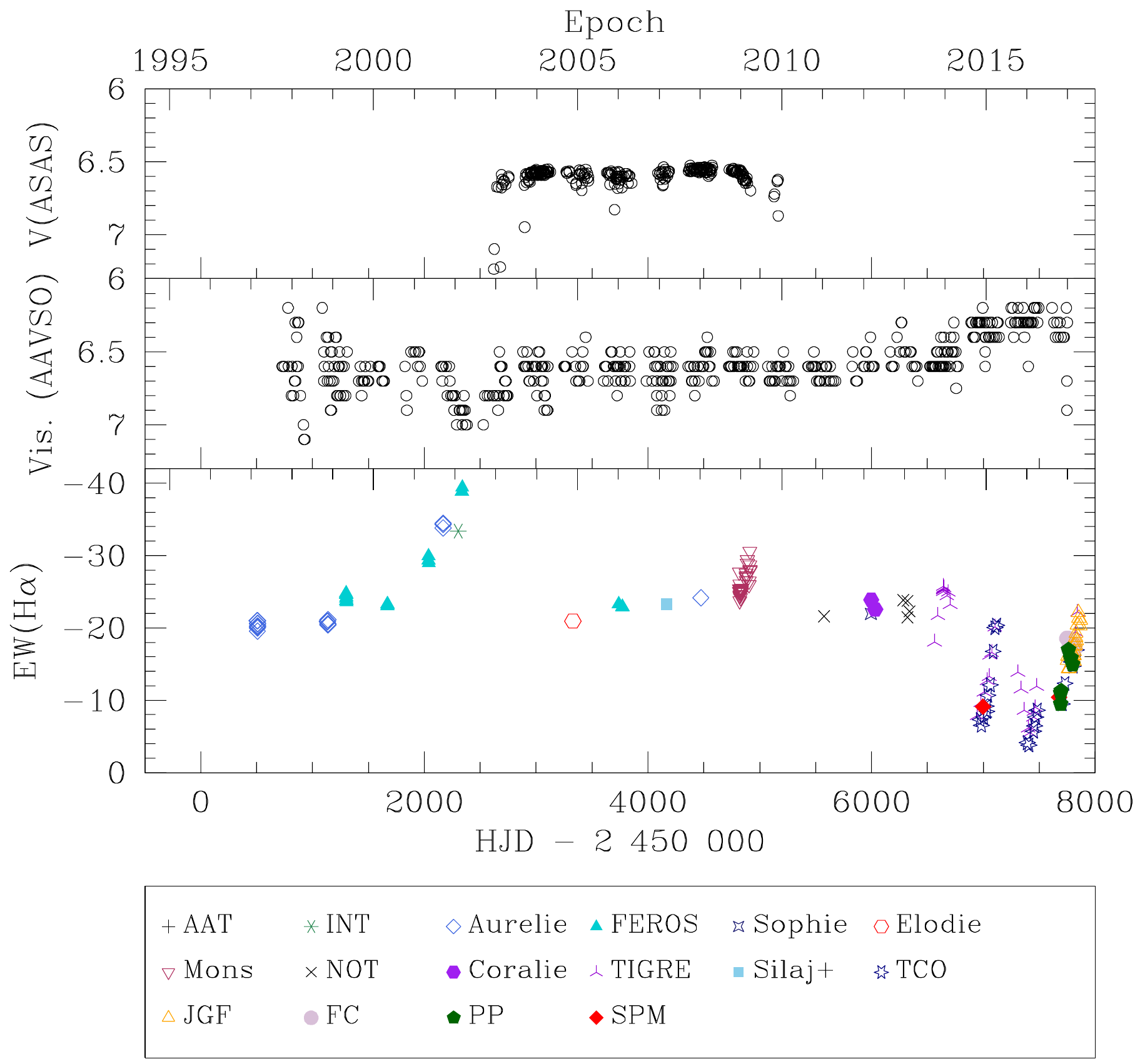}}
\end{center}
\caption{Photometric variations of HD~45314 along with the variations of the EW of the H$\alpha$ emission line. The shell event corresponds to the first minimum of EW(H$\alpha$) shortly before 2015. \label{EWHa}} 
\end{figure*}

\begin{figure}
\begin{center}
\resizebox{8cm}{!}{\includegraphics{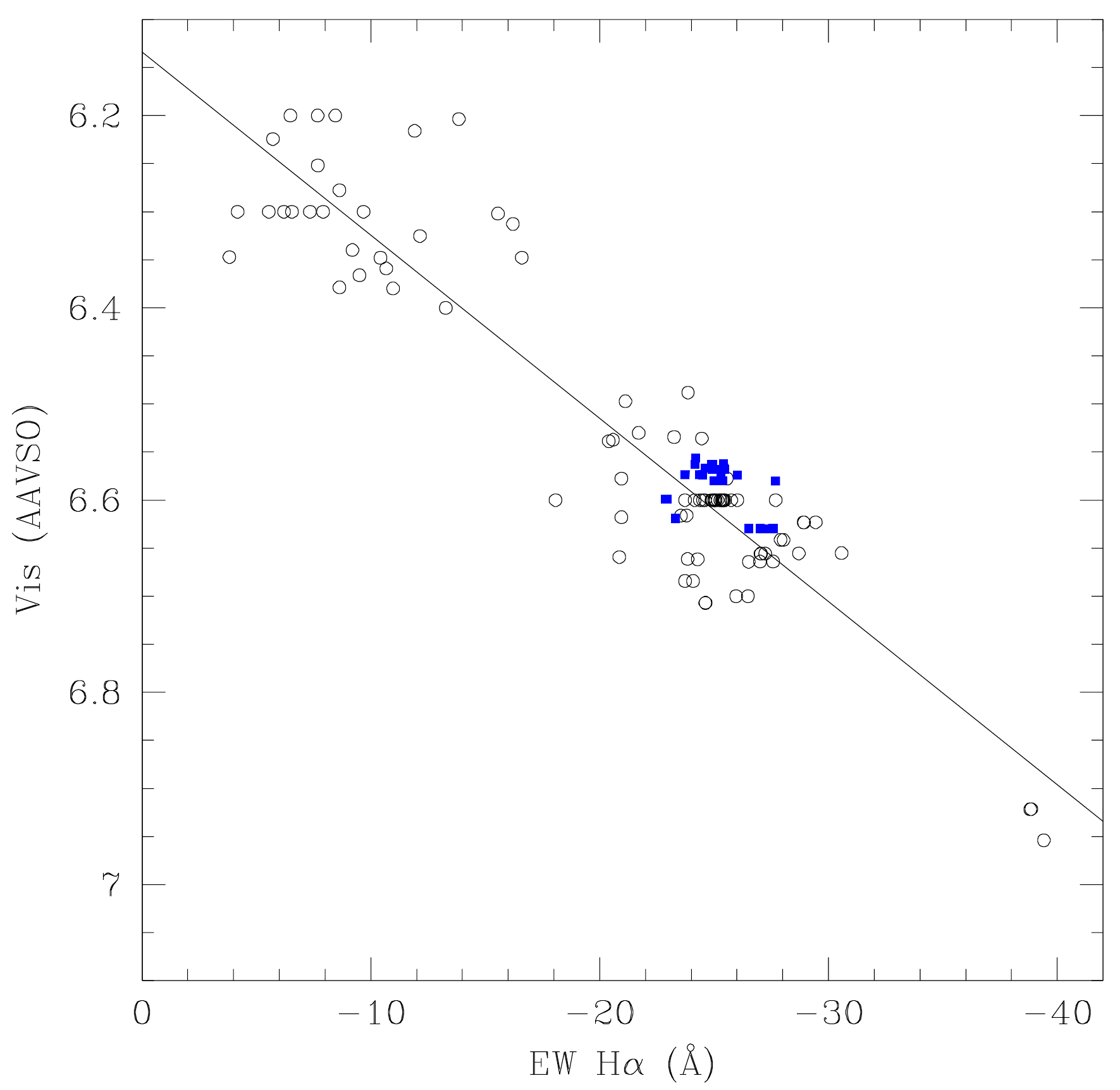}}
\end{center}
\caption{Anti-correlation between the EW of the H$\alpha$ emission and the visual magnitudes interpolated from the AAVSO data taken within less than ten\,days of the spectroscopic observation (open symbols). The solid line yields the best-fit linear relation. The filled symbols stand for the ASAS data over-plotted for comparison. \label{correlVHa}}
\end{figure}

Figure\,\ref{EWHa} displays the photometric data along with the EWs of the H$\alpha$ line. No periodic behaviour is apparent in the photometric data. Whilst there is a lot of dispersion in the AAVSO data, they nevertheless reveal an interesting anti-correlation between the optical brightness and the strength of the H$\alpha$ line (Fig.\,\ref{correlVHa}): when the H$\alpha$ emission reached a maximum strength near EW $\sim -40$\,\AA\ in 2002, the visual brightness of the star dropped by $\sim 0.5$\,mag (i.e.\ a reduction of the flux by 37\%). 
This behaviour is also supported by the trend seen at the beginning of the ASAS-3 lightcurve, which coincides with the recovery from this drop. Over the past three years, the star was apparently visually brighter, whilst the EW(H$\alpha$) was globally lower than in previous years. This relatively long-lived brightening notably includes the shell phase of autumn 2014 (Fig.\,\ref{EWHa}).

To quantify this behaviour, we computed the correlation coefficients between the sample of EW(H$\alpha$) and the AAVSO magnitudes. Since the measurements are not strictly simultaneous, we either considered the AAVSO data point that is nearest in time to the H$\alpha$ measurement, or we interpolated between the AAVSO measurements taken at dates bracketing that of our H$\alpha$ data point. In both cases, we only considered those points for which AAVSO measurements were obtained within at most ten\,days of the spectroscopic observation. The Pearson correlation coefficients amount to $r = -0.86$ and $-0.93$, respectively for the nearest neighbour or interpolated photometry methods. Using instead a Spearman rank method, we obtain $r_s = -0.80$ and $r_s =  -0.81$, respectively for the nearest neighbour or interpolated photometry cases. We thus conclude that the anti-correlation between EW(H$\alpha$) and visual magnitude is highly significant.

\citet{PR} noted that Be stars building-up emission lines often display a correlated brightening in the $V$ band.\footnote{A positive correlation was also found in $\gamma$~Cas, including during the 2010 outburst \citep{Smith12,Pollmann}.} In the case of HD~45314, we observe the opposite behaviour during the 2002 outburst. This suggests that the ejection of fresh material into the disk not only leads to a stronger line emission, but also to an enhanced continuum optical depth (possibly via a growth of the disk in the polar direction and ensuing enhanced free electron scattering). 

Finally, \citet{Gamen} report a shell-like event in the O8\,III:nep star HD~120678 that lasted less than four months. 
The ASAS lightcurve of HD~120678 revealed a slow brightening by 0.3\,mag prior to the shell event followed by a sharp decline by 0.85\,mag at the time of the shell event. HD~45314 again displays a different behaviour, as the star apparently brightened during the shell episode and remained bright afterwards (see Fig.\,\ref{EWHa}).

\subsection{V/R variations}
Variations of the ratio of the strengths of the violet and red emission peaks (V/R) are quite common among Be and Oe stars. Such variations can stem from one-armed spiral density pattern \citep{Okazaki,Okazaki16} or from binary effects \citep{Zharikov}. They occur on timescales that are much longer than the stellar or disk rotation period. 

\begin{figure}[h!tb]
\begin{center}
\resizebox{8.5cm}{!}{\includegraphics{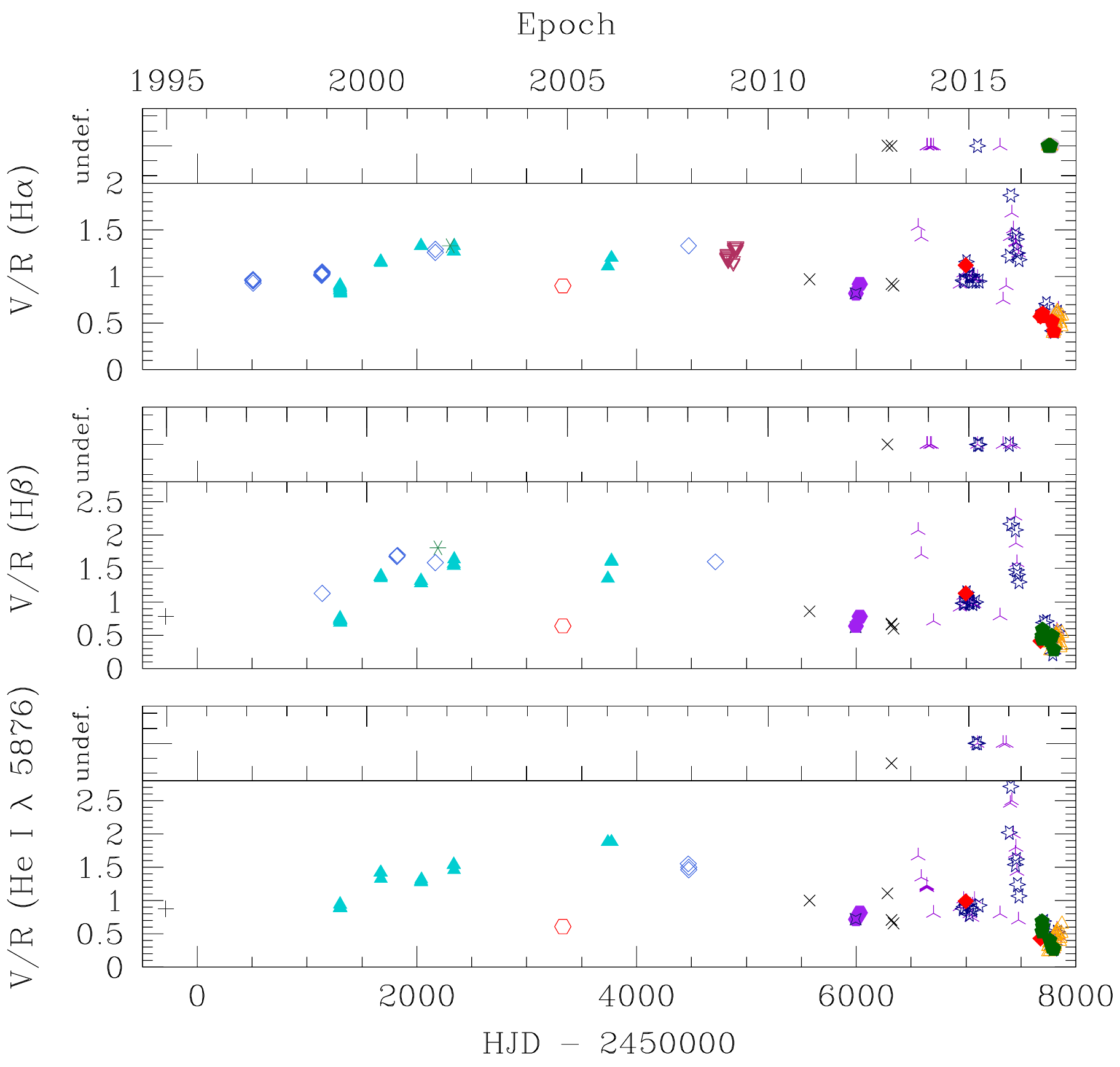}}
\end{center}
\caption{Variations of the V/R ratios of the H$\alpha$, H$\beta,$ and He\,{\sc i} $\lambda$\,5876 emission lines in the spectrum of HD~45314. Profiles either showing a single peak or displaying more than two peaks are labelled as undefined.\label{VoR}}
\end{figure}
In \citet{Oepaper}, we reported on the existence of such variations in the spectrum of HD~45314. Figure\,\ref{VoR} illustrates the V/R ratios of the three lines studied here. As can be seen, the amplitude of the variations dramatically increased during the episode of spectacular variations over the last four years, except during the shell episode where all three lines displayed V/R $\simeq 1$. After discarding those dates for which the V/R was undetermined either because the profiles exhibited a single peak or because of the presence of three or more peaks, we analysed the time series of V/R values using the Fourier method of \citet{HMM}. In this way, we found that since December 2012, the V/R variations display a rough timescale near 1000\,days in all three emission lines.

\subsection{H$\alpha$ line profile fitting \label{fitsHalpha}}
In \citet{Oepaper}, we attempted to fit the H$\alpha$ line profiles of two Oe stars, HD~45314 and HD~60848, using a simple model of an optically-thick Keplerian disk based on the model of \citet{HV}, which follows the formalism of \cite{HM} for accretion disks in cataclysmic variables. Whilst this approach allowed us to achieve consistent fits of reasonable quality for the H$\alpha$ line of HD~60848 (see also Appendix \ref{appendix}), we were unable to reach similar results in the case of HD~45314. For instance the shape of the wings of the emission lines could not be fit assuming a pure isothermal Keplerian disk combined with the most likely stellar parameters. We have thus modified our model by including two additional aspects. 

First we have dropped the assumption of a strictly isothermal disk with $T_{\rm disk} = 0.6\,T_{\rm eff}$. Instead, we assume that the temperature decreases outwards according to $T_{\rm disk} = 0.8\,T_{\rm eff}\,\left(\frac{R_*}{r}\right)^p$ with $p$ in the range between 0 and 0.5 as considered in the theoretical models of \citet{Kurfuerst}.
As a second step, we have adjusted our model by including an additional Gaussian velocity field to check whether including such an extra velocity field improves the quality of our fits. Whilst a turbulent velocity field could result from the magnetorotational instability in the disk, we stress that in our model the extra velocity field is purely empirical and does not necessarily reflect a genuine turbulence (see further discussion below).  

Let us recall that in our model the disk is assumed to extend from the stellar surface out to an outer radius $R_{\rm disk}$. The disk's particle number density follows a power law of index $\alpha$ in radius ($r$) and has a Gaussian profile in elevation ($z$) above the disk plane:
\begin{equation} 
n(r,z) = n_0\,\left(\frac{r}{R_*}\right)^{-\alpha}\,\exp{\left[-\frac{1}{2}\left(\frac{z}{H(r)}\right)^2\right]},
\end{equation}
where $H(r)$ is the local scale height.
We assume that the extra velocity field scales with distance as 
\begin{equation}
v_{\rm extra} = v_{\rm extra,0}\,\left(\frac{r}{R_*}\right)^{-3 + \alpha/2}.
\end{equation} 
The extra velocity at any position in the disk can then be obtained provided we specify its value at the inner edge of the disk $v_{\rm extra, 0}$.     

Once $v_{\rm extra}$ is specified, the line-of-sight optical depth ($\tau$) from any position in the disk is computed accounting for the line profile width due to the thermal velocity, the shear broadening, and the extra velocity. All other aspects of the problem, including the correction of the photospheric absorption line, are dealt with in the same way as done in \citet{Oepaper}. The models contain six free parameters: the disk's apparent inclination $i_{\rm disk}$, the outer disk radius $R_{\rm disk}$, the power-law index $\alpha$, as well as the particle number density $n_0$, the extra velocity $v_{\rm extra, 0}$ at the inner edge of the disk, and the $p$ exponent of the temperature law. 
 
We computed a grid of synthetic line profiles over a wide range of parameters: $i_{\rm disk}$ varying from 15 to $85^{\circ}$ with an increment of $5^{\circ}$, $R_{\rm disk}$ between 2 and 60\,$R_*$ by increments of 1\,$R_*$, $\alpha$ between 1.5 and 3.5 with an increment of 0.5, $n_0$ between 16 and 1600\,cm$^{-3}$ with an increment of 16\,cm$^{-3}$, and $v_{\rm extra,0}$ ranging from 0 to 250\,km\,s$^{-1}$ with increments of 50\,km\,s$^{-1}$. We note that $n_0$ stands for the number density of the particles that produce the line under consideration (i.e.\ the H\,{\sc i} atoms having their electron on the $n=3$ level). 

These synthetic spectra were then compared to the observed line profiles. For this purpose, we chose the same observed profiles as in \citet{Oepaper}, that is,\ mean FEROS spectra of May 1999, May 2000, March 2002, the Fiber-fed Echelle Spectrograph (FIES) spectrum of January 2011, and the HEROS spectrum of 20 December 2013. We further included HEROS spectra taken on 13 October 2014, 16 December 2014, 20 February 2015, 15 October 2015, 12 January 2016, and 21 March 2016, as well as the Coralie spectrum obtained on 4 April 2012. The latter spectra provide a representative sample of the line morphologies observed over recent years and further include the observations that are closest in time to our X-ray spectra.
 
\begin{figure*}
\begin{minipage}{4.5cm}
\resizebox{4.5cm}{!}{\includegraphics{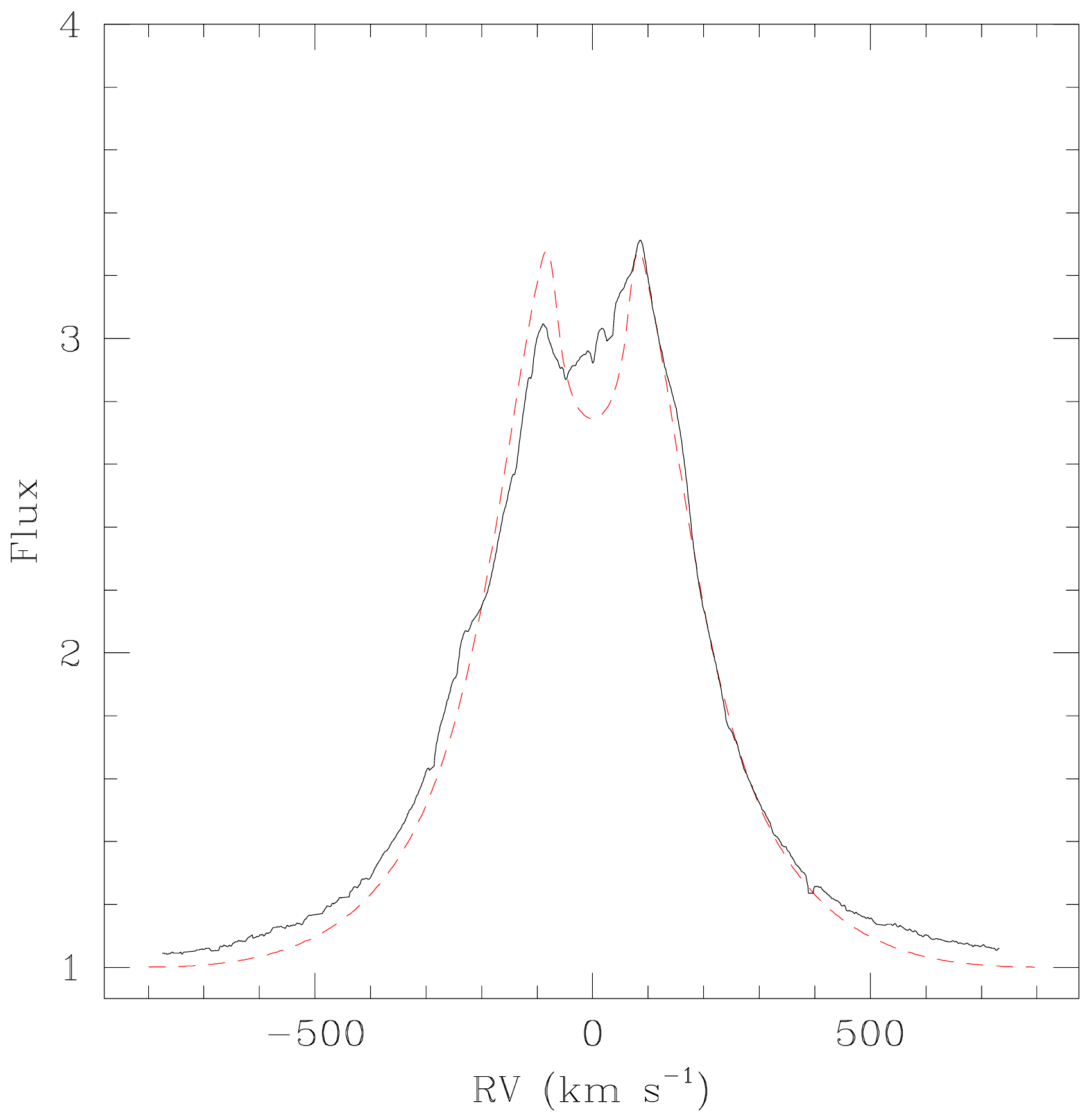}}
\end{minipage}
\hfill
\begin{minipage}{4.5cm}
\resizebox{4.5cm}{!}{\includegraphics{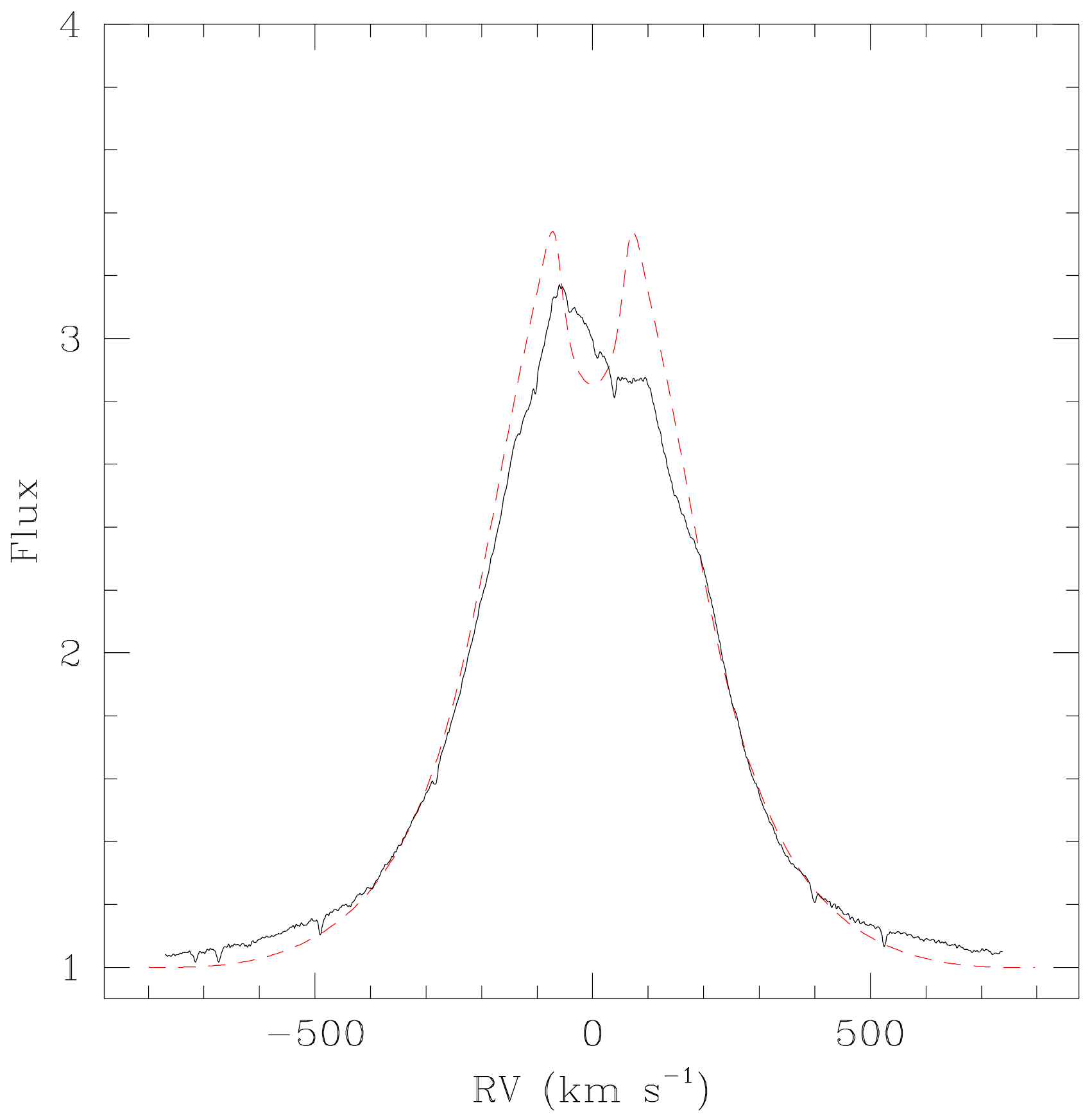}}
\end{minipage}
\hfill
\begin{minipage}{4.5cm}
\resizebox{4.5cm}{!}{\includegraphics{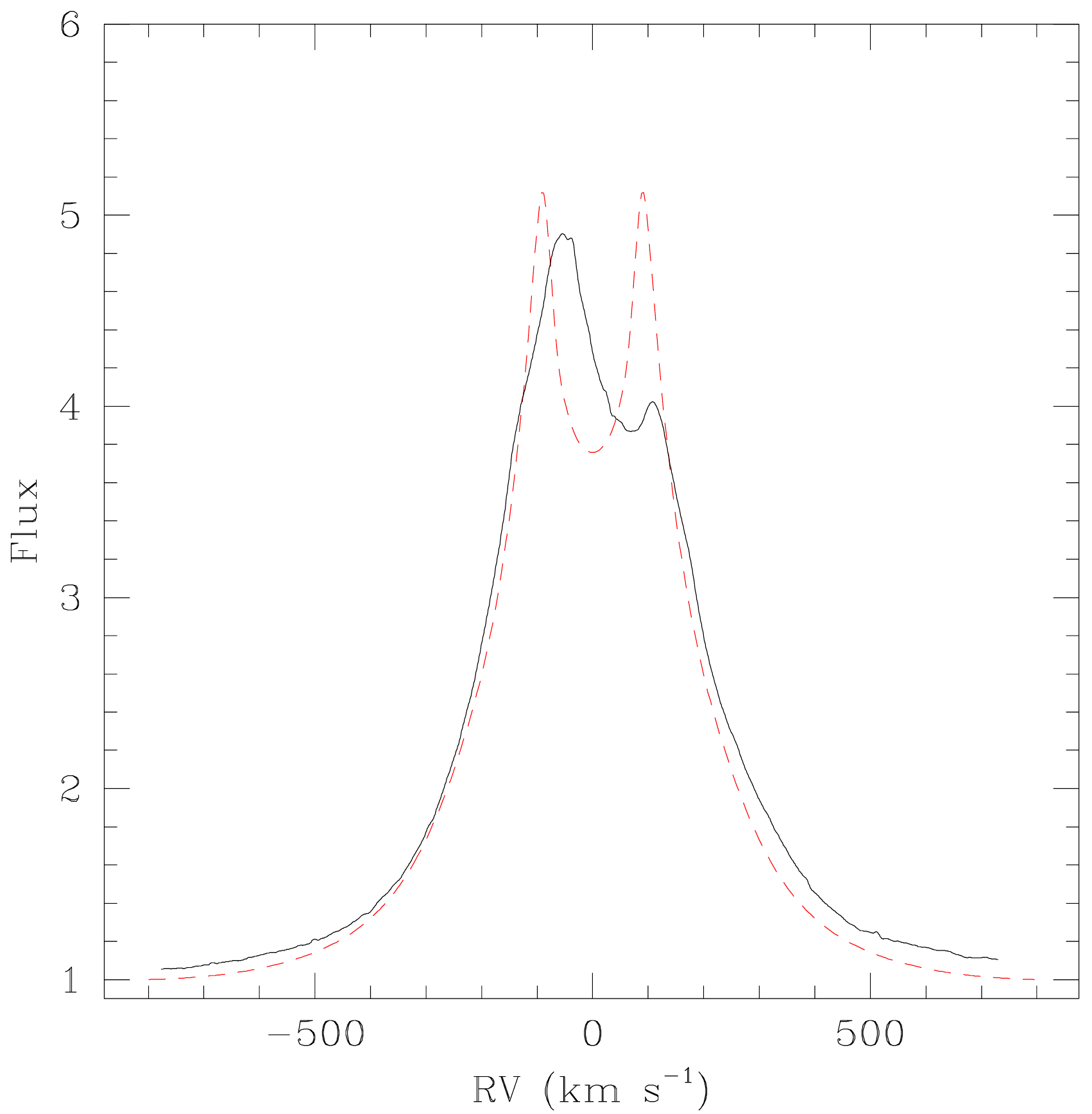}}
\end{minipage}
\hfill
\begin{minipage}{4.5cm}
\resizebox{4.5cm}{!}{\includegraphics{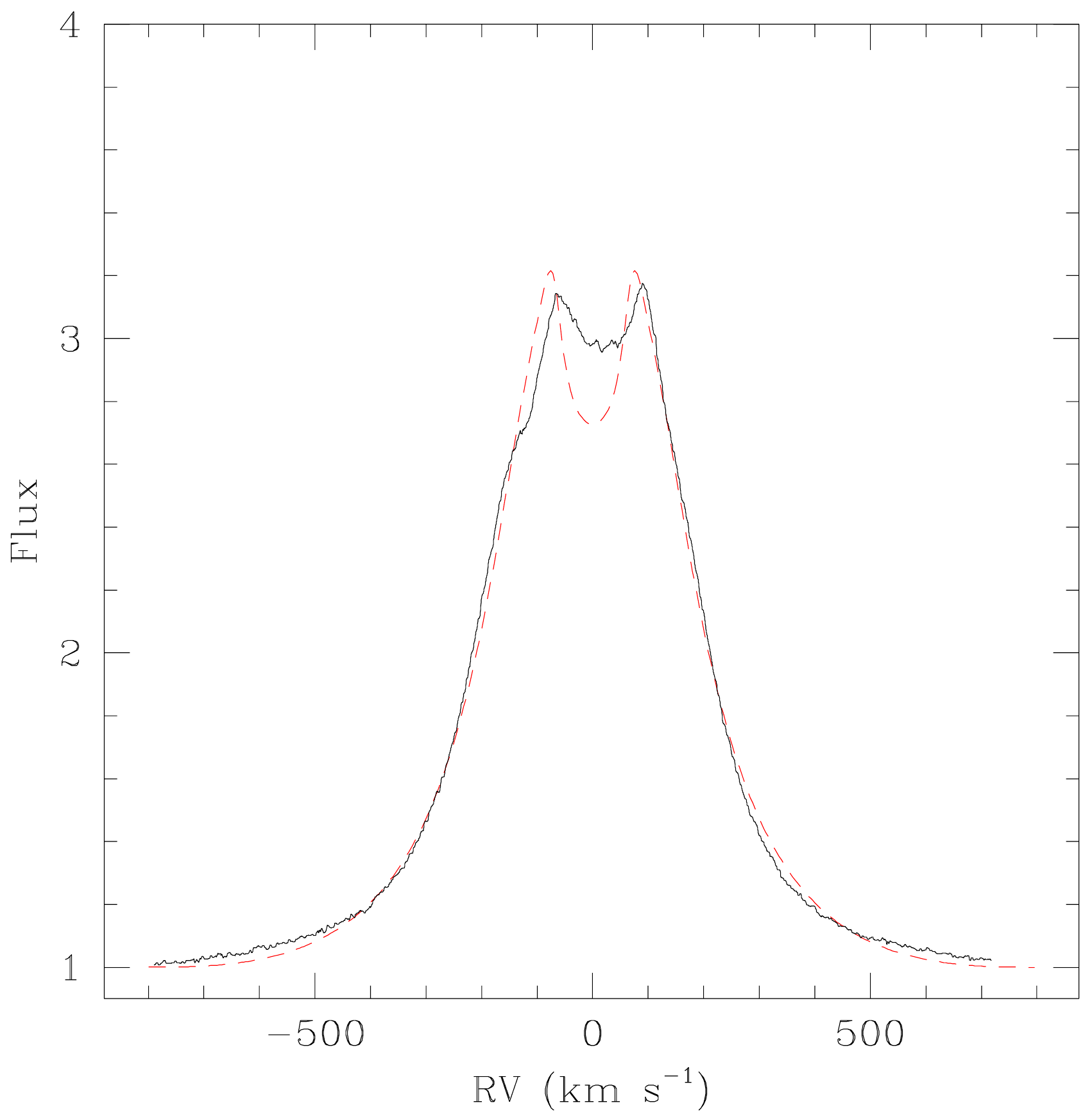}}
\end{minipage}
\begin{minipage}{4.5cm}
\resizebox{4.5cm}{!}{\includegraphics{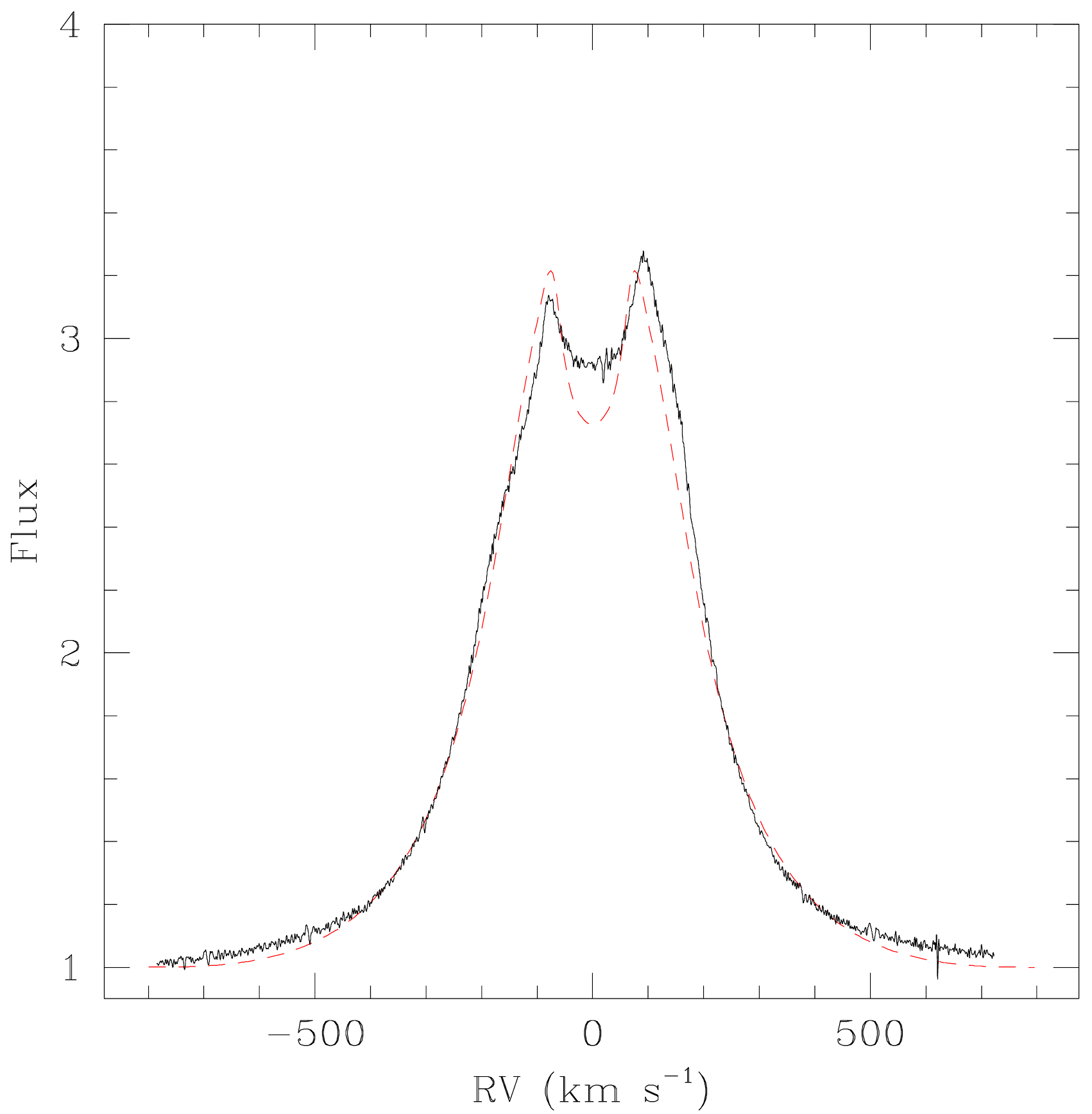}}
\end{minipage}
\hfill
\begin{minipage}{4.5cm}
\resizebox{4.5cm}{!}{\includegraphics{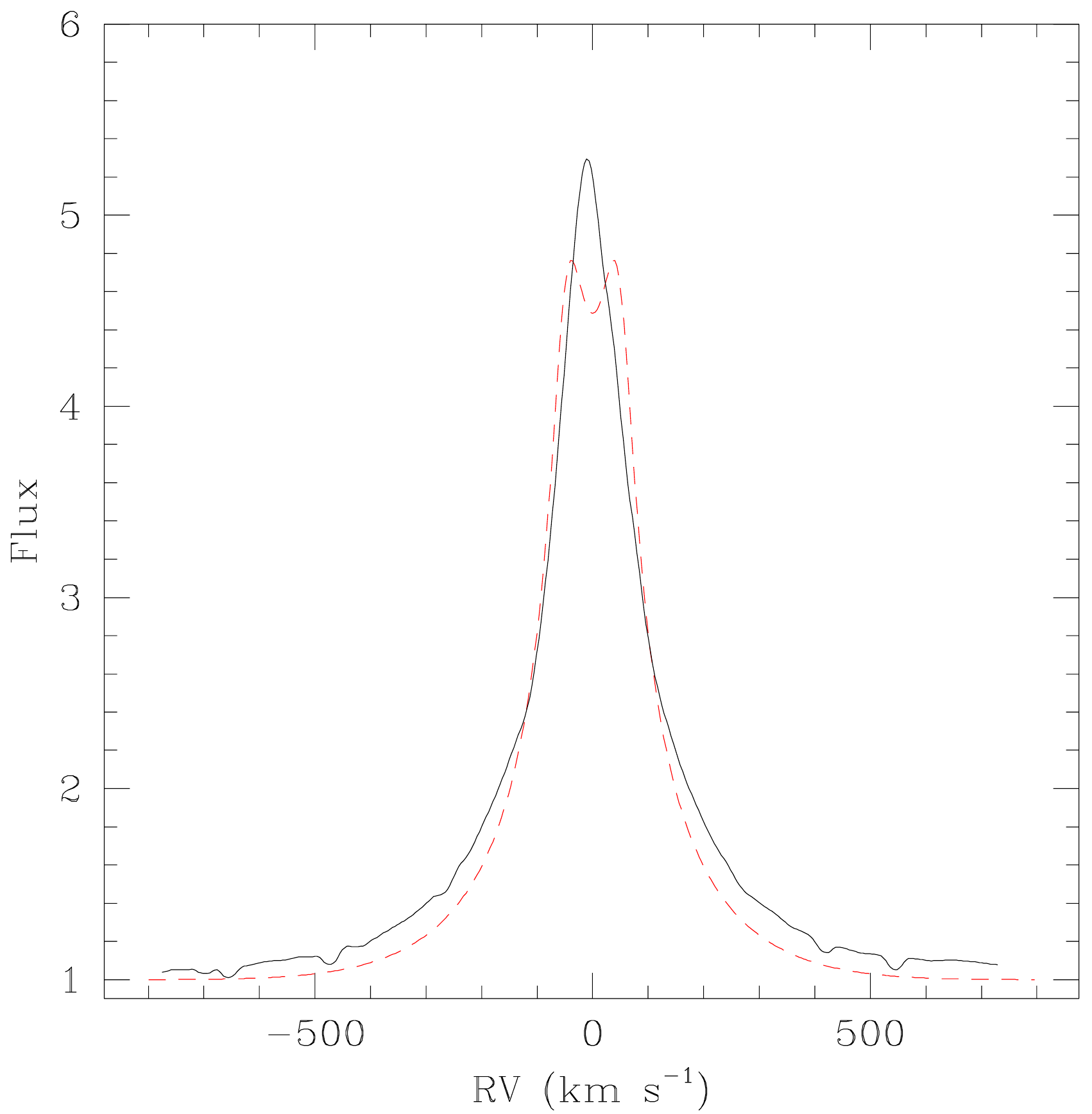}}
\end{minipage}
\hfill
\begin{minipage}{4.5cm}
\resizebox{4.5cm}{!}{\includegraphics{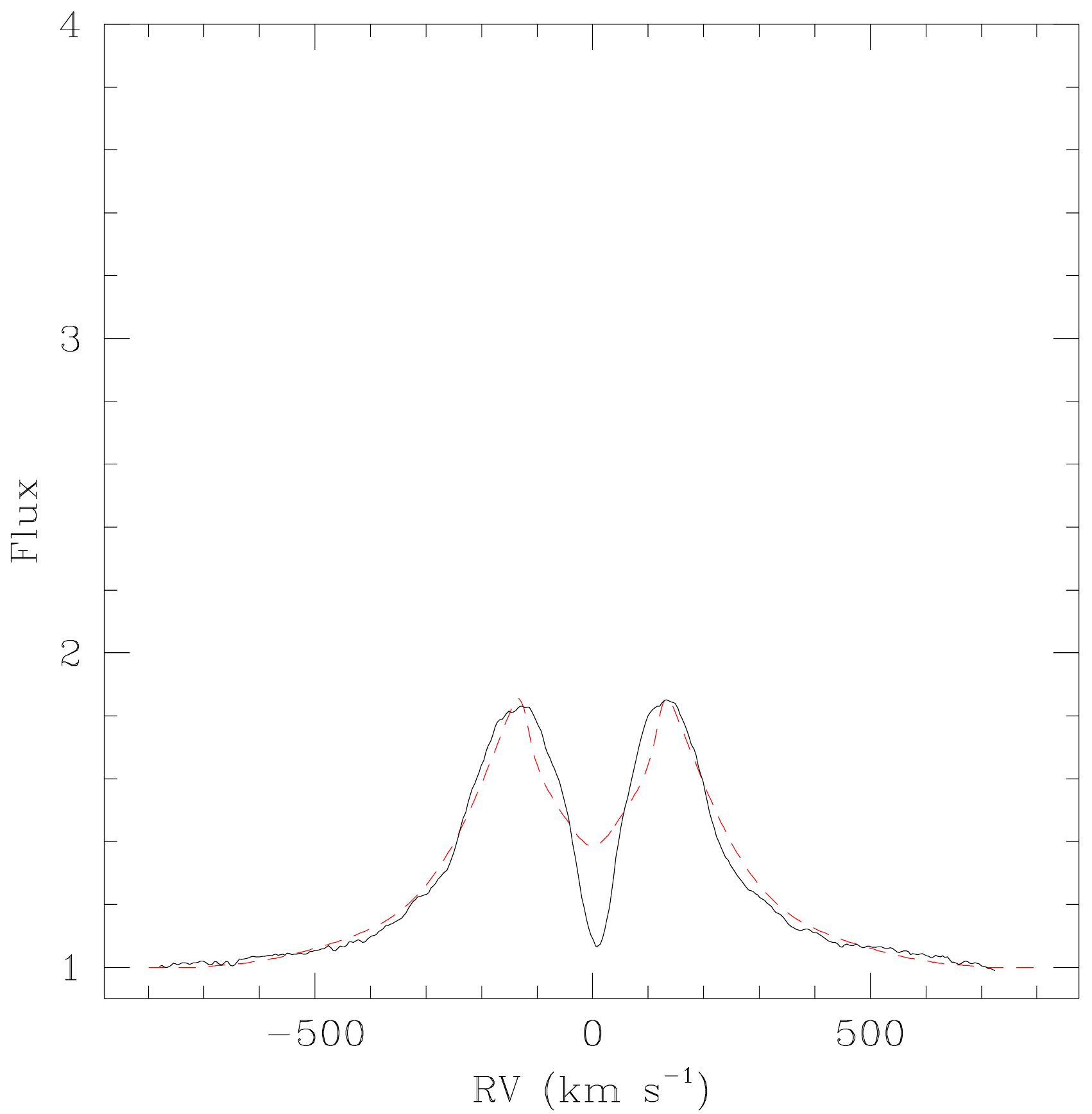}}
\end{minipage}
\hfill
\begin{minipage}{4.5cm}
\resizebox{4.5cm}{!}{\includegraphics{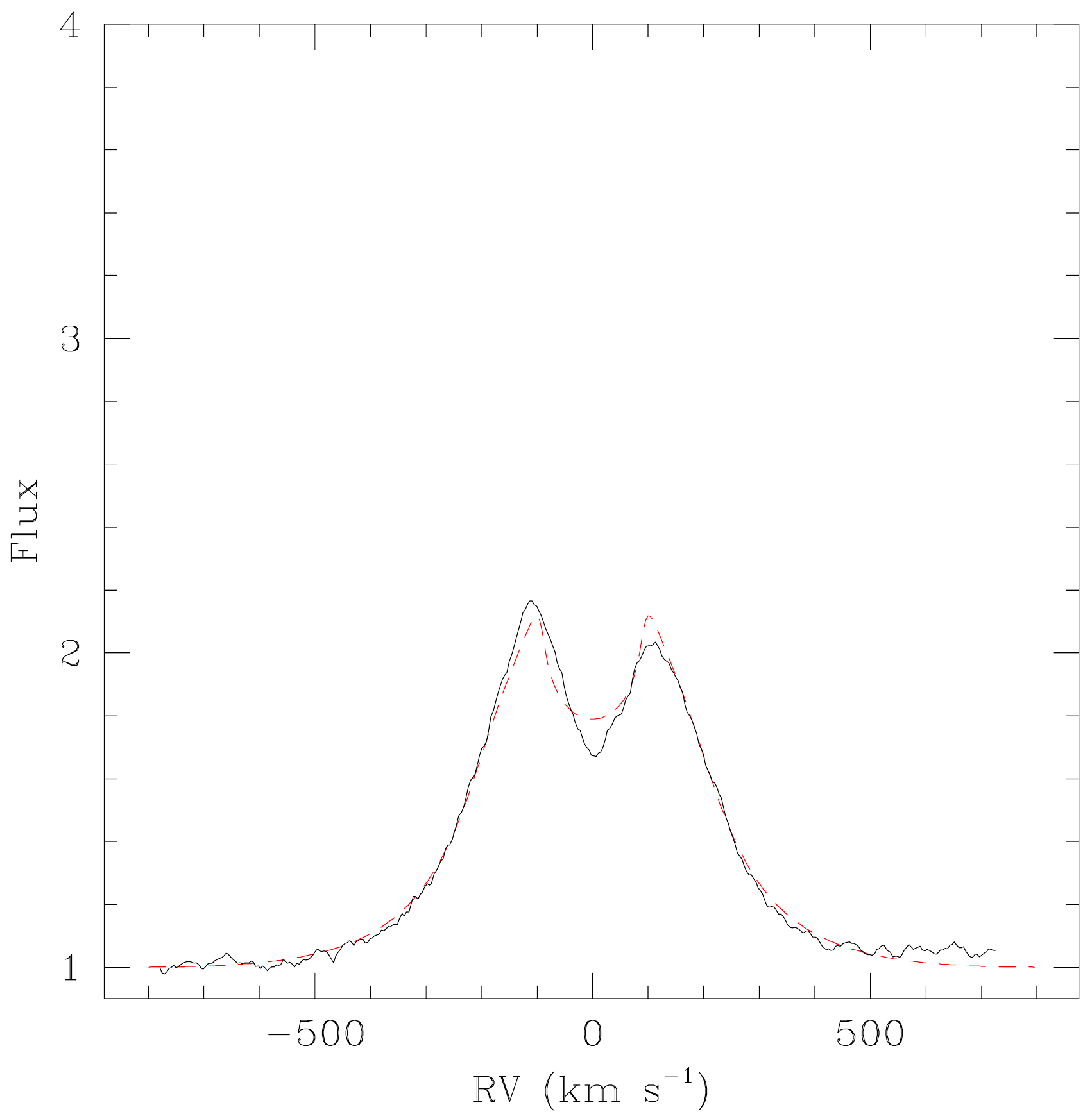}}
\end{minipage}
\begin{minipage}{4.5cm}
\resizebox{4.5cm}{!}{\includegraphics{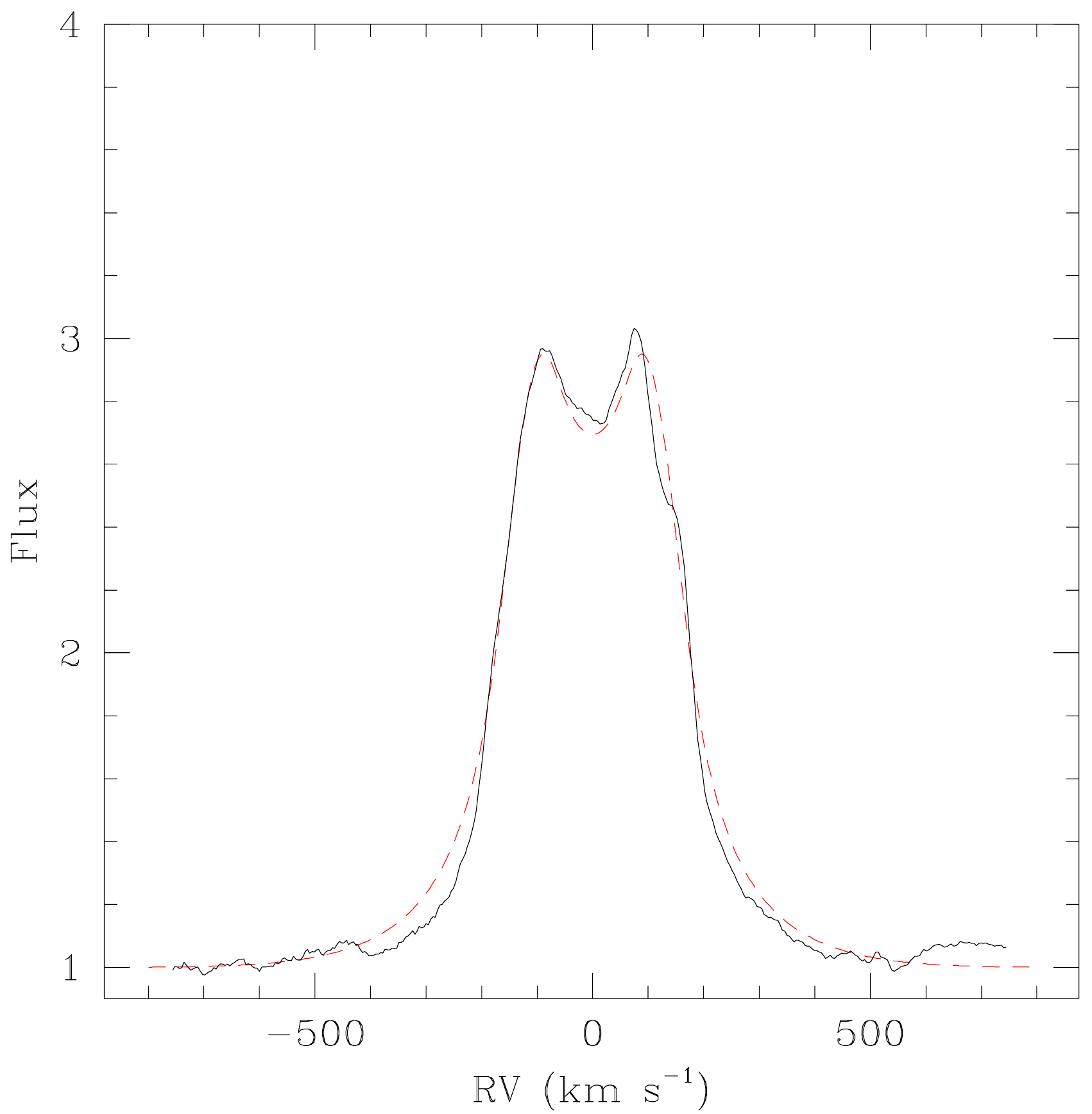}}
\end{minipage}
\hfill
\begin{minipage}{4.5cm}
\resizebox{4.5cm}{!}{\includegraphics{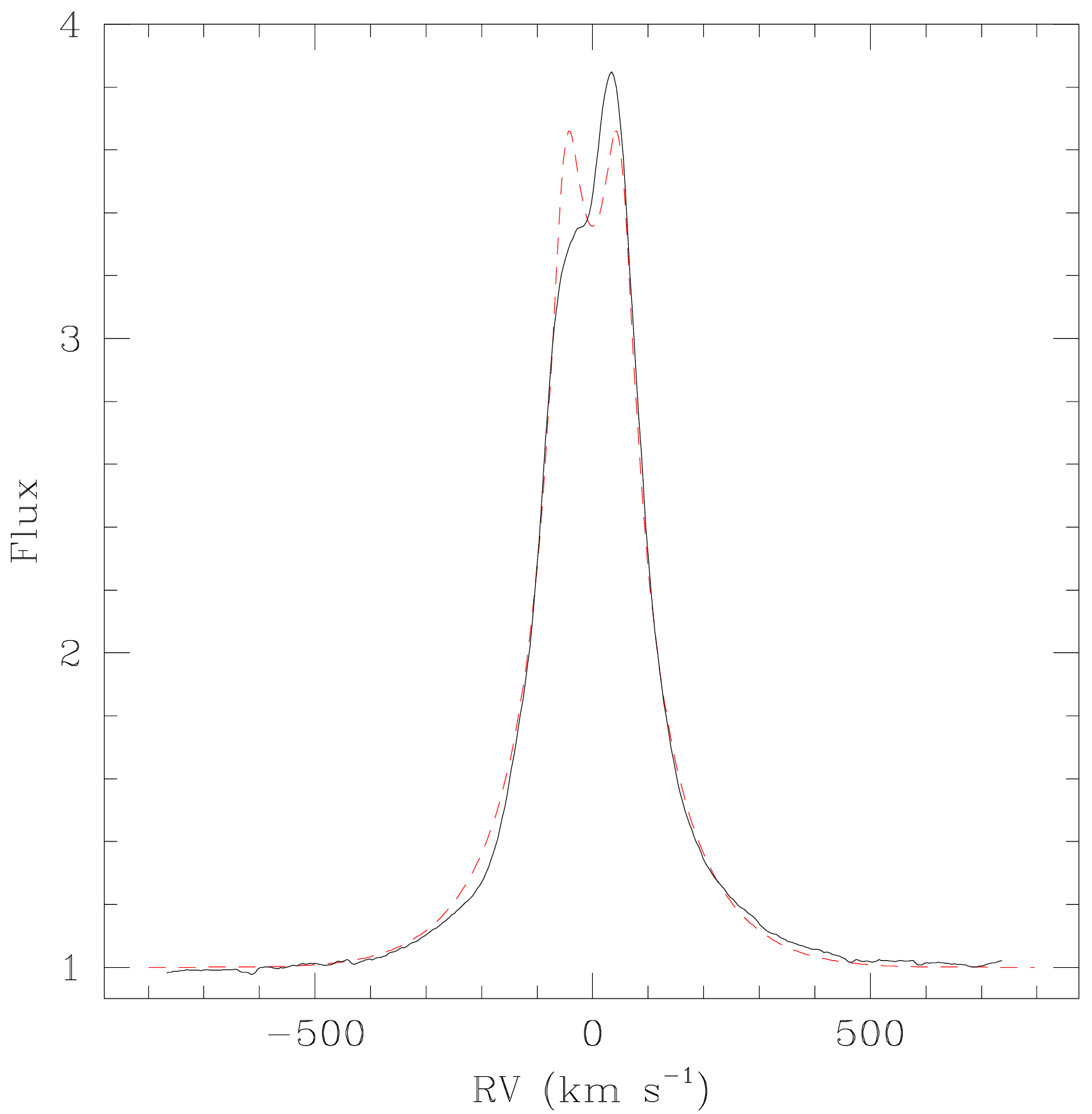}}
\end{minipage}
\hfill
\begin{minipage}{4.5cm}
\resizebox{4.5cm}{!}{\includegraphics{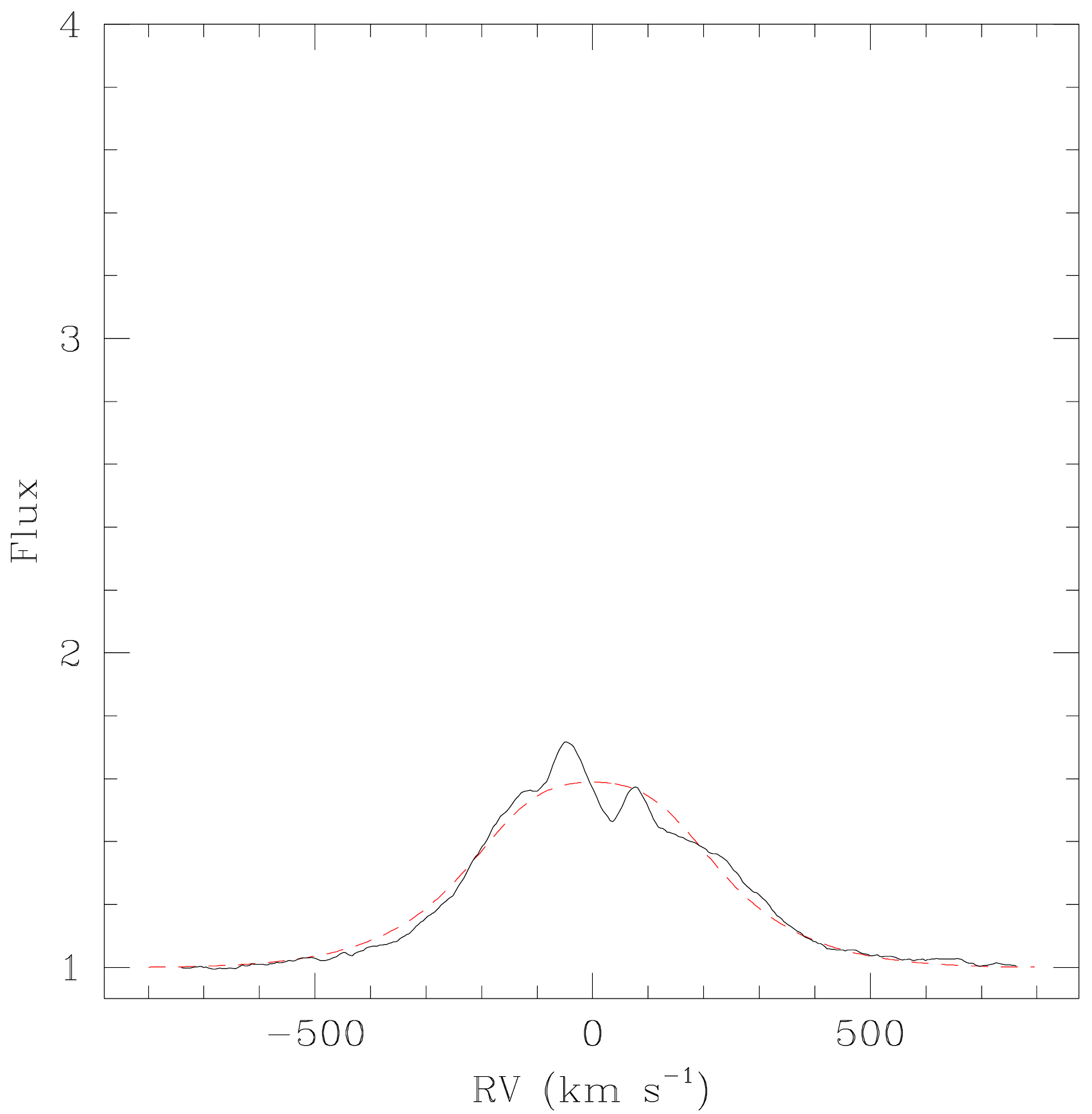}}
\end{minipage}
\hfill
\begin{minipage}{4.5cm}
\resizebox{4.5cm}{!}{\includegraphics{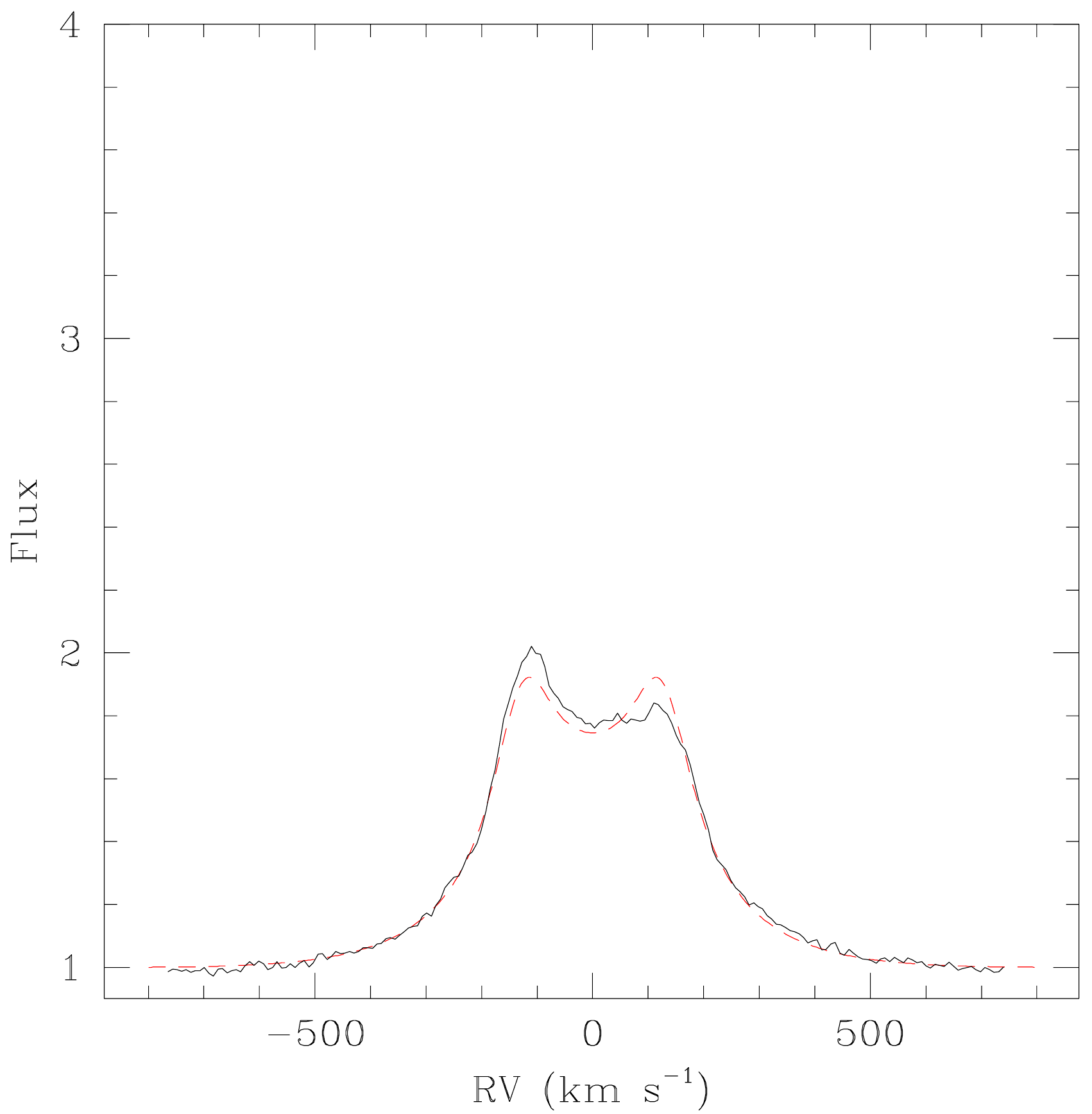}}
\end{minipage}
\caption{Observed (black solid line) and best-fit synthetic (red dashed line) H$\alpha$ profiles for different epochs. The top row illustrates (from left to right) the spectra of May 1999, May 2000, March 2002, and January 2011. The middle row corresponds to the data from April 2012, December 2013, October 2014, and December 2014. The bottom row yields the data from February 2015, October 2015, January 2016, and March 2016. All panels have the same vertical scale, except for the March 2002 and December 2013 spectra.\label{montage}}
\end{figure*}

By construction, our model implicitly assumes an axisymmetric disk. Hence, the synthetic line profiles are also symmetric about the central wavelength and the model cannot account for the V/R variations. Figure\,\ref{montage} illustrates the best fits that we obtained for each epoch. It shows that, apart from the shape and relative strength of the violet and red peaks, most of the spectra are now much better fitted than in our previous work. Yet, there remain a number of issues. For instance, our model has difficulties fitting the very narrow, single-peaked line core observed in 2013. The profile observed during the 2002 outburst is not well reproduced either. The parameters of these two epochs should thus be considered with caution. The model parameters corresponding to our best fits are listed in Table\,\ref{paramfits}.

\begin{table}
\caption{Model parameters of the fits in Fig.\,\ref{montage}.\label{paramfits}}
\tiny
\begin{tabular}{c c c c c c}
\hline
Epoch & $i_{\rm disk}$& $R_{\rm disk}$ & $n_0$ & $\alpha$ & $v_{\rm extra,0}$ \\
      & $(^{\circ})$ & $R_*$ & (cm$^{-3}$) &    & (km\,s$^{-1}$) \\
\hline
May 1999   & $43.3 \pm 3.4$ & $48.5 \pm 9.8$ & $1387 \pm 84$ & $3.5$ & $238 \pm 21$ \\
May 2000   & $43.4 \pm 6.6$ & $50.8 \pm 14.4$ & $1408 \pm 80$ & $3.4 \pm 0.3$ & $218 \pm 37$ \\
Mar.\ 2002 & $59.0 \pm 2.1$ & $49.9 \pm 3.9$ & $930 \pm 20$ & $3.0$ & $250$ \\
Jan.\ 2011 & $38.2 \pm 5.9$ & $47.2 \pm 15.8$ & $1396 \pm 81$ & $3.1 \pm 0.6$ & $222 \pm 37$ \\
Apr.\ 2012 & $38.5 \pm 5.7$ & $43.3 \pm 14.1$ & $1353 \pm 88$ & $3.0 \pm 0.6$ & $240 \pm 20$ \\
Dec.\ 2013 & $\leq 15.0$ & $13.5 \pm 1.5$ & $1200 \pm 36$ & $3.5$ & $250$ \\
Oct.\ 2014 & $80.0$ & $31.7 \pm 4.7$ & $1156 \pm 115$ & $3.5$ & $30 \pm 24$ \\
Dec.\ 2014 & $45.5 \pm 16.3$ & $29.6 \pm 12.9$ & $956 \pm 362$ & $3.1 \pm 0.4$ & $118 \pm 92$ \\
Feb.\ 2015 & $20.0$ & $4.5 \pm 0.5$ & $851 \pm 253$ & $2.8 \pm 0.5$ & $210 \pm 38$ \\
Oct.\ 2015 & $18.6 \pm 2.3$ & $41.6 \pm 17.6$ & $914 \pm 97$ & $3.5$ & $222 \pm 25$ \\
Jan.\ 2016 & $32.9 \pm 10.8$ & $31.0 \pm 22.2$ & $573 \pm 232$ & $3.3 \pm 0.4$ & $193 \pm 71$ \\
Mar.\ 2016 & $29.0 \pm 5.2$ & $25.2 \pm 17.5$ & $763 \pm 310$ & $3.1 \pm 0.4$ & $195 \pm 49$ \\
\hline
\end{tabular}
\tablefoot{Quoted numbers are the means and standard deviations of the parameters of the models that have a $\chi^2$ within the 90\% confidence range from the best-fit model. Values for which no uncertainties are provided correspond to situations where all models in the 90\% confidence range had the same value of the parameter.} 
\end{table}

Several problems are apparent when we consider the values of the parameters in Table\,\ref{paramfits}. The first problem concerns the values of the apparent disk inclination with respect to our line of sight. The results of our models suggest that the disk's apparent inclination with respect to our line of sight remained close to 40 -- $45^{\circ}$ until 2012. However, with the onset of the spectacular variations in 2013, the value of the disk's apparent inclination started changing. In 2013, we observed a narrow single-peaked profile that suggests a low apparent inclination (possibly lower than the lower limit of our grid of $15^{\circ}$). Ten months later, the star displayed a shell-like profile suggestive of a very high apparent inclination. In the subsequent months and years, the best-fit value of the apparent inclination oscillated.\\
For a disk located in the plane of the stellar equator, one does not expect genuine inclination changes, unless the stellar rotation axis itself is precessing. Such a precession of the stellar rotation axis would lead to large changes (by a factor $\sim 3.7$) of the projected stellar rotational velocity $v\,\sin{i}$. Such changes are clearly not seen in the behaviour of the He~{\sc ii} $\lambda$\,4686 absorption line of HD~45314, thereby ruling out precession as a possible origin of the observed behaviour of $i_{\rm disk}$. We will come back to this aspect in Sect.\,\ref{shell}.   

The second issue are the values of the best-fit extra velocities at the inner disk edge. Except for a few cases, most of the observed spectra require a very large extra velocity often exceeding 200\,km\,s$^{-1}$. Such a velocity amounts to more than one third of the Keplerian velocity at the stellar surface (690\,km\,s$^{-1}$) and exceeds the turbulent velocities predicted by the models of the magnetorotational instability of \citet{KKK}, which are rather of the order of the local sound speed. Part of these large values could stem from the fact that our model does not account for non-coherent electron scattering that could enhance the wings of the emission lines. Also our description of the distance dependence of the extra velocity is likely to be too simplistic. Nevertheless, what our results clearly show is that extra broadening, whatever its physical origin, is required to fit the observed line profiles.

\section{Discussion \label{discuss}}
\subsection{Variability of the optical spectrum \label{variabopt}}
The H$\alpha$ profile of HD~45314 exhibits an unusually strong range of activity, although this is  not unique for a classical Be star.  
Our observations indicate that the disk of HD~45314 underwent a shell episode in autumn 2014 and nearly disappeared in January 2016. Digging into the literature, we find that shell episodes or disk dissipation might have happened before in this star. \citet{CopelandHeard} noted that the spectrum of HD~45314 displayed double-peaked emission lines on their observations taken between 1939 and 1961, whilst their 1962 spectrum did not display emissions. \citet{AVD} presented a spectrum taken in February 1981 showing a very weak H$\alpha$ emission for which they determined EW(H$\alpha$) = $-7.3$\,\AA. Finally, from a spectrum taken on 19 October 1981, \citet{Andrillat} reported a weak (EW = $-4.7$\,\AA), broad ($\Delta\,v = 333$\,km\,s$^{-1}$), and symmetrical (V/R = 1) H$\alpha$ emission with a central absorption  and a He\,{\sc i} $\lambda$\,6678 line that was possibly in absorption. 
This latter description is very much reminiscent of the shell state observed in 2014. 

\subsubsection{Disk dissipation and build-up}
The Keplerian disks of Be stars can dissipate and build-up again over timescales of months to years. Stars of the $\gamma$~Cas type are no exceptions here. Indeed, the H$\alpha$ emission of $\gamma$\,Cas disappeared in the years 1942-1946 after a brief period of spectacular variations \citep[][and references therein]{Doazan,Hummel,gamCasrev}. 

\citet{Kee} theoretically studied the effects of line-driven ablation of Be and Oe disks. These authors show that the disks are removed from the inner edge outwards. Hence, the emission lines should decay most rapidly in their outer wings where the emission comes mostly from the highest orbital velocities, that is,\ from the innermost parts of the disk. \citet{Kee} infer disk destruction times on the order of months to years for B2 stars, but find much shorter destruction times for late O-type stars, such as HD~45314, where the strong radiation field should evacuate the disk material within a few days. Our data from the 2015-2016 epoch suggest a timescale of about two months (see Fig.\,\ref{montageHaprofile}) for the fading of the disk emission, somewhat longer than expected from these theoretical models. The spectra taken between October 2015 and January 2016 reveal a somewhat narrower emission than at other epochs. Since the larger radial velocities correspond to material closer to the star, this behaviour might reflect a (partial) dissipation of the inner disk that precedes the decay of the outer regions in line with the scenario of \citet{Kee}. We note, however, that the outermost parts of the wings (radial velocities beyond 300\,km\,s$^{-1}$) remain essentially unchanged. This relative constancy could indicate that the extra broadening mechanism responsible for these broad wings survives during the partial disk dissipation.   

Several observations suggest a strong connection between Be disk outbursts and stellar pulsations. Strong evidence for such a connection was obtained from {\it CoRoT} observations of the B0.5\,IVe star HD~49330 \citep{Huat}. These authors noted a correlation between amplitude changes and the presence or absence of certain pulsation frequencies, on the one hand, and the different phases of a small Be outburst on the other hand. In the B2\,Vnpe star $\mu$~Cen, cyclical outbursts were found to coincide in time with constructive interferences of two stellar pulsation modes \citep{muCen}. Further evidence comes from {\it BRITE} (BRIght Target Explorer) photometry that revealed a complex interplay between different pulsation frequencies \citep{Baade}. Hydrodynamic calculations by \citet{Kee2} support the scenario that the combination of rapid rotation and the dissipation of pulsational energy in the stellar atmosphere can eject material into the circumstellar disk. However, as pointed out by \citet{Huat}, the reverse could also be the case: the occurrence of the outburst itself could excite pulsation modes. In our case, we have no direct evidence for the presence of such pulsations, except for the erratic RV variations of the absorption lines reported in Sect.\,\ref{deltaRV}, and the low-significance line profile variability of the He~{\sc ii} $\lambda$~4686 absorption line. 

\subsubsection{The shell phase \label{shell}}
Transitions from a normal Be phase to a shell Be phase have been observed in several other objects including $\nu$~Gem \citep{Silaj14b}, HD~120678 \citep{Gamen}, Pleione, 59~Cyg, and $\gamma$~Cas \citep[see][and references therein]{PR}. In the conventional picture of Be stars that host a viscous decretion disk located in the plane of the star's equator, the difference between `normal' Be stars and shell stars is attributed to a difference in the inclination of the rotational axis. As a result, transitions between both morphologies are not expected in this simple model. \citet{Hummel} accordingly proposed that the changes in morphology of the emission lines of $\gamma$~Cas' optical spectrum in the first half of the twentieth century would be due to a Be disk temporarily inclined with respect to the stellar equator and slowly precessing around the star's rotation axis. \citet{Hummel} favoured the action of a secondary star on a non-coplanar eccentric orbit as the cause of the tilt of the disk. Yet, as an alternative to the binary scenario, \citet{Porter} proposed that radiatively-induced warps may develop in Be disks if they are optically thick at the wavelength at which the disk emission is maximum. The typical timescales of the growing warp modes are on the order of days, but \citet{Porter} cautioned that the observable signature of such features might rather develop on timescales of months. The expected precession rates of the warps are on the order of months to years. 

The appearance and disappearance of multiple peaks in the H$\alpha$ line profiles such as observed for HD~45314 in December 2015 is nowadays interpreted as a manifestation of a warped disk \citep{Okazaki16}. The most likely explanation of the spectacular variations that we reported in the previous sections is thus a warped disk probably combined with, or as a result of, several (partially failed) mass ejection events occuring over a range of stellar latitudes. Evidence that mass ejection does not always occur in the equatorial plane was reported for instance by \citet{deltaSco} during the disk build-up observed in $\delta$~Sco and by \citet{Huat} for HD~49330.

Material ejected from different latitudes might form a transient ring around the star that would not be in the equatorial plane, but would progressively settle down to the equator under the action of viscous forces. Mass ejection outside the plane of the stellar equator would thus temporarily alter the mean plane of the circumstellar material, giving the impression of a change in apparent inclination. Such situations would lead to a complex disk geometry, possibly also including the presence of multiple, nested, disk structures \citep{Clark}. There are also indications that during disk build-up, part of the ejected mass might actually fall back onto the star whilst other parts are advected outwards by viscous forces \citep[e.g.\ the case of the B2\,IVne star $\lambda$~Eri,][]{lamEri}. A combination of these processes might explain the strange behaviour exhibited by HD~45314 over the last few years.

\subsection{Constraints on the origin of the $\gamma$~Cas phenomenon in HD~45314}
Different scenarios are discussed in the literature to explain the $\gamma$~Cas phenomenon. Some of them involve the presence of a compact companion. Indeed, $\gamma$\,Cas itself is a binary with a period of 203.55\,days \citep{Mirosh,Smith12}. The accreting neutron-star companion scenario, originally proposed by \citet{White} was subsequently excluded when it became clear that the X-ray luminosity was much lower than that of typical Be/X-ray binaries, and when no evidence for non-thermal X-ray emission \citep[including at very high energies, e.g.][]{Shrader,Postnov2} or X-ray pulsations was found. The alternative possibility of accretion onto a white dwarf companion \citep{Murakami} appears at odds with the observed correlation between the variations of the UV and X-ray emission of $\gamma$~Cas \citep{Smith17}. 

The neutron star (NS) scenario was, however, revived by \citet{Postnov1} who suggest that the rapidly rotating magnetic neutron star companion might be in a propeller stage. In this scenario, the centrifugal barrier created by the neutron star's rigidly co-rotating magnetosphere can prevent the matter captured by the neutron star's gravitational field from reaching its surface. According to \citet{Postnov1}, this material will emit thermal X-rays via bremsstrahlung. 

Recently, \citet{Smith17} presented a series of arguments against this propeller scenario. Based on evolutionary considerations, these authors find that the propeller phase is too short-lived to account for the observed number of $\gamma$~Cas stars. Another problem comes from the observed correlations between the X-ray flux of $\gamma$~Cas and the cyclical variations of its $V$-band flux. The $V$-band flux likely reflects the conditions of the inner Be disk \citep{Robinson,Motch}.\footnote{\citet{Robinson} and \citet{Smith12} further noted that there exists no correlation between the X-ray flux variations of $\gamma$~Cas and the orbital phase of its companion.} This and other correlations and anti-correlations between various UV spectral features and the X-ray flux rather suggest a direct link of the X-ray emission to the Be star \citep{Smith17}. Most recently, \citet{NRC} showed that $\pi$~Aqr, which is known to have a non-degenerate binary companion, is a $\gamma$~Cas star. This finding is at odds with the accreting compact companion scenario as there is no room for such a compact companion between the Be disk and the non-degenerate companion.   

An alternative to the compact companion scenario was proposed by \citet{Smith98} and \citet{Robinson}. Based on the magnetorotational instability (MRI) of accretion disks predicted by \citet{BH} and \citet{HB}, \citet{Robinson} suggested that localized magnetic fields emanating from the star trigger a shear instability and turbulent motions within the Be disk. 
This MRI results in a positive feedback in which the turbulence amplifies the disk's magnetic field, which in turn increases the level of turbulence. The stretching and reconnection of the magnetic field lines lead to particle acceleration and plasma heating.

\citet{gamCasrev} emphasize the role of temporal correlations of the X-ray properties and optical and UV diagnostics in searching to understand the causes of the $\gamma$~Cas phenomenon. Long-term variations of the X-ray flux of $\gamma$~Cas amount to no more than tens of percent over a few years and X-ray and optical/UV emissions are correlated without a significant delay \citep[see][]{gamCasrev}. The variations observed in the case of HD~45314 are much larger and can thus provide even more sensitive diagnostics for the origin of the $\gamma$~Cas phenomenon.\\ 

In the compact companion scenario, we need to keep in mind that the compact companion orbits the Be or Oe star at a distance which is (most of the time) much larger than the radius of the decretion disk. In fact, \citet{ON} and \citet{Okazaki+02} showed that the Be disk in a Be/X-ray binary with a low eccentricity is resonantly truncated well inside the radius of the Roche lobe at periastron. \citet{Okazaki+02} further showed that the potential created by the neutron star can lead to an eccentric mode in the Be disk. This mode might help some fraction of the disk material to cross the gap between the outer edge of the truncated disk and the first Lagrangian point. For more eccentric systems, the disk radius could be as large as the Roche lobe radius at periastron \citep{ON}. In such situations, however, one expects that the compact companion efficiently accretes material from the disk around periastron passage triggering periodic Type-I X-ray outbursts where the X-ray luminosity reaches $10^{36}$ -- $10^{37}$\,erg\,s$^{-1}$. Such outbursts have not been observed for $\gamma$~Cas stars \citep{gamCasrev}. Instead, prior to our observation of the low state of HD~45314, $\gamma$~Cas stars were usually observed at a relatively constant level of $L_X \simeq 10^{32}$ -- $10^{33}$\,erg\,s$^{-1}$. The lack of such outbursts would then suggest a compact companion on a wide, nearly circular, orbit. This sort of configuration applies to persistent Be/X-ray binaries, such as X~Per. The latter objects feature a relatively constant X-ray luminosity of $L_X \simeq 10^{34}$ -- $10^{35}$\,erg\,s$^{-1}$ , which is interpreted as the neutron star accreting material from the Be or Oe-wind via the Bondi-Hoyle-Littleton mechanism \citep{Reig}. 

\citet{Postnov1} argue that $\gamma$~Cas features a neutron star companion in the propeller regime. Such a situation occurs if the corotation radius of the neutron star 
$$R_c = \left[G\,M_{\rm NS}\,\left(\frac{P_{\rm rot, NS}}{2\,\pi}\right)^2\right]^{1/3}$$ 
is smaller than the neutron star's Alfv\`en radius $R_A$. The gravitationally captured material then accumulates in a roughly spherical shell extending from $R_A$ to the Bondi radius \citep{BonHoy},
$$R_G = 2\,\frac{G\,M_{\rm NS}}{v_{\rm rel}^2}.$$ 
This shell should power a gravitational luminosity 
$$L_{\rm X} \simeq \frac{G\,M_{\rm NS}\,\dot{M}_C}{R_A},$$
where 
$$\dot{M}_C = \pi\,\rho\,v_{\rm rel}\,R_G^2$$
is the rate of material captured by the Bondi-Hoyle-Littleton mechanism. The latter directly scales with the wind density $\rho$ at the position of the neutron star. The X-ray luminosity would thus be modulated by the density of the flow from the Be star. Since the Be disk should be well inside the Roche lobe of the Be or Oe star, the flow captured by the neutron star's gravitational field should consist of a mix of material from the outer parts of the disk and from the Be or Oe wind. Material coming from the outer parts of the disk reaches the vicinity of the neutron star with a velocity of a few 10\,km\,s$^{-1}$ comparable to the speed of sound \citep{ON}. The corresponding Bondi radius is quite large, on the order of several thousand R$_{\odot}$, that is,\ exceeding the radius of the neutron star's Roche lobe ($\sim 50$\,R$_{\odot}$ for a period near 100\,days) by a large factor. 
In comparison, for the polar wind of the Oe star, the flow velocity is on the order of 1000\,km\,s$^{-1}$, thus leading to a much smaller Bondi radius. Therefore, we do not expect the polar wind to contribute significantly to the flow of material that reaches the neutron star. As a result, one could expect $\dot{M}_C$ and thus $L_{\rm X}$ to scale with the mass of the Be disk. However, in the propeller scenario, the response of the X-ray emission would be delayed with respect to the change in the disk properties. This delay includes the free-fall time into the neutron star potential well \citep{Postnov1}, but also the time it takes for the disk material to drift from the outer edge of the disk to the L$_1$ point \citep{Smith17}. \cite{Motch} showed that in $\gamma$~Cas, the delay between X-ray and $V$-band flux variations, if any, is less than one month. On the other hand, for Be + NS systems (including X~Per) delays of typically a few years have been observed between an outburst of the Be star and the ensuing increase of the X-ray emission due to the increased amount of material available for accretion \citep{Motch}. Our observations of HD~45314 clearly show that decreases in X-ray and optical emissions occur concurrently, with delays of less than a month,\ well below the empirically determined delays in accreting Be + NS systems. 

In the propeller scenario, the X-ray plasma temperature is mainly set by the gravitational potential of the compact companion ($\frac{G\,M_{\rm NS}}{R_A}$). Whilst the numerator of this expression remains constant when the accretion rate changes, the Alfv\`en radius has a mild dependence on $\dot{M}_C$. \citet{Postnov1} show that this results in a dependence of $kT$ on $L_{\rm X}^{2/15}$ (their Equation 10). Between the high and low emission states, the observed X-ray flux of HD~45314 changed by a factor 0.11. We would thus expect a reduction of the plasma temperature by a factor 0.75. The observed reduction of $kT_h$ is larger (a factor 0.5 or 0.3 compared to the 2-T or 1-T fit of the high state spectrum, respectively).
In view of the above considerations, it seems unlikely that accretion by a compact companion can explain the X-ray emission of HD~45314.\\
 
Although the model for the X-ray emission from magnetically-torqued disks of Be or Oe stars proposed by \citet{Li} was not explicitly designed for $\gamma$~Cas stars, let us briefly consider this model. In this scenario, the X-ray emission arises from the collision between the winds from the upper and lower stellar hemisphere and the dense equatorial disk. The corresponding plasma temperature would reach $kT = 1.24\,v_8^2$, where $kT$ is given in keV and $v_8$ is the wind velocity expressed in 1000\,km\,s$^{-1}$ \citep{Li}. To reach $kT \simeq 15$\,keV with this scenario would require a wind velocity of 3500\,km\,s$^{-1}$ , which seems unlikely for such stars which instead have polar wind velocities of $\sim 1000$\,km\,s$^{-1}$ \citep{PR}. Moreover, the magnetically-torqued disk model predicts $\frac{ L_{\rm X}}{L_{\rm bol}}$ values that are usually well below the one observed for HD~45314 and $\gamma$~Cas stars. Approaching $\log{\frac{ L_{\rm X}}{L_{\rm bol}}} = -6.1$ would require both a fast rotation and a very strong magnetic field (several kG). However, strong large-scale dipolar magnetic fields do not exist in Be stars. Such fields, if they did exist, would efficiently spin down the fast rotation of the Be star. Moreover, spectropolarimetric observations of a sample of 85 Be stars yielded no evidence for strong, organized stellar magnetic fields \citep{Wade}. From the upper limits on the field strengths, \citet{Wade} found that 80\% of their sample are consistent with a field strength of less than 105\,G. Furthermore, hydrodynamical simulations of \citet{Asif} indicate that large-scale organized stellar magnetic fields with a polar field strength of $\geq 100$\,G lead to the rapid disruption of a Keplerian circumstellar decretion disk. Even dipolar field strengths of 10\,G affect the disk structure though the disk is not totally disrupted in this case. For a field of 1\,G, \citet{Asif} found only some minor perturbations of the disk structure. Such weak fields cannot be directly detected with current spectropolarimetric instrumentations, but are clearly not sufficient to explain the observed X-ray luminosity by means of the mechanism proposed by \citet{Li}.

Finally, let us turn to the magnetic star-disk interaction scenario proposed by \citet{Robinson}. These authors actually predict that once the Be disk dissipates, the X-ray emission should fade as well. This is precisely what we have observed. Contrary to the \citet{Li} scenario, the lack of strong, large-scale magnetic fields with a strength of more than 100\,G in Be stars is not a major issue for the \citet{Robinson} scenario. Actually, \citet{BH} pointed out that weak magnetic seed fields are the most efficient in triggering the disk instability. Moreover, the Be stars might feature local magnetic fields, rather than strong, global fields. Such localized magnetic fields could stem from a sub-surface convective zone caused by a peak in the opacity associated with iron-group elements \citep{Cantiello}. These fields would reach the stellar surface as star spots, possibly triggering the variability of some massive stars \citep[e.g.\ $\lambda$~Cep, see][]{SH}. The expected magnetic field strength decreases along the main sequence and for a 20\,M$_{\odot}$ main-sequence star, \citet{Cantiello} predict minimum surface field strengths of about 10\,G. In the case of Be stars, the rapid rotation leads to rotational flattening, which in turn implies that the star is cooler near the equator than at higher stellar latitudes. For the rapidly rotating A7 star Altair, \citet{Robrade} argued that the gravity darkening is sufficient to cool the equatorial regions to the typical temperature of an early F-type star. This situation then leads to the formation of an equatorial corona. In our case, gravity darkening is not sufficient to lower the equatorial temperature by a sufficiently large amount to produce a local corona. Yet, a reduction of the equatorial temperature might impact on the sub-surface convection zone, reducing the expected field strength to values that would be sufficient to trigger the \citet{BH} mechanism while simultaneously preserving the presence of a stable decretion disk. In the future, it would be most interesting to build a self-consistent model of the magnetic star-disk interaction scenario, including magneto-hydrodynamical calculations.  

\section{Conclusion \label{conclusion}}
Our X-ray observations and optical monitoring of the $\gamma$~Cas analog HD~45314 have shed new light on the behaviour of this star and on the origin of the $\gamma$~Cas phenomenon as a whole. In 2012--2013, the circumstellar disk of HD~45314 entered a phase of spectacular variations including a shell phase and a phase of near-dissipation. The X-ray data that were taken when the disk was slowly rebuilding revealed an unprecedented change in the X-ray behaviour of a $\gamma$~Cas star: the X-ray flux was reduced by an order of magnitude compared to when the star was at its nominal level of circumstellar disk emission. These results point towards a direct link between the $\gamma$~Cas phenomenon and the density or mass of the decretion disk, making a magnetic star-disk interaction the most likely scenario. 

\begin{acknowledgement}
G.R.\ and Y.N.\ acknowledge support through an ARC grant for Concerted Research Actions, financed by the French Community of Belgium (Wallonia-Brussels Federation), from the Fonds de la Recherche Scientifique (FRS/FNRS), as well as through an XMM PRODEX contract (Belspo). The TIGRE facility is funded and operated by the universities of Hamburg, Guanajuato, and Li\`ege. S.V.Zh.\ and A.S.M.\ acknowledge support from DGAPA/PAPIIT Project IN100617. This research has made use of the SIMBAD database, operated at CDS, Strasbourg, France. We further acknowledge with thanks the variable star observations from the AAVSO International Database contributed by observers worldwide and used in this research.
\end{acknowledgement}

\appendix
\section{Revised fits of the H$\alpha$ line of HD~60848 \label{appendix}}
\begin{figure*}
\begin{minipage}{4.5cm}
\resizebox{4.5cm}{!}{\includegraphics{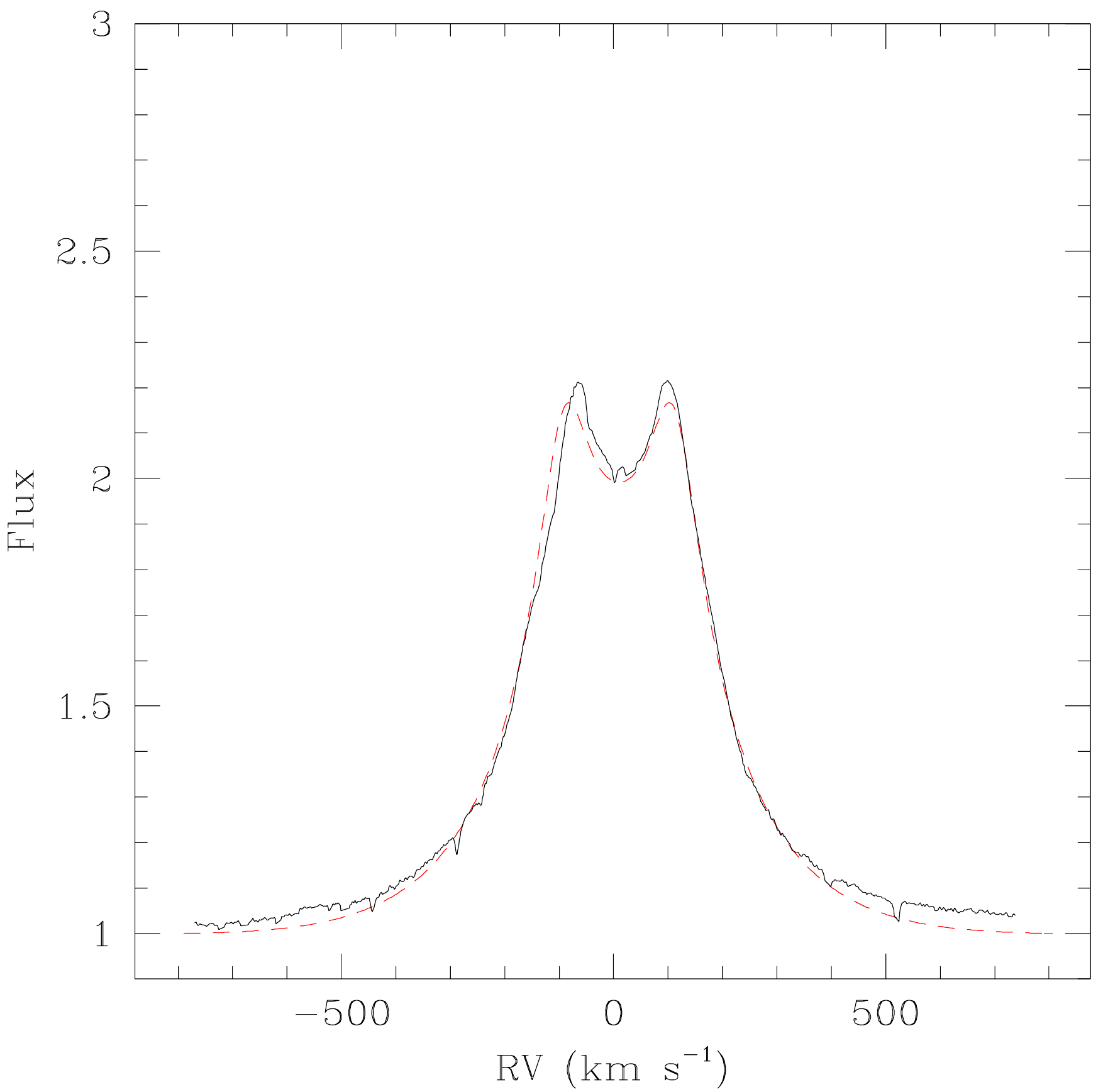}}
\end{minipage}
\hfill
\begin{minipage}{4.5cm}
\resizebox{4.5cm}{!}{\includegraphics{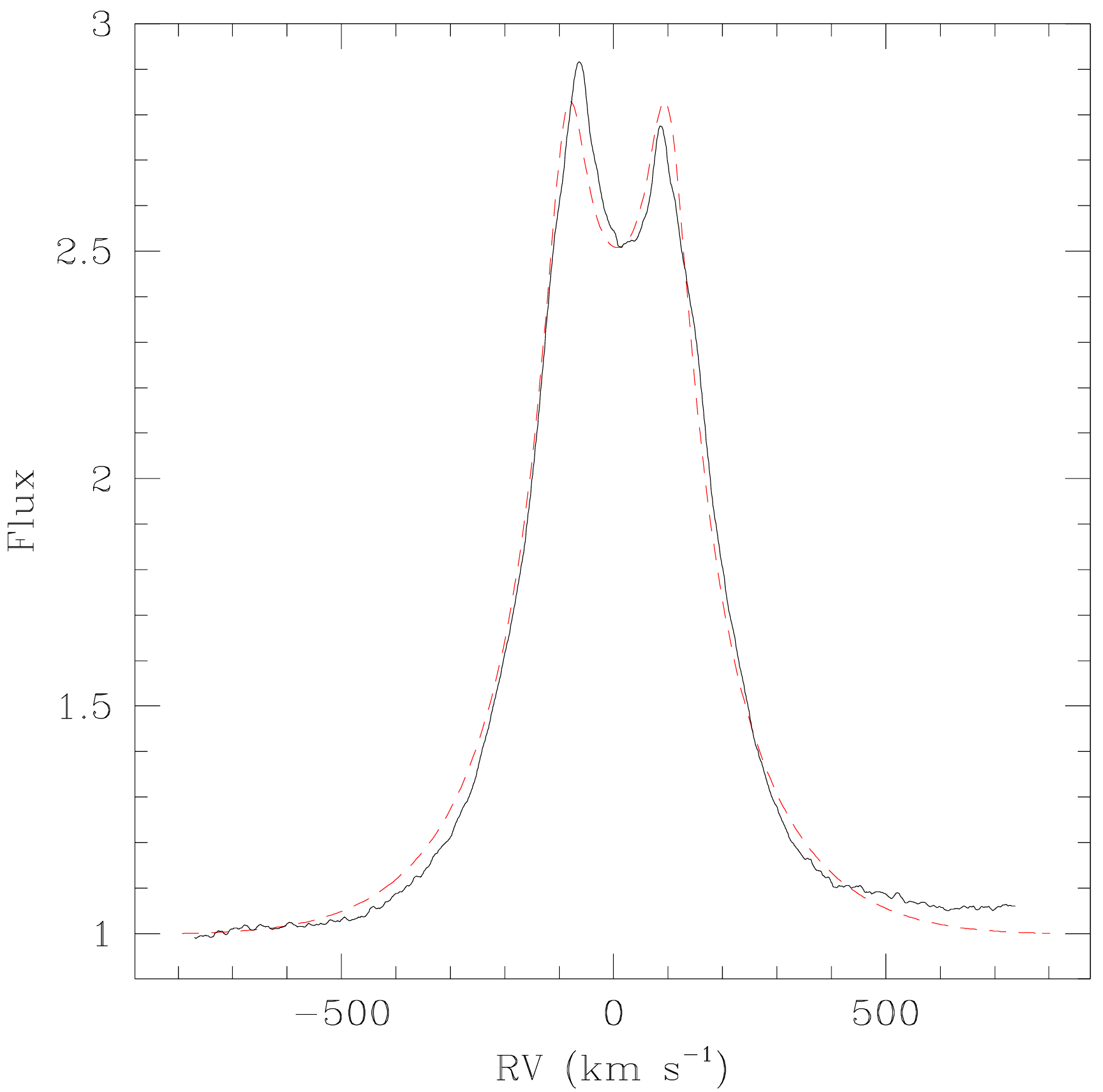}}
\end{minipage}
\hfill
\begin{minipage}{4.5cm}
\resizebox{4.5cm}{!}{\includegraphics{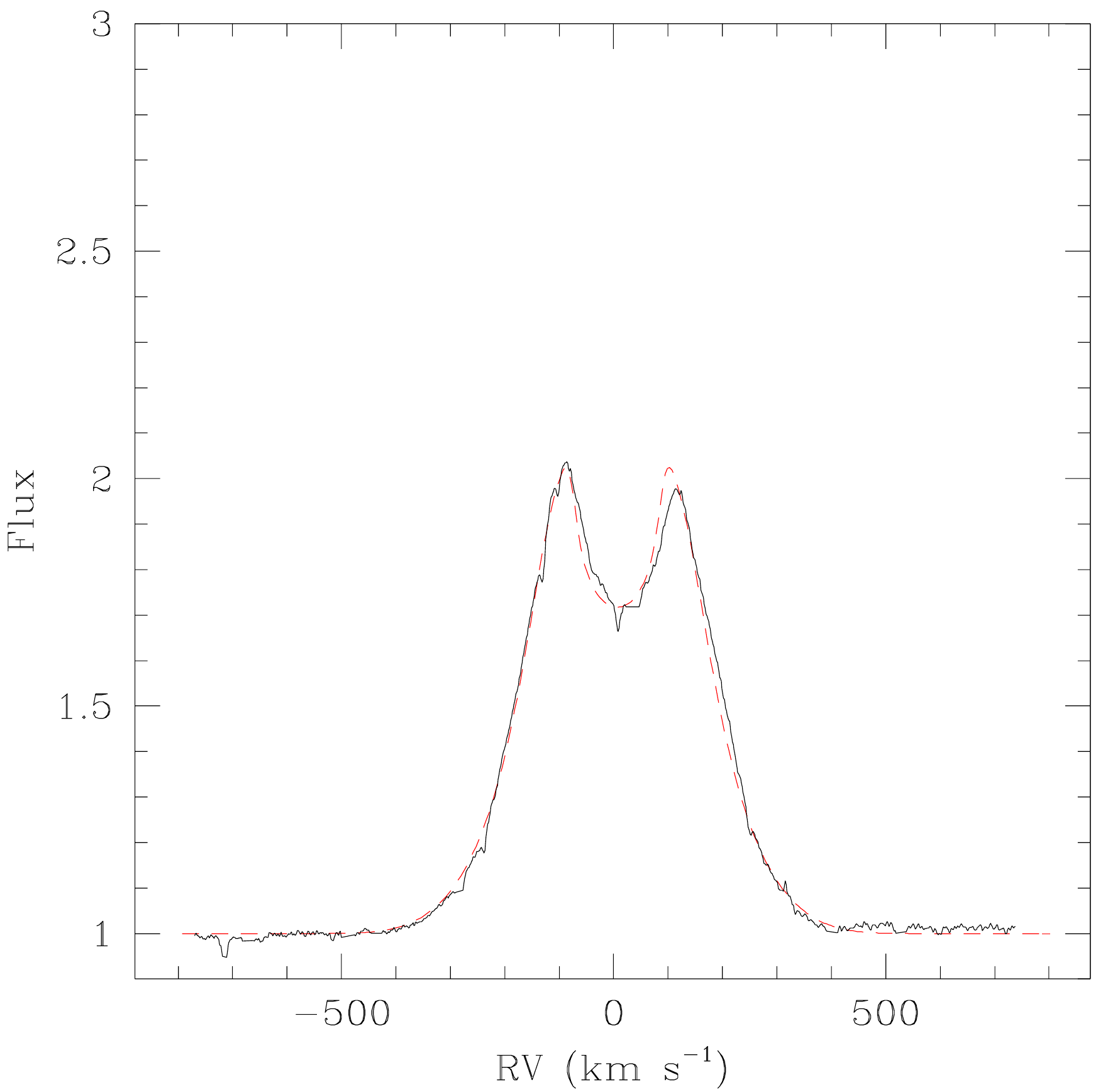}}
\end{minipage}
\hfill
\begin{minipage}{4.5cm}
\resizebox{4.5cm}{!}{\includegraphics{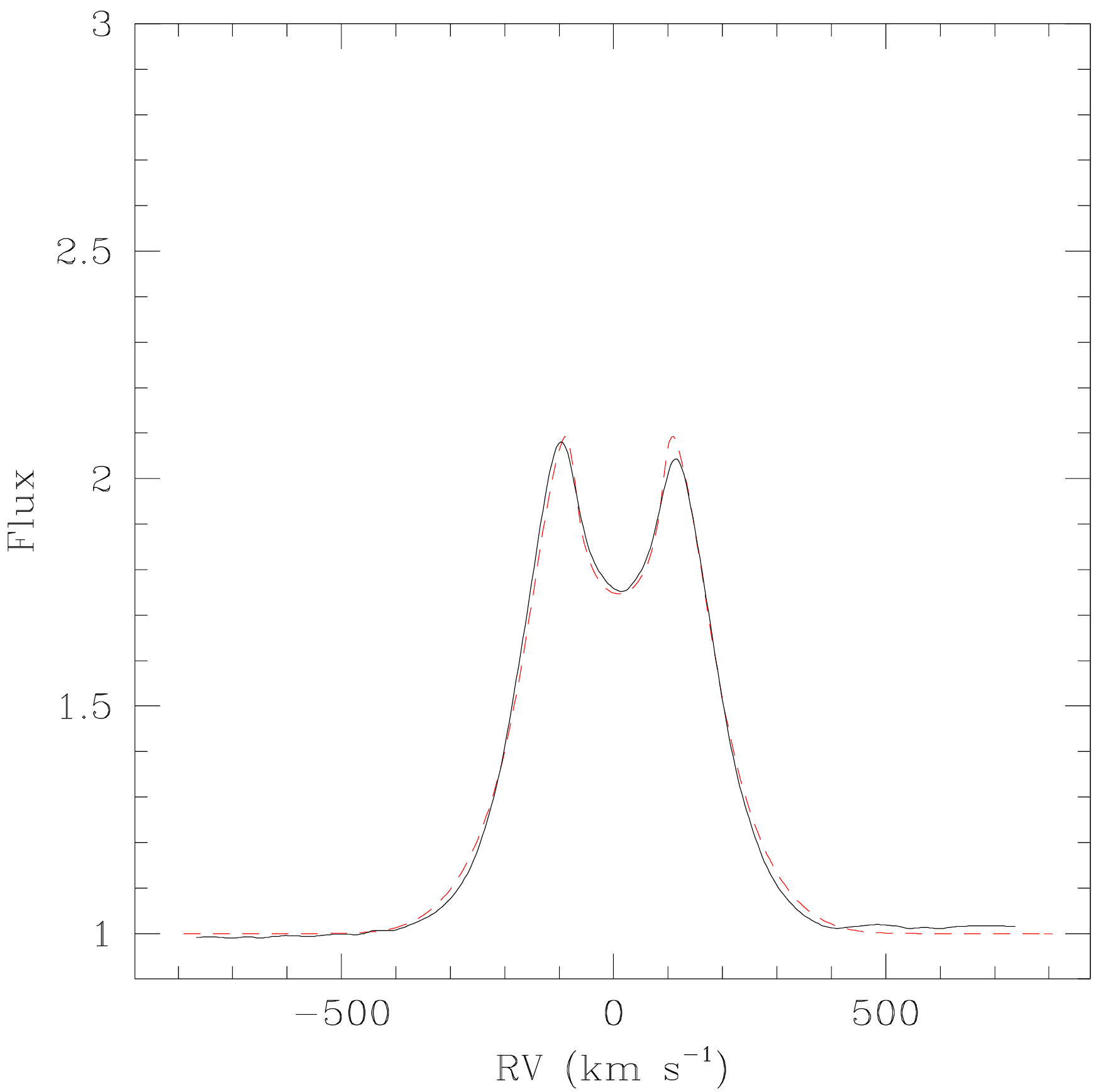}}
\end{minipage}
\caption{Observed (black solid line) and synthetic (red dashed line) H$\alpha$ profiles of HD~60848 for different epochs. The panels illustrate (from left to right) the spectra of May 1999, May 2000, March 2002, observing season 2013--14. \label{montage60848}}
\end{figure*}

We have taken advantage of the more extensive grid of Oe disk models that we have computed here to revise the fitting parameters of the Oe star HD~60848 studied in \citet{Oepaper}. The resulting new best-fit parameters are quoted in Table\,\ref{bestfit60848} and are illustrated in Fig.\,\ref{montage60848}. From these fits, we see that the disk inclination is consistently found to be close to $29^{\circ}$, in agreement with our conclusion in \citet{Oepaper}. 

\begin{table}[h]
\caption{Model parameters of the fits of the H$\alpha$ line of HD~60848 in Fig.\,\ref{montage60848}.\label{bestfit60848}}
\tiny
\begin{tabular}{c c c c c c}
\hline
Epoch & $i_{\rm disk}$& $R_{\rm disk}$ & $n_0$ & $\alpha$ & $v_{\rm extra,0}$ \\
      & $(^{\circ})$ & $R_*$ & (cm$^{-3}$) &    & (km\,s$^{-1}$) \\
\hline
May 1999   & $30.2 \pm 5.1$ & $29.0 \pm 14.7$ & $677 \pm 123$ & $3.3 \pm 0.3$ & $239 \pm 20$ \\
May 2000   & $29.6 \pm 4.3$ & $27.4 \pm 12.5$ & $869 \pm 49$ & $3.4 \pm 0.2$ & $220 \pm 37$ \\
Mar.\ 2002 & $28.7 \pm 4.2$ & $23.9 \pm 13.8$ & $664 \pm 185$ & $3.0 \pm 0.4$ & $92 \pm 59$ \\
2013--14 & $26.2 \pm 4.5$ & $12.4 \pm 9.8$ & $589 \pm 185$ & $2.8 \pm 0.5$ & $100 \pm 65$ \\
\hline
\end{tabular}
\tablefoot{The quoted numbers are the means and standard deviations of the parameters of the models that have a $\chi^2$ within the 90\% confidence range from the best-fit model. Values for which no uncertainties are provided correspond to situations where all models in the 90\% confidence range had the same value as the parameter.} 
\end{table}

\end{document}